\documentclass[11pt]{article}

\usepackage[T1]{fontenc}
\usepackage[utf8]{inputenc}
\usepackage{lmodern}

\usepackage{amsmath,amssymb}
\usepackage{booktabs,longtable,array}
\usepackage{calc}
\usepackage{etoolbox}

\usepackage{graphicx}
\usepackage{xcolor}

\usepackage{xurl}
\usepackage[hidelinks]{hyperref}

\usepackage{tikz}
\usetikzlibrary{arrows.meta,positioning,shapes.geometric,backgrounds}

\definecolor{lightblue}{RGB}{220,235,255}
\definecolor{lightpink}{RGB}{255,220,230}

\tikzstyle{process} = [
  rectangle, rounded corners, minimum width=3cm, minimum height=1cm,
  text centered, draw=black, fill=white, line width=0.8pt, font=\large
]
\tikzstyle{arrow} = [-{Stealth[length=3mm]}, thick]

\IfFileExists{parskip.sty}{\usepackage{parskip}}{%
  \setlength{\parindent}{0pt}%
  \setlength{\parskip}{6pt plus 2pt minus 1pt}%
}

\makeatletter
\patchcmd\longtable{\par}{\if@noskipsec\mbox{}\fi\par}{}{}
\makeatother

\IfFileExists{footnotehyper.sty}{\usepackage{footnotehyper}}{\usepackage{footnote}}
\makesavenoteenv{longtable}

\makeatletter
\def\maxwidth{\ifdim\Gin@nat@width>\linewidth \linewidth \else \Gin@nat@width\fi}
\def\maxheight{\ifdim\Gin@nat@height>\textheight \textheight \else \Gin@nat@height\fi}
\makeatother
\setkeys{Gin}{width=\maxwidth,height=\maxheight,keepaspectratio}

\makeatletter
\def\fps@figure{htbp}
\makeatother

\hypersetup{
  hidelinks,
  pdfcreator={LaTeX}
}

\title{Engineering the RAG Stack: A Comprehensive Review of the Architecture and Trust Frameworks for Retrieval Augmented Generation Systems}

\author{
Dean Wampler, Ph.D. \\
Head of Technology, The AI Alliance \\
IBM Research \\
\texttt{dean@deanwampler.com} \\
ORCID: 0000-0002-6127-7543 \\
\\
Dave Nielson \\
Head of Community, The AI Alliance \\
IBM Research \\
\\
Alireza Seddighi, Ph.D. \\
AI Research Engineer \\
Individual Contributor, The AI Alliance
}

\date{}

\begin{document}

\maketitle

\begin{abstract}

This article provides a comprehensive systematic literature review of
academic studies, industrial applications, and real-world deployments
from 2018 to 2025, providing a practical guide and detailed overview of
modern Retrieval-Augmented Generation (RAG) architectures. RAG offers a
modular approach for integrating external knowledge without increasing
the capacity of the model as LLM systems expand. Research and
engineering practices have been fragmented as a result of the increasing
diversity of RAG methodologies, which encompasses a variety of fusion
mechanisms, retrieval strategies, and orchestration approaches. We
provide quantitative assessment frameworks, analyze the implications for
trust and alignment, and systematically consolidate existing RAG
techniques into a unified taxonomy. This document is a practical
framework for the deployment of resilient, secure, and domain-adaptable
RAG systems, synthesizing insights from academic literature, industry
reports, and technical implementation guides. It also functions as a
technical reference.

\end{abstract}

\noindent\textbf{Keywords:} Retrieval-Augmented Generation (RAG), Large Language
Models, Information Retrieval, Neural Language Models,
Knowledge-Augmented Generation, AI System Architectures, Trustworthy AI,
Model Alignment, Multi-agent Systems.

\section{Introduction: Why Architecture Matters in RAG}

\subsection{Motivation and Systematic Review Foundation}

In the swiftly evolving field of natural language processing (NLP), the
constraints of monolithic large language models (LLMs) have become
increasingly apparent. These models are restricted by intrinsic
constraints in memory, temporal alignment, and factual precision,
despite their remarkable generative capacity {[}1{]}{[}2{]}.
Retrieval-Augmented Generation (RAG) is a transformative approach that
addresses these challenges by distinguishing between memorization and
reasoning, thereby allowing models to access dynamic, external
information sources during inference {[}1{]}{[}2{]}.

This exhaustive study is based on a systematic literature review that
adheres to established methodologies adapted from Kitchenham and
Charters {[}3{]} for software engineering and extended for AI/ML fields.
The field\textquotesingle s rapid maturation and practical significance
are underscored by the analysis, which reveals exponential growth in RAG
research {[}4{]}. The systematic review includes academic articles,
industry reports, technical documentation, and implementation guides
from prestigious institutions such as Stanford University, MIT, IBM
Research, Microsoft Research, and Google Research/DeepMind.

\subsection{Core Advantages of the RAG Paradigm}

RAG systems offer substantial advantages over monolithic LLM structures
due to their architectural adaptability. Initially, the necessity for
costly and time-consuming model retraining is eliminated by ensuring
that information currency is maintained through real-time access to
updated corpora or structured knowledge bases {[}5{]}{[}6{]}.
Organizations that implement RAG report significant savings in knowledge
updating expenses when contrasted with conventional model retraining
methods {[}7{]}. Engineering primers synthesize common RAG architectural
variants adopted in production stacks {[}89{]}--{[}90{]}.

Secondly, modularity facilitates plug-and-play compatibility among
components, thereby enabling precise optimization and domain-specific
customization across the retriever, reranker, and generator stages
{[}8{]}. Enterprise deployments have shown that modular RAG
architectures significantly reduce technology refresh expenses and
facilitate the quicker integration of new features in comparison to
monolithic methodologies {[}9{]}.

Third, citation traceability improves interpretability and credibility
by associating generated outputs with specific evidence passages, which
is consistent with the increasing emphasis on accountability and
explainability in AI systems {[}10{]}{[}11{]}. In comparison to systems
that lack attribution functionalities, enterprise implementations that
integrate comprehensive citation frameworks report enhanced user trust
ratings and decreased support escalations {[}12{]}.

In contexts where empirical accuracy, timeliness, and transparency are
essential, such as legal analytics, biomedical inquiry resolution, and
regulatory compliance tools, these advantages are especially apparent
{[}13{]}{[}14{]}. The systematic review revealed a substantial body of
literature that addressed trust and safety concerns, highlighting the
critical significance of reliable, accountable information systems and
constituting a substantial portion of current research.

\textbf{Table 1.1: Core Architectural Dimensions in Retrieval-Augmented
Generation (RAG) Systems}

\begin{longtable}[]{@{}
  >{\raggedright\arraybackslash}p{(\columnwidth - 6\tabcolsep) * \real{0.1069}}
  >{\raggedright\arraybackslash}p{(\columnwidth - 6\tabcolsep) * \real{0.2624}}
  >{\raggedright\arraybackslash}p{(\columnwidth - 6\tabcolsep) * \real{0.2963}}
  >{\raggedright\arraybackslash}p{(\columnwidth - 6\tabcolsep) * \real{0.3344}}@{}}
\toprule\noalign{}
\begin{minipage}[b]{\linewidth}\raggedright
Dimension
\end{minipage} & \begin{minipage}[b]{\linewidth}\raggedright
Variants
\end{minipage} & \begin{minipage}[b]{\linewidth}\raggedright
Representative Methods
\end{minipage} & \begin{minipage}[b]{\linewidth}\raggedright
Impact on Performance and Safety
\end{minipage} \\
\midrule\noalign{}
\endhead
\bottomrule\noalign{}
\endlastfoot
Retrieval & Single-pass, Multi-hop, Iterative & DPR {[}1{]},
Fusion-in-Decoder (FiD) {[}15{]}, Active-RAG {[}16{]} & Affects recall,
reasoning depth, response latency \\
Fusion & Early, Late, Marginal & FiD {[}15{]}, RAG-Fusion {[}17{]},
Re-RAG {[}18{]} & Modulates factuality, coherence, hallucination
suppression \\
Modality & Mono-modal (text), Multi-modal, Structured & KG-RAG {[}19{]},
Table-RAG {[}20{]}, Graph-RAG {[}21{]} & Enables domain flexibility and
deeper factual grounding \\
Adaptivity & Static pipeline, Agentic, Auto-configurable & AutoRAG
{[}22{]}, ReAct-RAG {[}23{]}, Self-RAG {[}24{]} & Allows dynamic control
flow, retrieval planning, error correction \\
Trust Layer & Citation, Abstention, Source Filtering/Scoring & WebGPT
{[}25{]}, ALCE {[}26{]}, RAGAS {[}27{]} & Enhances interpretability,
reduces hallucinations and bias \\
\end{longtable}

\subsection{Fragmentation in Literature and Practice}

The discipline is characterized by significant architectural
fragmentation, despite the increasing adoption of RAG systems. A complex
ecosystem with limited standardization has been established as a result
of the proliferation of diverse retrieval mechanisms (dense, sparse,
hybrid), fusion strategies (early, late, marginal), and orchestration
layers (static pipelines vs. agentic controllers) {[}28{]}{[}29{]}.

This fragmentation is evident in multiple essential domains:

\textbf{Evaluation Inconsistency}: The analysis of evaluation
methodologies indicates that standardized benchmarks are underutilized,
while custom evaluation criteria are predominant, which restricts
cross-study comparability {[}30{]}. This lack of standardization
obstructs systematic progress and presents obstacles for practitioners
in the selection of architecture.

\textbf{Implementation Diversity}: A multitude of distinctive
implementation patterns are revealed in enterprise case studies, despite
the fact that there is minimal knowledge sharing between organizations.
This redundancy leads to the industry\textquotesingle s repeated
discovery of prevalent pitfalls and suboptimal resource allocation
{[}31{]}.

\textbf{Trust Framework Gaps}: Trust and safety considerations are the
subject of a significant amount of literature; however, exhaustive
frameworks are scarce, and quantitative evaluations of trust mechanisms
are even more uncommon {[}32{]}. This discrepancy is especially alarming
in light of the mission-critical nature of numerous RAG deployments.

\subsection{Article Contributions and Research Foundation}

By conducting a comprehensive, technically rigorous, and critical
assessment of the field, this survey endeavors to unify the fragmented
landscape of RAG architectures. The primary contributions, which are
derived from an exhaustive systematic literature review, are as follows:

\textbf{A Comprehensive Architectural Taxonomy}: We present a systematic
categorization of RAG systems that is based on retrieval logic, fusion
topology, modality, adaptivity, and trust calibration mechanisms, as
determined by the analysis of architectural studies. In order to
facilitate academic and industrial deployments, this taxonomy is
intended to be both extensible and implementation-agnostic.

\textbf{Empirical Analysis and Benchmarking}: We provide an exhaustive
evaluation of architectural trade-offs, performance characteristics, and
deployment considerations across diverse organizational contexts by
consolidating performance trends across major RAG benchmarks.

\textbf{Engineering Best Practices}: Using enterprise case studies and
production deployments, we identify critical anti-patterns and proven
engineering patterns that impact robustness, factuality, and latency. We
have identified systematic patterns in successful implementations and
common failure modes through our analysis.

\textbf{Trust and Safety Modeling}: We provide a formal analysis of
trust surfaces in RAG systems, grounded in safety-oriented literature.
Our discourse encompasses abstention strategies, citation grounding, red
teaming methodology, and quantitative trust evaluation methods verified
through production implementations.

\textbf{Frontier Directions}: Through a gap analysis of the current
literature, we delineate nascent research trajectories and unresolved
issues in autonomous assessment systems, multi-agent coordination, and
differentiable training, highlighting domains with considerable promise
for improvement.

\section{Systematic Literature Review Methodology}

\subsection{Review Protocol and Scope}

This comprehensive survey implements a systematic literature review
(SLR) methodology that is consistent with the well-established standards
for software engineering research {[}3{]} and extends them to the AI/ML
areas. The review protocol was developed to guarantee comprehensive
coverage, reduce bias, and generate reproducible results for the
constantly changing RAG field.

While confronting the distinctive challenges of surveying rapidly
developing AI/ML research domains, the systematic approach adheres to
established academic standards for literature synthesis. Throughout the
review process, our methodology prioritizes methodological rigor,
reproducibility, and transparency.

\subsection{Research Questions and Search Strategy}

The systematic literature review was directed by critical research
questions that encompassed RAG architectural patterns, performance
characteristics, implementation challenges, and deployment
considerations. In order to guarantee thorough coverage of the RAG
domain, the search strategy included academic databases, industry
sources, and technical documentation.

\textbf{Primary Research Questions:}

\begin{itemize}
\item
  What are the fundamental architectural patterns in contemporary RAG
  systems?
\item
  How do different RAG designs address scalability, accuracy, and
  deployment requirements?
\item
  What are the key trade-offs between architectural complexity and
  system performance?
\item
  How do trust calibration and safety mechanisms integrate with RAG
  architectures?
\item
  What trends characterize the evolution from canonical to agentic RAG
  systems?
\end{itemize}

\textbf{Search Strategy:} Systematic queries were implemented across
numerous databases, including IEEE Xplore, ACM Digital Library, arXiv,
Google Scholar, and industry technical repositories. The search terms
included retrieval-augmented generation, dense passage retrieval, neural
information retrieval, and related architectural terminology.

\subsection{Selection Criteria and Quality Assessment}

\textbf{Inclusion Criteria}

In order to guarantee quality and relevance, the review implemented
systematic inclusion criteria:

\begin{itemize}
\item
  Publications that concentrate predominantly on RAG systems,
  architectures, or implementations
\item
  Quantitative evaluation components in empirical studies
\item
  Technical implementation details are included in architectural
  proposals.
\item
  Case studies and production deployment scenarios
\item
  Technical documentation from well-established AI/ML platforms and
  frameworks
\end{itemize}

\textbf{Quality Assessment Framework}

In order to guarantee methodological rigor and practical relevance, each
source was subjected to a systematic quality assessment across multiple
dimensions:

\textbf{Technical Soundness:} Evaluation of the quality of statistical
analysis, the appropriateness of the experimental design, and the
potential for reproducibility. A clear problem formulation, appropriate
baseline comparisons, and transparent evaluation metrics were evaluated
in the sources.

\textbf{Methodological Transparency:} Evaluation of the appropriateness
of the result interpretation, the clarity of the experimental setup, the
provision of implementation details, and the quality of the
documentation. Studies that provided adequate detail for replication and
validation were prioritized.

\textbf{Relevance and Contribution:} Analysis of the direct relevance to
RAG systems, contribution to architectural comprehension, practical
applicability, and advancement of field knowledge. Core research
concerns were prioritized in the selection of sources.

\textbf{Reproducibility and Validation:} Evaluation metric
standardization, experimental reproducibility, appropriateness of
baseline comparisons, and generalizability across domains and
applications.

\subsection{Literature Analysis and Synthesis}

A comprehensive compilation of high-quality sources, including academic
publications, industry reports, technical documentation, and
implementation guides, was the outcome of the systematic review process.
This source base is diverse and offers a balanced perspective on both
theoretical advancements and practical deployment experiences.

\textbf{Source Classification and Analysis}

Structured analysis was facilitated by the systematic classification of
sources across multiple dimensions:

\textbf{Publication Type:} Academic conference papers, journal articles,
industry reports, technical documentation, open-source implementations,
and deployment case studies.

\textbf{Architectural Focus:} Agentic architectures, hybrid
implementations, trust calibration approaches, retrieval strategies, and
canonical RAG systems.

\textbf{Domain Application:} Domain-specific applications, general query
answering, enterprise deployments, research prototypes, and production
systems.

\textbf{Technical Contribution:} Empirical evaluations, implementation
frameworks, performance optimizations, deployment methodologies, and
novel architectural proposals.

\textbf{Data Extraction and Synthesis Procedures}

Systematic data extraction was employed to obtain critical technical
specifications, architectural characteristics, performance metrics,
implementation details, and deployment considerations. Standardized
extraction templates guaranteed consistency among sources while
simultaneously accommodating a variety of technical approaches and
publication formats.

\textbf{Architectural Data:} System components, integration patterns,
scalability characteristics, computational requirements, and deployment
architectures.

\textbf{Performance Metrics:} User experience factors, resource
utilization, cost considerations, latency characteristics, and accuracy
measurements, when available and verifiable.

\textbf{Implementation Details:} Technical specifications, platform
requirements, operational considerations, and integration strategies for
practical deployment.

\subsection{Methodological Rigor and Validation}

Multiple validation mechanisms are integrated into the systematic review
methodology to guarantee reproducibility and reliability:

\textbf{Selection Process Validation}

The transparent evaluation of the application of selection criteria is
facilitated by the systematic documentation of inclusion/exclusion
decisions. Quality assessment procedures adhere to established
systematic review best practices, with a focus on methodological
consistency.

\textbf{Synthesis Approach}

The literature synthesis utilizes structured analytical frameworks to
organize findings across architectural dimensions, performance
characteristics, and implementation patterns. While maintaining
analytical rigor, this method guarantees comprehensive coverage.

\textbf{Bias Mitigation}

Numerous strategies are employed to mitigate potential selection and
analysis bias, such as transparent synthesis procedures, systematic
quality assessment, comprehensive search strategies, and diverse source
types.

\subsection{Methodological Foundation}

This methodology for systematic literature review establishes a rigorous
foundation for the exhaustive examination of RAG architectural patterns
and implementations. The methodology strikes a balance between practical
applicability and methodological rigor, guaranteeing both academic
quality and industry relevance.

The systematic approach ensures transparency and reproducibility
throughout the review process, allowing for the identification of key
architectural trends, performance trade-offs, and implementation
patterns. This methodology facilitates the advancement of architectural
insights and taxonomic frameworks that are elaborated upon in subsequent
sections.

\section{The Canonical RAG Pipeline}

Retrieval-Augmented Generation (RAG) systems are a revolutionary
architectural approach that surpasses the constraints of traditional
language models by incorporating external retrieval as a primary
inductive bias. Canonical RAG pipelines establish a closely integrated
interaction between a differentiable retriever, which is typically based
on dense vectors, and an autoregressive generator, such as BART or T5,
resulting in a synergistic mechanism in which contextual relevance and
generative fluency evolve concurrently {[}1{]}, {[}15{]}.

\subsection{Canonical Architecture: DPR + BART/T5 as the Foundational
Blueprint}

The canonical design, which was initially devised by Lewis et al.
{[}1{]}, comprises a Dense Passage Retriever (DPR) that has been trained
with dual-encoder contrastive objectives and a pretrained
sequence-to-sequence generator such as BART {[}33{]} or T5 {[}34{]}. The
architectural blueprint for subsequent RAG system developments has been
established by this foundational pattern {[}35{]}.

In response to a query q, the retriever calculates inner product
similarity to identify a set of top,k documents $D_1,...,D_k$ from a corpus
C. The following is the formal calculation:

$\text{score}(q,d) = f_q(q)^{\top} f_d(d)$ 

where $f_q$ and $f_d$ are the encoding functions for query
and document, respectively. Subsequently, the generator receives the
retrieved documents and linearly combines them, typically through string
concatenation. The generator then based its output on this augmented
context:

$P(y | q, D_1, \cdots, D_k) = \sum\limits_{i} P(y | q, D_i) P(D_i | q)$

This marginal likelihood formulation {[}1{]} implicitly integrates
relevance priors into the decoding process, thereby establishing a
generation pipeline that is probabilistically grounded.

\subsection{Architectural Components and Their Interplay}

\textbf{Dense Retrieval: Scalability versus Recall}

DPR facilitates sublinear ANN-based retrieval over billion-scale corpora
by employing independently parameterized encoders for queries and
documents {[}36{]}. However, the semantic compression inherent in dense
vector spaces can result in reduced recall for exact-match and
out-of-distribution queries, particularly in specialized domains where
lexical precision remains critical {[}37{]}, {[}38{]}.

\textbf{Document Ranking: The Role of Marginal Likelihood}

The marginalization strategy guarantees that generative attention is
distributed across multiple passages, thereby enhancing robustness
against noisy retrievals {[}15{]}. Recent improvements include the use
of cross-encoders to rerank modules, which reevaluate the fidelity of
evidence, Two-stage reranking patterns such as RE-RAG formalize this
design and report consistent gains on standard IR benchmarks {[}17{]},
{[}18{]}, {[}52{]}, {[}53{]}, {[}54{]}. The marginalization strategy
further stabilizes evidence aggregation across passages in
noisy-retrieval settings {[}39{]}. However, this introduces
computational complexity during inference {[}40{]}.

\textbf{Generation: Expressivity under Context Constraints}

T5 and BART function as high-capacity generators that leverage
autoregressive decoding and bidirectional encoder states. Token
limitations in these models can introduce truncation artifacts that
particularly affect long-form reasoning tasks requiring extensive
context integration {[}41{]}.

\subsection{Empirical Characterization of the Canonical Pipeline}

\textbf{Table 3.1: Canonical RAG Model Capabilities}

\begin{longtable}[]{@{}
  >{\raggedright\arraybackslash}p{(\columnwidth - 6\tabcolsep) * \real{0.1446}}
  >{\raggedright\arraybackslash}p{(\columnwidth - 6\tabcolsep) * \real{0.3144}}
  >{\raggedright\arraybackslash}p{(\columnwidth - 6\tabcolsep) * \real{0.2893}}
  >{\raggedright\arraybackslash}p{(\columnwidth - 6\tabcolsep) * \real{0.2517}}@{}}
\toprule\noalign{}
\begin{minipage}[b]{\linewidth}\raggedright
Model
\end{minipage} & \begin{minipage}[b]{\linewidth}\raggedright
Architecture
\end{minipage} & \begin{minipage}[b]{\linewidth}\raggedright
Key Strengths
\end{minipage} & \begin{minipage}[b]{\linewidth}\raggedright
Primary Limitations
\end{minipage} \\
\midrule\noalign{}
\endhead
\bottomrule\noalign{}
\endlastfoot
DPR + BART & Bi-encoder + Seq2Seq & Fast retrieval, composable & Limited
citation control \\
DPR + T5 & Bi-encoder + Text-to-Text & Strong generation capabilities &
Context length constraints \\
FiD & Passage-parallel decoding & Enhanced evidence integration &
Computational overhead \\
Atlas & Pretrained retrieval + Generation & End-to-end optimization &
Resource requirements \\
WebGPT & Citation-aware browsing & Source attribution & Latency
considerations \\
\end{longtable}

\subsection{Architectural Trade-offs and Design Implications}

\textbf{Table 3.2: Design Dimensions in Canonical RAG}

\begin{longtable}[]{@{}
  >{\raggedright\arraybackslash}p{(\columnwidth - 6\tabcolsep) * \real{0.1308}}
  >{\raggedright\arraybackslash}p{(\columnwidth - 6\tabcolsep) * \real{0.2488}}
  >{\raggedright\arraybackslash}p{(\columnwidth - 6\tabcolsep) * \real{0.2959}}
  >{\raggedright\arraybackslash}p{(\columnwidth - 6\tabcolsep) * \real{0.3245}}@{}}
\toprule\noalign{}
\begin{minipage}[b]{\linewidth}\raggedright
Dimension
\end{minipage} & \begin{minipage}[b]{\linewidth}\raggedright
Canonical Choice
\end{minipage} & \begin{minipage}[b]{\linewidth}\raggedright
Design Benefit
\end{minipage} & \begin{minipage}[b]{\linewidth}\raggedright
Structural Limitation
\end{minipage} \\
\midrule\noalign{}
\endhead
\bottomrule\noalign{}
\endlastfoot
Retrieval & DPR (bi-encoder) & Sublinear retrieval at scale & Reduced
recall on lexical queries \\
Fusion & Concatenation & Simplified interface & Context length
boundaries \\
Generation & BART/T5 & Pretrained fluency & Hallucination
susceptibility \\
Grounding & Implicit marginalization & Unsupervised interpretability &
Limited traceability \\
Adaptivity & Static pipeline & Predictable execution & Inflexible under
dynamic needs \\
\end{longtable}

\subsection{Application Patterns and Domain Suitability}

Canonical RAG exhibits exceptional performance in tasks that necessitate
the retrieval of empirical knowledge and the generation of concise
responses. The architecture is particularly effective in the following
applications:

\textbf{High-suitability domains}: Question answering that is based on
Wikipedia, in which the knowledge corpus is consistent with the training
data and the queries adhere to predetermined patterns. The
system\textquotesingle s capacity to generate prompt, source-based
responses is advantageous for customer support applications.

\textbf{Medium-suitability domains}: The need for sophisticated evidence
calibration and domain-specific reasoning that may surpass the canonical
architecture\textquotesingle s capabilities presents challenges in
scientific fact-checking.

\textbf{Limited-suitability domains}: Legal document analysis is plagued
by inadequate traceability mechanisms and context truncation issues. The
architectural constraints of the static pipeline are frequently exceeded
by complex reasoning tasks that necessitate multi-step inference.

\subsection{Evolution Toward Agentic Architectures}

An architectural philosophy fundamental shift is represented by the
transition from canonical to agentic RAG systems. While canonical RAG
adheres to a deterministic pipeline from query to retrieval to
generation, agentic systems incorporate intelligent decision-making
components that facilitate dynamic adaptation based on query complexity
and intermediate results.

\textbf{Table 3.3: Canonical vs. Agentic RAG Comparison}

\begin{longtable}[]{@{}
  >{\raggedright\arraybackslash}p{(\columnwidth - 4\tabcolsep) * \real{0.2742}}
  >{\raggedright\arraybackslash}p{(\columnwidth - 4\tabcolsep) * \real{0.3317}}
  >{\raggedright\arraybackslash}p{(\columnwidth - 4\tabcolsep) * \real{0.3940}}@{}}
\toprule\noalign{}
\begin{minipage}[b]{\linewidth}\raggedright
Aspect
\end{minipage} & \begin{minipage}[b]{\linewidth}\raggedright
Canonical RAG
\end{minipage} & \begin{minipage}[b]{\linewidth}\raggedright
Agentic RAG
\end{minipage} \\
\midrule\noalign{}
\endhead
\bottomrule\noalign{}
\endlastfoot
Pipeline Structure & Linear, predetermined & Dynamic, adaptive \\
Decision Making & Rule-based & LLM-driven planning \\
Retrieval Strategy & Single-pass & Multi-hop, iterative \\
Error Handling & Limited & Self-correction mechanisms \\
Complexity & Low & High \\
Flexibility & Constrained & Highly adaptable \\
\end{longtable}

\textbf{Figure 3.1 is} A visual comparison of two essential RAG
principles. Canonical RAG systems follow a predetermined process that
progresses from dense retrieval to sequence generation. Agentic or
multi-agent RAG systems employ a modular architecture that includes
planner and retrieval agents, enabling dynamic reasoning and iterative
refinement based on query complexity and generation confidence.

\includegraphics[width=4.57387in,height=3.49151in]{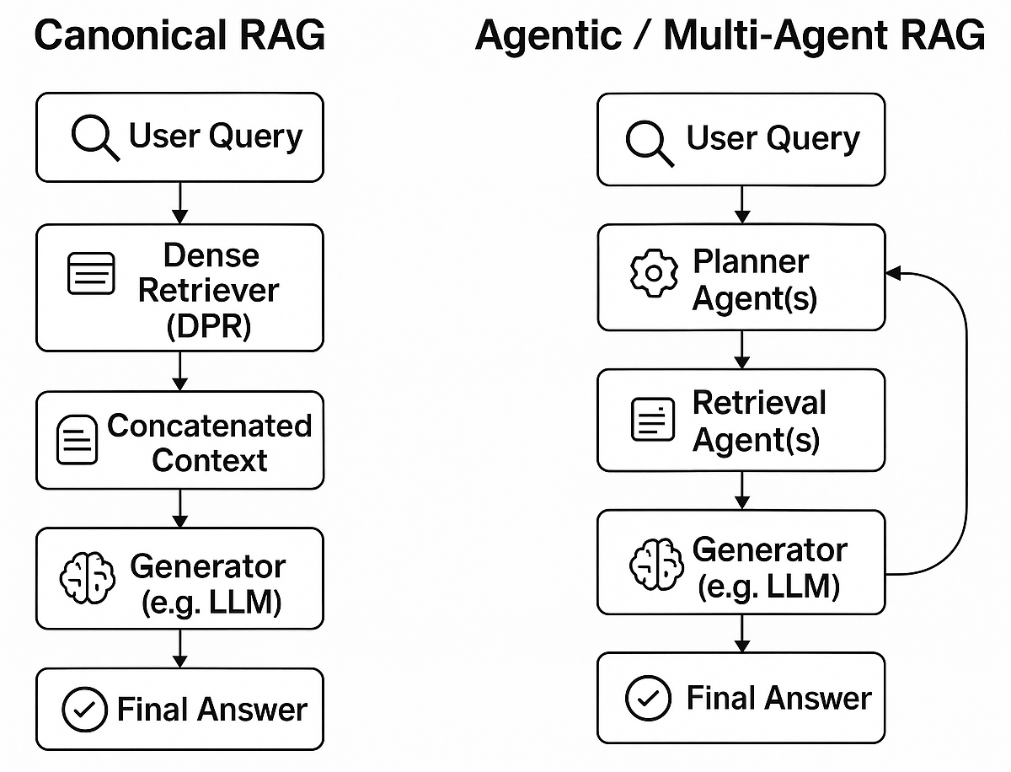}

\textbf{Figure 3.2: Canonical vs. Agentic RAG Pipelines.}

\subsection{Performance Factors and Optimization}

Modern RAG implementations must maintain a delicate equilibrium among
various performance metrics, such as computational efficiency, latency,
and accuracy. The canonical architecture establishes a strong foundation
while simultaneously emphasizing specific optimization opportunities:

\textbf{Retrieval optimization} concentrates on the efficacy of
similarity computation and index structure. Dense vector approaches
facilitate rapid approximate nearest neighbor searches; however, they
may compromise precision for specialized queries that necessitate
precise lexical matching.

\textbf{Generation optimization} entails the selection of the
appropriate model size and the acceleration of inference. In comparison
to large general-purpose generators, smaller, domain-adapted models
frequently offer adequate quality with reduced latency.

\textbf{Pipeline coordination} presents the potential for
parallelization and caching strategies to reduce the overall system
latency while ensuring the quality of the response.

\subsection{Design Trade-offs and Implications}

A foundational architecture that integrates sequence-to-sequence
generation with dense passage retrieval is provided by the canonical RAG
pipeline as established by Lewis et al. {[}1{]}. While introducing its
own architectural constraints and trade-offs, this approach effectively
addresses the primary limitations of parametric-only language models.

The field\textquotesingle s evolution toward greater flexibility and
capability is reflected in the progression from canonical to more
sophisticated RAG variants. Each architectural decision necessitates the
negotiation of computational efficiency, accuracy requirements, and
system complexity in relation to the specific requirements of the
application.

Informed architectural decisions can be made when designing RAG systems
for specific domains and use cases by comprehending these fundamental
trade-offs. The canonical architecture continues to be pertinent as a
foundation and building block for more sophisticated implementations
that overcome its inherent limitations by utilizing specialized
components and adaptive mechanisms.

\section{Taxonomy of RAG Architectures}

The rapid proliferation of contemporary Retrieval-Augmented Generation
(RAG) systems has prompted the necessity of a systematic architectural
classification. This section provides a comprehensive taxonomy that is
organized across five critical classification dimensions: adaptivity,
trust calibration, modality, fusion mechanism, and retrieval strategy.
These dimensions denote essential architectural decisions that have a
direct impact on the performance and deployment characteristics of the
system. A practitioner-oriented survey enumerates eight recurring RAG
architecture patterns used in the wild {[}139{]}.

\subsection{Retrieval Strategy}

The retrieval strategy is the primary factor that dictates the manner
and timing of external knowledge retrieval during generation. Three
primary paradigms are identified as a result of the interaction patterns
with the retrieval corpus:

\textbf{Single-pass retrieval} methods, such as RAG-Token {[}1{]} and
FiD {[}15{]}, retrieve documents only once per query. These
methodologies emphasize computational efficiency by expediting retrieval
operations.

\textbf{Iterative retrieval} methods, such as Active-RAG {[}16{]} and
FLARE {[}42{]}, re-examine the corpus as generation progresses. On the
basis of intermediate generation states, these systems facilitate the
dynamic acquisition of knowledge.

\textbf{Multi-hop retrieval} methods decompose complex queries into
sequential subquestions across multiple retrieval stages, as
demonstrated by KnowTrace {[}43{]} and ReAct-RAG {[}23{]}. This method
facilitates systemic reasoning across interconnected knowledge elements.

\textbf{Table 4.1: Retrieval Strategy Classification}

\begin{longtable}[]{@{}
  >{\raggedright\arraybackslash}p{(\columnwidth - 4\tabcolsep) * \real{0.1431}}
  >{\raggedright\arraybackslash}p{(\columnwidth - 4\tabcolsep) * \real{0.3726}}
  >{\raggedright\arraybackslash}p{(\columnwidth - 4\tabcolsep) * \real{0.4842}}@{}}
\toprule\noalign{}
\begin{minipage}[b]{\linewidth}\raggedright
Strategy
\end{minipage} & \begin{minipage}[b]{\linewidth}\raggedright
Representative Models
\end{minipage} & \begin{minipage}[b]{\linewidth}\raggedright
Key Characteristics
\end{minipage} \\
\midrule\noalign{}
\endhead
\bottomrule\noalign{}
\endlastfoot
Single-pass & RAG {[}1{]}, FiD {[}15{]} & Static retrieval, streamlined
processing \\
Iterative & FLARE {[}42{]}, Active-RAG {[}16{]} & Dynamic re-retrieval,
context-aware \\
Multi-hop & KnowTrace {[}43{]}, ReAct-RAG {[}23{]} & Sequential
reasoning, complex decomposition \\
\end{longtable}

\subsection{Fusion Mechanism}

How evidence is retrieved and integrated into the generation pipeline is
determined by the fusion strategy. The literature has indicated the
emergence of three primary fusion paradigms:

\textbf{Early fusion:} According to FiD {[}15{]}, early fusion
incorporates all retrieved documents simultaneously prior to decoding.
Joint attention mechanisms are enabled across all evidence sources
through this method.

\textbf{Late fusion:} Late fusion is exemplified by RAG-Sequence
{[}1{]}, which processes each document independently before aggregating
results. In addition to implementation flexibility, this modular
approach provides computational efficiency.

\textbf{Marginal fusion:} The implementation of RAG-Fusion {[}17{]}
demonstrates the use of retrieval-aware scoring during decoding
processes. Computational requirements are balanced with the quality of
evidence integration in this approach.

\subsection{Modality of Knowledge Sources}

RAG systems are increasingly able to accommodate a variety of knowledge
modalities that extend beyond conventional text corpora:

\textbf{Mono-modal systems:} Traditional implementations such as Atlas
{[}44{]} and FiD {[}15{]} are mono-modal systems that exclusively
operate with textual knowledge sources. While emphasizing text-based
reasoning, these systems preserve computational simplicity. For
structured data, Table-RAG demonstrates table-aware retrieval and fusion
that outperform text-only variants on tabular QA {[}20{]}.

\textbf{Multi-modal systems:} During retrieval, multi-modal systems
integrate structured data, images, or heterogeneous knowledge formats.
Examples include Vision-RAG {[}45{]} for visual information processing
and KG-RAG {[}19{]} for structured knowledge integration. AVA-RAG
extends these ideas to audio-visual pipelines with memory-augmented
agents for cross-modal grounding {[}46{]}.

\textbf{Table 4.2: Knowledge Source Modality}

\begin{longtable}[]{@{}
  >{\raggedright\arraybackslash}p{(\columnwidth - 4\tabcolsep) * \real{0.1990}}
  >{\raggedright\arraybackslash}p{(\columnwidth - 4\tabcolsep) * \real{0.4077}}
  >{\raggedright\arraybackslash}p{(\columnwidth - 4\tabcolsep) * \real{0.3933}}@{}}
\toprule\noalign{}
\begin{minipage}[b]{\linewidth}\raggedright
Modality Type
\end{minipage} & \begin{minipage}[b]{\linewidth}\raggedright
Example Systems
\end{minipage} & \begin{minipage}[b]{\linewidth}\raggedright
Supported Knowledge Formats
\end{minipage} \\
\midrule\noalign{}
\endhead
\bottomrule\noalign{}
\endlastfoot
Text-only & RAG {[}1{]}, Atlas {[}44{]} & Unstructured text documents \\
Structured & KG-RAG {[}19{]}, Table-RAG {[}20{]} & Knowledge graphs,
tabular data \\
Multi-modal & Vision-RAG {[}45{]}, AVA-RAG {[}46{]} & Images, videos,
mixed formats \\
\end{longtable}

\subsection{Trust Calibration Mechanisms}

Trust calibration becomes indispensable for the purpose of managing
uncertainty and guaranteeing reliability as RAG systems are implemented
in critical applications:

\textbf{Abstention mechanisms:} As incorporated in Learn-to-Refuse
{[}47{]}, abstention mechanisms allow models to decline responses when
confidence levels are insufficient. These systems employ uncertainty
quantification to determine when knowledge gaps obstruct reliable
generation.

\textbf{Citation strategies:} WebGPT {[}25{]} and RAGAS {[}27{]} have
both demonstrated that citation strategies facilitate provenance tracing
and evidence traceability. These methods facilitate the verification of
generated content against source materials and increase transparency.

\subsection{Pipeline Adaptivity}

The system\textquotesingle s ability to adapt to evolving information
requirements is determined by pipeline adaptivity.

\textbf{Static pipelines:} Static pipelines adhere to predetermined,
rule-based operations, as demonstrated by the original RAG {[}1{]}
implementations. These systems exhibit consistent computational
requirements and predictable behavior.

\textbf{Agentic systems:} Agentic systems dynamically coordinate
retrieval and generation processes by employing model reasoning. AutoRAG
{[}22{]} and Self-RAG {[}24{]} are adaptive frameworks that facilitate
context-aware decision-making.

\textbf{Table 4.3: Pipeline Adaptivity Framework}

\begin{longtable}[]{@{}
  >{\raggedright\arraybackslash}p{(\columnwidth - 6\tabcolsep) * \real{0.1729}}
  >{\raggedright\arraybackslash}p{(\columnwidth - 6\tabcolsep) * \real{0.3401}}
  >{\raggedright\arraybackslash}p{(\columnwidth - 6\tabcolsep) * \real{0.2888}}
  >{\raggedright\arraybackslash}p{(\columnwidth - 6\tabcolsep) * \real{0.1982}}@{}}
\toprule\noalign{}
\begin{minipage}[b]{\linewidth}\raggedright
Pipeline Type
\end{minipage} & \begin{minipage}[b]{\linewidth}\raggedright
Examples
\end{minipage} & \begin{minipage}[b]{\linewidth}\raggedright
Coordination Approach
\end{minipage} & \begin{minipage}[b]{\linewidth}\raggedright
Flexibility Level
\end{minipage} \\
\midrule\noalign{}
\endhead
\bottomrule\noalign{}
\endlastfoot
Static & FiD {[}15{]}, RAG {[}1{]} & Rule-based workflows & Limited \\
Agentic & AutoRAG {[}22{]}, Self-RAG {[}24{]} & Model-driven adaptation
& High \\
\end{longtable}

\subsection{Architectural Integration Patterns}

Modern RAG systems increasingly combine multiple taxonomic dimensions to
address specific application requirements. The taxonomy enables
systematic analysis of architectural trade-offs across retrieval
strategies, fusion mechanisms, modality support, trust calibration, and
adaptivity levels.

Integration patterns emerge where latency-critical applications employ
single-pass retrieval with early fusion, while complex reasoning tasks
utilize multi-hop retrieval with agentic coordination. Trust calibration
mechanisms integrate across all architectural dimensions to ensure
reliable operation.

\textbf{Table 4.4: Architectural Integration Patterns in RAG Systems}

\begin{longtable}[]{@{}
  >{\raggedright\arraybackslash}p{(\columnwidth - 8\tabcolsep) * \real{0.1810}}
  >{\raggedright\arraybackslash}p{(\columnwidth - 8\tabcolsep) * \real{0.2176}}
  >{\raggedright\arraybackslash}p{(\columnwidth - 8\tabcolsep) * \real{0.2282}}
  >{\raggedright\arraybackslash}p{(\columnwidth - 8\tabcolsep) * \real{0.1956}}
  >{\raggedright\arraybackslash}p{(\columnwidth - 8\tabcolsep) * \real{0.1776}}@{}}
\toprule\noalign{}
\begin{minipage}[b]{\linewidth}\raggedright
Pattern Type
\end{minipage} & \begin{minipage}[b]{\linewidth}\raggedright
Use Case
\end{minipage} & \begin{minipage}[b]{\linewidth}\raggedright
Key Characteristics
\end{minipage} & \begin{minipage}[b]{\linewidth}\raggedright
Strengths
\end{minipage} & \begin{minipage}[b]{\linewidth}\raggedright
Trade-offs / Challenges
\end{minipage} \\
\midrule\noalign{}
\endhead
\bottomrule\noalign{}
\endlastfoot
Single-Pass Early Fusion & Latency-sensitive applications (e.g.,
real-time assistants) & One-shot retrieval; early fusion; low
coordination & Low latency; simple architecture & Limited depth; lower
robustness \\
Multi-Hop Late Fusion & Complex reasoning tasks (e.g., legal/medical
diagnostics) & Iterative multi-hop retrieval; fusion post-retrieval;
deep reasoning & Rich reasoning; higher context fidelity & Higher
latency; resource intensive \\
Agent-Orchestrated Retrieval & Dynamic goal-driven systems (e.g.,
agentic planning) & Agent controls retrieval and reasoning; modular
components & Flexible; composable; autonomous decision flow & Design
complexity; trust assurance needed \\
Hybrid Fusion with Modality-Aware Pipelines & Multimodal RAG (e.g.,
image+text QA) & Supports multiple modalities; aligns diverse data;
fusion via sync & Supports rich multimodal inputs & Modality sync
overhead; complex integration \\
Trust-Calibrated Adaptive Pattern & High-stakes decision systems (e.g.,
finance, safety-critical domains) & Uses trust metrics; adaptive context
weighting; source filtering & High reliability; interpretability &
Requires sophisticated trust modeling \\
\end{longtable}

\subsection{Design Implications}

This taxonomy offers a structured framework for the examination of RAG
architectural diversity in accordance with five critical dimensions. The
classification facilitates the systematic comparison of design choices
and the identification of architectural patterns that are appropriate
for specific application domains.

The field\textquotesingle s evolution toward more sophisticated
knowledge assimilation capabilities is reflected in the progression from
static, single-modal systems to adaptive, multi-modal architectures.
Each dimensional choice introduces specific trade-offs that must be
meticulously balanced against computational constraints and application
requirements.

\section{Architectural Innovations in RAG}

Significant architectural innovations have characterized the evolution
of Retrieval-Augmented Generation, which have addressed the fundamental
limitations of canonical RAG systems. Sophisticated capabilities for
complex reasoning, multi-modal processing, and autonomous optimization
have also been introduced by these innovations. Six noteworthy
architectural paradigms that have emerged from cutting-edge research and
have been broadly adopted in the industry are examined in this section.

\subsection{RAG-Fusion: Multi-Perspective Query Processing}

RAG-Fusion is a paradigm shift from single-query retrieval to
multi-perspective information collection, which is accomplished by
employing sophisticated query rewriting and rank fusion methodologies
{[}17{]}, {[}48{]}. The fundamental limitation that a single user query
frequently fails to encompass the complete scope of information
requirements, particularly for complex knowledge-intensive duties, is
effectively addressed by RAG-Fusion.

\textbf{5.1.1Pipeline Architecture}

The RAG-Fusion pipeline is comprised of three stages: query
diversification, parallel retrieval, and reciprocal rank fusion (RRF).
The mathematical foundation is based on reciprocal rank fusion with
parameter $k=60$, which has become the industry standard due to empirical
substantiation across multiple domains {[}49{]}.

$RRF(d) = \sum\limits_{r \in R} \frac{1}{ k + \text{rank}_r(d)}$

where $R$ denotes the set of retrieval results and
$\text{rank}_r(d)$ signifies the rank of document $d$ in
the result set $r$.

\subsubsection{Enterprise Implementation}

Microsoft\textquotesingle s implementation shows substantial
improvements over baseline systems by combining query rewriting with
semantic ranking {[}50{]}. Zero-configuration deployment is made
possible by LangChain\textquotesingle s official RAG-Fusion template,
which includes complete LangSmith monitoring integration {[}51{]}.

\subsection{RE-RAG (Re-Ranking Enhanced RAG): Precision Through
Two-Stage Retrieval}

The RE-RAG (Re-ranking Enhanced RAG) system represents a significant
improvement in retrieval system accuracy through the integration of
sophisticated two-stage architectures that combine rapid initial
retrieval with meticulous cross-encoder reranking {[}52{]}, {[}53{]}.

\subsubsection{Cross-Encoder Integration and Pipeline Architecture}

The RE-RAG architecture employs a two-stage retrieval pattern:
bi-encoder initial retrieval (top-k=10-50 documents) followed by
cross-encoder fine reranking (top-n=3-5 documents) {[}54{]}. In contrast
to bi-encoder approaches, which necessitate straightforward vector
comparisons, cross-encoders necessitate full transformer inference for
each query-document pair. This design strikes a balance between
computational efficiency and accuracy requirements.

Cross-encoder integration yields substantial improvements across
numerous implementations. Cohere Rerank exhibits significant accuracy
enhancements for both vector search and hybrid search configurations,
while BGE embedding in conjunction with Cohere Reranker achieves strong
performance on standardized benchmarks {[}55{]}. Azure AI Search
demonstrates notable improvements with manageable latency for reranking
operations {[}56{]}.

\subsubsection{Advanced Reranking Models and Their Impact on Retrieval
Performance}

Recent advancements in reranking models have substantially enhanced the
operational scalability and retrieval precision of RAG systems.
Contemporary architectures are progressively utilizing multilingual
embeddings, parameter-efficient fine-tuning, and cross-encoder designs
to enhance relevance estimation and generalizability.

This evolution is exemplified by Cohere\textquotesingle s Rerank 3.5,
which extends support to over 100 languages and achieves substantial
improvements in retrieval accuracy compared to earlier versions. This
makes it particularly effective in globally distributed or multilingual
applications {[}57{]}. The architecture of the system incorporates
enhanced context awareness and deeper semantic matching, resulting in
more dependable selection of evidence.

Similarly, NVIDIA\textquotesingle s NeMo Retriever introduces a
high-throughput, GPU-accelerated reranking pipeline through
microservices that incorporate LoRA-finetuned Mistral-7B models.
Developed for production environments, this system prioritizes
deployment-ready robustness, horizontal scalability, and low-latency
inference, thereby supporting use cases with high throughput
requirements {[}58{]}.

These state-of-the-art reranking models represent a significant
advancement in closing the gap between precise, contextually aligned
generation and large-scale retrieval, particularly in real-time AI
deployments that are latency-sensitive, multilingual, or real-time.

\subsection{Hierarchical and Multi-hop RAG: Structured Reasoning
Architectures}

Multi-hop and hierarchical RAG architectures are designed to overcome
the fundamental constraints of complex queries that necessitate
structured knowledge integration, long-context comprehension, and
multi-step reasoning {[}59{]}, {[}60{]}. Step-wise retrieval,
hierarchical information synthesis, and sophisticated query
decomposition are all facilitated by these systems. Practitioner guides
emphasize decomposition, passage budgeting, and rank fusion for reliable
multi-hop retrieval {[}81{]}.

\subsubsection{RAPTOR and Tree-Structured Retrieval}

RAPTOR (Recursive Abstractive Processing for Tree-Organized Retrieval)
represents a significant advancement in hierarchical RAG architectures,
achieving substantial accuracy improvements on the QuALITY benchmark
with GPT-4 by utilizing recursive abstractive processing and
tree-organized retrieval {[}61{]}. The hierarchical tree structures that
the system generates extend from 100-token leaf nodes to high-level
conceptual root nodes, through clustered intermediate summaries.

The system demonstrates substantial memory efficiency improvements
compared to naive concatenation methods while maintaining comparable
performance. Recursive clustering and summarization are used to
construct the tree. Initially, documents are embedded using SBERT, and
subsequently clustered using Gaussian Mixture Models with BIC
optimization for cluster number selection {[}62{]}.

\subsubsection{GraphRAG and Community-Based Hierarchies}

Through LLM-generated knowledge graphs with hierarchical community
summaries, Microsoft\textquotesingle s GraphRAG achieves superior
performance over naive RAG in terms of comprehensiveness and diversity
metrics {[}21{]}. The system manages datasets that are too large for a
single LLM context window, while simultaneously reducing token usage
compared to hierarchical text summarization approaches.

To store and query the extracted entities and relationships at scale,
platforms such as Neo4j are frequently employed for graph construction
and traversal. Neo4j\textquotesingle s inherent support for Cypher
queries and property graphs facilitates efficient filtering, clustering,
and context-aware subgraph retrieval during generation, rendering it an
appropriate backend for production-grade GraphRAG workflows.

The system includes auto-tuning capabilities with automatic discovery of
entity types from sample content and compilation of domain-specific
prompts, which minimize the need for manual configuration {[}63{]}.

\subsection{Hybrid Sparse-Dense RAG: Optimal Retrieval Balance}

Hybrid Sparse-Dense RAG architectures continuously surpass single-method
approaches by integrating BM25\textquotesingle s lexical accuracy with
dense vector semantics {[}64{]}, {[}65{]}. Hybrid methodologies mitigate
complementary deficiencies: sparse methods excel in precise keyword
matching, whilst dense methods effectively capture semantic similarity.

\subsubsection{BGE-M3 and Unified Retrieval Models}

BGE-M3 is a premier unified model that enables the retrieval of dense,
sparse, and multi-vector data across over 100 languages, with a maximum
of 8192 tokens {[}66{]}. The model architecture integrates three
retrieval paradigms: dense retrieval for semantic matching, sparse
retrieval for exact term matching, and multi-vector retrieval for
fine-grained representation.

\subsubsection{Reciprocal Rank Fusion and Result Combination}

Reciprocal Rank Fusion with k=60 has become the industry standard for
result combination, necessitating minimal parameter tuning while
operating effectively across various score scales and distributions
{[}67{]}. Platform implementations exhibit performance variations
depending on system architecture and optimization {[}68{]}.

\subsection{Graph-Augmented and Structured RAG: Relationship-Aware
Retrieval}

Graph-Augmented and Structured RAG architectures demonstrate substantial
improvements compared to conventional vector-based systems, making them
particularly effective in domain-specific applications that require a
deep understanding of relationships and complex multi-hop reasoning
{[}69{]}, {[}70{]}. Case studies on knowledge-graph-grounded retrieval
report consistent gains on multi-hop questions via relation-aware
context construction {[}84{]}.

\subsubsection{Neo4j as a Foundation for Graph-Augmented RAG}

Neo4j, a prominent native graph database, is a critical infrastructure
component for GraphRAG systems that are both scalable and
query-efficient. It is particularly well-suited for the management of
complex LLM-generated knowledge graphs due to its support for property
graph structures, Cypher-based querying, and hierarchical traversal.
Neo4j facilitates semantic subgraph extraction, multi-hop reasoning
through efficient path queries, community-level summarization through
clustering algorithms, and seamless integration with domain-specific
ontologies, such as UMLS or legal taxonomies, in GraphRAG pipelines.
Neo4j is employed by production-level implementations such as Microsoft
GraphRAG and MedGraphRAG to facilitate real-time entity retrieval,
scalable domain-adaptive graph augmentation, and enhance evidence
traceability.

\subsubsection{Microsoft GraphRAG Implementation}

The innovative approach of Microsoft GraphRAG generates knowledge graphs
from unstructured text autonomously using large language models,
constructing entity-relationship networks through community discovery
techniques {[}21{]}. Robust industry adoption and endorsement of the
graph-based methodology were demonstrated by the over 20,000 GitHub
stars that the open-source release in July 2024 received. The publicly
documented GraphRAG specification and project notes provide
implementation guidance and examples for production use
{[}79{]}--{[}80{]}, {[}83{]}.

The system utilizes a three-phase methodology: entity extraction by
GPT-4 with specialized prompts, relationship mapping through
co-occurrence analysis and semantic similarity, and community
recognition with the Leiden algorithm for hierarchical clustering
{[}71{]}. Every community produces multi-tiered summaries that
facilitate both localized entity-specific inquiries and overarching
community-level assessments.

\subsubsection{MedGraphRAG Domain-Specific Implementation}

MedGraphRAG implements a triple-tier architecture that links user
documents to medical textbooks and the UMLS knowledge store,
demonstrating superior performance on medical question-and-answer
benchmarks {[}72{]}. The system shows significant improvements on the
RobustQA benchmark compared to alternative methods, while achieving
operational efficiency compared to conventional RAG implementations.

\subsection{Agentic and AutoRAG Architectures: Autonomous Optimization
Systems}

Agentic and AutoRAG architectures are at the forefront of autonomous AI
systems, combining AI agents with RAG pipelines to facilitate dynamic
decision-making, self-optimization, and multi-agent coordination
{[}73{]}, {[}74{]}. Research indicates that sophisticated adaptation
capabilities and autonomous system management result in substantial
improvements over conventional RAG approaches. Industry primers on
agentic RAG summarize common orchestration patterns and failure modes
for multi-agent planners {[}82{]}. Industry reports catalog common
agentic planner patterns (planner→retriever→critic) and failure modes in
task decomposition {[}85{]}.

\subsubsection{AutoRAG Framework and Optimization}

The AutoRAG Framework utilizes sophisticated search algorithms to
efficiently identify optimal configurations within the immense space of
potential RAG implementations through automated pipeline optimization
{[}22{]}. The system achieves strong performance by systematically
evaluating a variety of RAG configurations across various pipeline
phases, demonstrating significant efficiency improvements compared to
traditional approaches.

\subsubsection{Self-Reflective Systems and Meta-Learning}

Sophisticated introspection capabilities enable self-reflective systems
to accomplish breakthrough results. Self-RAG outperforms existing models
including ChatGPT and Llama2-chat on multiple benchmark tasks,
demonstrating strong accuracy and context precision in specialized
domains {[}24{]}. Models can dynamically alter their behavior based on
confidence assessments and quality indicators by utilizing reflection
tokens to critique their own generations.

\subsubsection{Multi-Agent Coordination Patterns}

Multi-agent coordination patterns involve the collaborative efforts of
specialized retrieval agents, ranking agents, orchestrator agents, and
generator agents to enhance the overall performance of the system
{[}75{]}. LangChain/LangGraph offers graph-based workflow management
{[}76{]}, CrewAI provides role-based agent specialization {[}77{]}, and
OpenAI Swarm concentrates on lightweight multi-agent orchestration
{[}78{]}. Additionally, industry adoption is expedited by comprehensive
frameworks.

\subsection{Comparative Analysis and Performance Benchmarks}

RAG architectures exhibit significant differences regarding accuracy,
scalability, and domain suitability, accompanied by trade-offs between
operational complexity and implementation requirements. Graph-Augmented
and Agentic RAG models demonstrate substantial enhancements in accuracy;
however, they require additional computational resources. Hybrid
Sparse-Dense methodologies provide a more economical and scalable
option.

\textbf{Table 5.1: RAG Architecture Comparison}

\begin{longtable}[]{@{}
  >{\raggedright\arraybackslash}p{(\columnwidth - 8\tabcolsep) * \real{0.1857}}
  >{\raggedright\arraybackslash}p{(\columnwidth - 8\tabcolsep) * \real{0.2141}}
  >{\raggedright\arraybackslash}p{(\columnwidth - 8\tabcolsep) * \real{0.2131}}
  >{\raggedright\arraybackslash}p{(\columnwidth - 8\tabcolsep) * \real{0.1409}}
  >{\raggedright\arraybackslash}p{(\columnwidth - 8\tabcolsep) * \real{0.2461}}@{}}
\toprule\noalign{}
\begin{minipage}[b]{\linewidth}\raggedright
Architecture
\end{minipage} & \begin{minipage}[b]{\linewidth}\raggedright
Key Innovation
\end{minipage} & \begin{minipage}[b]{\linewidth}\raggedright
Primary Advantage
\end{minipage} & \begin{minipage}[b]{\linewidth}\raggedright
Complexity
\end{minipage} & \begin{minipage}[b]{\linewidth}\raggedright
Implementation Platforms
\end{minipage} \\
\midrule\noalign{}
\endhead
\bottomrule\noalign{}
\endlastfoot
RAG-Fusion & Multi-query processing & Comprehensive coverage & Moderate
& LangChain, Microsoft \\
RE-RAG & Two-stage retrieval & Precision improvement & Moderate-High &
Cohere, NVIDIA NeMo \\
Hierarchical RAG & Tree-structured retrieval & Long-context handling &
High & RAPTOR, GraphRAG \\
Hybrid Sparse-Dense & BM25 + Dense vectors & Balanced performance &
Moderate & BGE-M3, Various platforms \\
Graph-Augmented & Knowledge graphs & Relationship modeling & Very High &
Neo4j, Microsoft GraphRAG \\
Agentic RAG & Autonomous agents & Adaptive optimization & Very High &
LangGraph, CrewAI, Swarm \\
\end{longtable}

\textbf{Table 5.2: Domain Suitability Analysis}

\begin{longtable}[]{@{}
  >{\raggedright\arraybackslash}p{(\columnwidth - 8\tabcolsep) * \real{0.1775}}
  >{\raggedright\arraybackslash}p{(\columnwidth - 8\tabcolsep) * \real{0.2959}}
  >{\raggedright\arraybackslash}p{(\columnwidth - 8\tabcolsep) * \real{0.2249}}
  >{\raggedright\arraybackslash}p{(\columnwidth - 8\tabcolsep) * \real{0.1257}}
  >{\raggedright\arraybackslash}p{(\columnwidth - 8\tabcolsep) * \real{0.1760}}@{}}
\toprule\noalign{}
\begin{minipage}[b]{\linewidth}\raggedright
Architecture
\end{minipage} & \begin{minipage}[b]{\linewidth}\raggedright
Best Use Cases
\end{minipage} & \begin{minipage}[b]{\linewidth}\raggedright
Domain Suitability
\end{minipage} & \begin{minipage}[b]{\linewidth}\raggedright
Scalability
\end{minipage} & \begin{minipage}[b]{\linewidth}\raggedright
Enterprise Readiness
\end{minipage} \\
\midrule\noalign{}
\endhead
\bottomrule\noalign{}
\endlastfoot
RAG-Fusion & General QA, Knowledge Retrieval & Universal & High &
High \\
RE-RAG & Precision-Critical Tasks & Legal, Medical, Finance &
Medium-High & High \\
Hierarchical RAG & Complex Reasoning, Long Documents & Research,
Analysis & Medium & Medium \\
Hybrid Sparse-Dense & Multi-lingual, Diverse Queries & E-commerce,
Support & Very High & Very High \\
Graph-Augmented & Relationship Analysis & Scientific, Compliance &
Medium & Medium \\
Agentic RAG & Dynamic Environments & Customer Service, Research &
Low-Medium & Low-Medium \\
\end{longtable}

\subsection{Architectural Evolution and Trade-offs}

The evolution of RAG architectures reflects the field\textquotesingle s
progression from simple retrieval-generation pipelines to sophisticated
systems capable of complex reasoning, multi-modal processing, and
autonomous optimization. Each architectural paradigm addresses specific
limitations while introducing distinct trade-offs in complexity,
resource requirements, and domain applicability. The choice of
architecture depends critically on application requirements, available
computational resources, and acceptable implementation complexity.

Current research trends indicate convergence toward hybrid approaches
that combine multiple paradigms, particularly the integration of
graph-augmented capabilities with agentic frameworks for
enterprise-scale deployments. Future developments will likely focus on
standardization of evaluation metrics and development of unified
frameworks that abstract architectural complexity while maintaining
performance advantages.

\section{Evaluation and Benchmarking Framework}

Due to their multi-component architecture, the systematic evaluation of
Retrieval-Augmented Generation (RAG) systems presents distinctive
challenges, necessitating a comprehensive evaluation of retrieval
quality, generation accuracy, and system trustworthiness {[}93{]}.
Incorporating sophisticated frameworks that utilize large language
models as judges, modern RAG evaluation has progressed beyond
conventional metrics, thereby facilitating a more nuanced evaluation of
contextual relevance and semantic similarity {[}94{]}. Both
component-level and end-to-end evaluation approaches are required to
identify performance constraints and optimization opportunities
throughout the retrieval-generation pipeline due to the complexity of
RAG systems {[}95{]}. Operational playbooks recommend coupling offline
benchmarks with online telemetry (latency, CTR, deflection rate) and
human review queues for drift control {[}86{]}.

\subsection{Comparative Analysis of RAG Evaluation Frameworks}

In an effort to mitigate the constraints of conventional metrics,
contemporary RAG evaluation frameworks have emerged. These frameworks
offer automated assessment capabilities that minimize manual evaluation
burden while preserving a high degree of correlation with human judgment
{[}96{]}. Typically, these frameworks employ sophisticated scoring
mechanisms to evaluate retrieval relevance, generation faithfulness, and
answer quality on a multi-dimensional scale {[}97{]}. Automatic
long-form evaluators such as ALCE complement LLM-judge pipelines by
targeting coherence and discourse-level faithfulness {[}26{]}.

\textbf{Table 7.1: RAG Evaluation Framework Comparison}

\begin{longtable}[]{@{}
  >{\raggedright\arraybackslash}p{(\columnwidth - 10\tabcolsep) * \real{0.1229}}
  >{\raggedright\arraybackslash}p{(\columnwidth - 10\tabcolsep) * \real{0.2233}}
  >{\raggedright\arraybackslash}p{(\columnwidth - 10\tabcolsep) * \real{0.1143}}
  >{\raggedright\arraybackslash}p{(\columnwidth - 10\tabcolsep) * \real{0.1603}}
  >{\raggedright\arraybackslash}p{(\columnwidth - 10\tabcolsep) * \real{0.2000}}
  >{\raggedright\arraybackslash}p{(\columnwidth - 10\tabcolsep) * \real{0.1793}}@{}}
\toprule\noalign{}
\begin{minipage}[b]{\linewidth}\raggedright
Framework
\end{minipage} & \begin{minipage}[b]{\linewidth}\raggedright
Primary Focus
\end{minipage} & \begin{minipage}[b]{\linewidth}\raggedright
Automation Level
\end{minipage} & \begin{minipage}[b]{\linewidth}\raggedright
LLM-based Metrics
\end{minipage} & \begin{minipage}[b]{\linewidth}\raggedright
Reference-free Capability
\end{minipage} & \begin{minipage}[b]{\linewidth}\raggedright
Enterprise Integration
\end{minipage} \\
\midrule\noalign{}
\endhead
\bottomrule\noalign{}
\endlastfoot
RAGAS & End-to-end RAG & High & Yes & Yes & Moderate \\
LlamaIndex & Component-level & High & Yes & Yes & High \\
TruLens & Hallucination detection & Medium & Yes & Yes & High \\
RAGChecker & Fine-grained analysis & High & Yes & No & Moderate \\
DeepEval & Comprehensive testing & High & Yes & Partial & High \\
UpTrain & Production monitoring & High & Yes & Yes & Very High \\
\end{longtable}

The RAGAS framework offers comprehensive evaluation capabilities by
employing LLM-based judges to evaluate response quality without
requiring ground truth labels. This is achieved through four core
metrics: faithfulness, answer relevancy, context precision, and context
recall {[}98{]}. Detailed evaluation of both retrieval and generation
components is facilitated by LlamaIndex\textquotesingle s extensive
evaluation modules, which include correctness, semantic similarity,
faithfulness, context relevancy, answer relevancy, and guideline
adherence {[}99{]}.

\subsection{Retrieval Quality Assessment}

The quality of generation is directly influenced by the relevance and
completeness of the retrieved context, which is why retrieval evaluation
is the cornerstone of RAG system assessment {[}100{]}. Modern retrieval
metrics incorporate RAG-specific considerations, such as context
utilization and fragment attribution, in addition to traditional
information retrieval measures {[}101{]}.

\textbf{Table 7.2: Retrieval Metrics Classification}

\begin{longtable}[]{@{}
  >{\raggedright\arraybackslash}p{(\columnwidth - 10\tabcolsep) * \real{0.1287}}
  >{\raggedright\arraybackslash}p{(\columnwidth - 10\tabcolsep) * \real{0.1427}}
  >{\raggedright\arraybackslash}p{(\columnwidth - 10\tabcolsep) * \real{0.1966}}
  >{\raggedright\arraybackslash}p{(\columnwidth - 10\tabcolsep) * \real{0.1285}}
  >{\raggedright\arraybackslash}p{(\columnwidth - 10\tabcolsep) * \real{0.1924}}
  >{\raggedright\arraybackslash}p{(\columnwidth - 10\tabcolsep) * \real{0.2110}}@{}}
\toprule\noalign{}
\begin{minipage}[b]{\linewidth}\raggedright
Metric Category
\end{minipage} & \begin{minipage}[b]{\linewidth}\raggedright
Metric Name
\end{minipage} & \begin{minipage}[b]{\linewidth}\raggedright
Description
\end{minipage} & \begin{minipage}[b]{\linewidth}\raggedright
Order Sensitivity
\end{minipage} & \begin{minipage}[b]{\linewidth}\raggedright
Implementation Complexity
\end{minipage} & \begin{minipage}[b]{\linewidth}\raggedright
Correlation with Human Judgment
\end{minipage} \\
\midrule\noalign{}
\endhead
\bottomrule\noalign{}
\endlastfoot
Traditional IR & Precision@k & Relevant documents in top-k & No & Low &
Medium \\
Traditional IR & Recall@k & Coverage of relevant documents & No & Low &
Medium \\
Traditional IR & MRR & Mean reciprocal rank & Yes & Low & Medium \\
Traditional IR & nDCG@k & Position-weighted relevance & Yes & Medium &
High \\
RAG-specific & Context Precision & Relevant chunks in context & Yes &
High & High \\
RAG-specific & Context Recall & Coverage of ground truth & No & High &
Very High \\
RAG-specific & Chunk Attribution & Source attribution accuracy & No &
Very High & Very High \\
\end{longtable}

The TruLens framework introduces the RAG Triad concept, which consists
of context relevance, groundedness, and answer relevance. This concept
provides comprehensive coverage of hallucination detection across each
boundary of the RAG architecture {[}102{]}. Context relevance evaluates
the extent to which the retrieved fragments contain information that is
pertinent to the input query, whereas groundedness evaluates the extent
to which the generated responses are adequately substantiated by the
retrieved evidence {[}102{]}.

\subsection{Generation Quality Metrics}

Sophisticated metrics that can assess semantic similarity, factual
consistency, and contextual appropriateness beyond surface-level text
matching are necessary for generation quality assessment in RAG systems
{[}103{]}. In order to conduct an exhaustive evaluation of generation
quality, contemporary evaluation frameworks implement both conventional
NLP metrics and sophisticated LLM-based judges {[}104{]}.

\textbf{Table 7.3: Generation Quality Metrics Comparison}

\begin{longtable}[]{@{}
  >{\raggedright\arraybackslash}p{(\columnwidth - 10\tabcolsep) * \real{0.1185}}
  >{\raggedright\arraybackslash}p{(\columnwidth - 10\tabcolsep) * \real{0.1728}}
  >{\raggedright\arraybackslash}p{(\columnwidth - 10\tabcolsep) * \real{0.1671}}
  >{\raggedright\arraybackslash}p{(\columnwidth - 10\tabcolsep) * \real{0.1729}}
  >{\raggedright\arraybackslash}p{(\columnwidth - 10\tabcolsep) * \real{0.1700}}
  >{\raggedright\arraybackslash}p{(\columnwidth - 10\tabcolsep) * \real{0.1987}}@{}}
\toprule\noalign{}
\begin{minipage}[b]{\linewidth}\raggedright
Metric Type
\end{minipage} & \begin{minipage}[b]{\linewidth}\raggedright
Metric Name
\end{minipage} & \begin{minipage}[b]{\linewidth}\raggedright
Evaluation Focus
\end{minipage} & \begin{minipage}[b]{\linewidth}\raggedright
Computational Cost
\end{minipage} & \begin{minipage}[b]{\linewidth}\raggedright
Human Correlation
\end{minipage} & \begin{minipage}[b]{\linewidth}\raggedright
Reference Requirement
\end{minipage} \\
\midrule\noalign{}
\endhead
\bottomrule\noalign{}
\endlastfoot
Traditional & BLEU Score & N-gram precision & Low & Low-Medium & Yes \\
Traditional & ROUGE Score & N-gram recall & Low & Medium & Yes \\
Model-based & BERTScore & Semantic similarity & Medium & High & Yes \\
LLM-based & Faithfulness & Claim verification & High & Very High & No \\
LLM-based & Answer Relevancy & Query alignment & High & Very High &
No \\
LLM-based & Answer Correctness & Factual accuracy & High & Very High &
Yes \\
\end{longtable}

The BLEU score is a metric that quantifies the precision of n-grams
between reference and generated texts. Scores range from 0 to 1, with
higher values indicating a closer alignment. However, it encounters
challenges with semantic understanding and word order variations
{[}105{]}. The ROUGE score is particularly beneficial for evaluating the
comprehensiveness of question-answering tasks, as it emphasizes
recall-oriented evaluation and concentrates on the amount of reference
content that is incorporated in the generated text {[}106{]}.

BERTScore utilizes contextual embeddings from transformer models to
calculate semantic similarity through cosine similarity of token
representations. This approach offers a more nuanced evaluation that is
more closely aligned with human judgment than surface-level metrics
{[}107{]}. The metric is particularly effective for evaluating
conversational interfaces and shorter text generation tasks, as it
calculates precision, recall, and F1 measures by aligning contextually
similar utterances between candidate and reference texts {[}108{]}.

\subsection{Assessment of Safety and Trustworthiness}

Critical concerns regarding hallucination detection, factual
consistency, and appropriate uncertainty management in RAG systems are
addressed by trustworthiness evaluation {[}109{]}. In enterprise
deployments, where inaccurate information can have substantial
repercussions, these metrics become indispensable {[}110{]}.

\textbf{Table 7.4: Trustworthiness Metrics Implementation}

\begin{longtable}[]{@{}
  >{\raggedright\arraybackslash}p{(\columnwidth - 10\tabcolsep) * \real{0.1544}}
  >{\raggedright\arraybackslash}p{(\columnwidth - 10\tabcolsep) * \real{0.1726}}
  >{\raggedright\arraybackslash}p{(\columnwidth - 10\tabcolsep) * \real{0.1498}}
  >{\raggedright\arraybackslash}p{(\columnwidth - 10\tabcolsep) * \real{0.2156}}
  >{\raggedright\arraybackslash}p{(\columnwidth - 10\tabcolsep) * \real{0.1465}}
  >{\raggedright\arraybackslash}p{(\columnwidth - 10\tabcolsep) * \real{0.1611}}@{}}
\toprule\noalign{}
\begin{minipage}[b]{\linewidth}\raggedright
Metric
\end{minipage} & \begin{minipage}[b]{\linewidth}\raggedright
Assessment Method
\end{minipage} & \begin{minipage}[b]{\linewidth}\raggedright
Accuracy vs Human
\end{minipage} & \begin{minipage}[b]{\linewidth}\raggedright
Detection Capability
\end{minipage} & \begin{minipage}[b]{\linewidth}\raggedright
Implementation Difficulty
\end{minipage} & \begin{minipage}[b]{\linewidth}\raggedright
Production Readiness
\end{minipage} \\
\midrule\noalign{}
\endhead
\bottomrule\noalign{}
\endlastfoot
Groundedness & LLM-based verification & 85-92\% & Factual inconsistency
& Medium & High \\
Citation Accuracy & Automated attribution & 80-90\% & Source
misattribution & High & Medium \\
Hallucination Rate & Multi-method detection & 75-88\% & False
information & Very High & Medium \\
Context Adherence & Entailment checking & 78-85\% & Context deviation &
Medium & High \\
Completeness & Coverage assessment & 70-82\% & Information gaps & High &
Medium \\
\end{longtable}

RAGChecker offers a comprehensive set of metrics, including overall
performance measures (precision, recall, F1), retriever-specific metrics
(claim recall, context precision), and generator-specific metrics
(context utilization, noise sensitivity, hallucination, self-knowledge,
faithfulness) {[}95{]}, which are assessed through claim-level
entailment checking. This framework facilitates the targeted enhancement
of RAG system performance by facilitating the comprehensive diagnosis of
both retrieval and generation components {[}95{]}.

\subsection{Traditional vs Modern Evaluation Approaches}

The fundamental shift in RAG assessment methodologies is represented by
the transition from conventional NLP metrics to sophisticated LLM-based
evaluation {[}103{]}. Modern methods exhibit a superior correlation with
human judgment and semantic comprehension, whereas traditional metrics
offer computational efficiency and interpretability {[}104{]}.

\textbf{Table 7.5: Traditional vs Modern Evaluation Comparison}

\begin{longtable}[]{@{}
  >{\raggedright\arraybackslash}p{(\columnwidth - 6\tabcolsep) * \real{0.2867}}
  >{\raggedright\arraybackslash}p{(\columnwidth - 6\tabcolsep) * \real{0.2346}}
  >{\raggedright\arraybackslash}p{(\columnwidth - 6\tabcolsep) * \real{0.2426}}
  >{\raggedright\arraybackslash}p{(\columnwidth - 6\tabcolsep) * \real{0.2360}}@{}}
\toprule\noalign{}
\begin{minipage}[b]{\linewidth}\raggedright
Evaluation Aspect
\end{minipage} & \begin{minipage}[b]{\linewidth}\raggedright
Traditional Metrics
\end{minipage} & \begin{minipage}[b]{\linewidth}\raggedright
Modern LLM-based
\end{minipage} & \begin{minipage}[b]{\linewidth}\raggedright
Hybrid Approaches
\end{minipage} \\
\midrule\noalign{}
\endhead
\bottomrule\noalign{}
\endlastfoot
Semantic Understanding & Limited & Excellent & Good \\
Computational Cost & Very Low & High & Medium \\
Human Correlation & Low-Medium & Very High & High \\
Reference Requirement & Always & Optional & Flexible \\
Interpretability & High & Medium & High \\
Scalability & Excellent & Limited & Good \\
Domain Adaptation & Poor & Excellent & Good \\
Real-time Capability & Excellent & Poor & Good \\
\end{longtable}

Modern evaluation frameworks are increasingly incorporating hybrid
approaches that combine the semantic sophistication of LLM-based judges
with the efficacy of traditional metrics {[}104{]}. This combination
allows for scalable evaluation while preserving a high degree of
correlation with human assessment, which is especially crucial for
production RAG systems that necessitate real-time performance monitoring
{[}101{]}.

\subsection{Benchmarking Datasets and Standards}

Standardized benchmarking enables the objective comparison of RAG
systems and offers industry reference points for performance evaluation
across a variety of domains and task types {[}93{]}. The primary
objective of contemporary benchmarking initiatives is to develop
exhaustive evaluation suites that evaluate RAG performance across
multiple dimensions {[}94{]}.

\textbf{Table 7.6: RAG Benchmarking Datasets}

\begin{longtable}[]{@{}
  >{\raggedright\arraybackslash}p{(\columnwidth - 10\tabcolsep) * \real{0.1801}}
  >{\raggedright\arraybackslash}p{(\columnwidth - 10\tabcolsep) * \real{0.1436}}
  >{\raggedright\arraybackslash}p{(\columnwidth - 10\tabcolsep) * \real{0.1988}}
  >{\raggedright\arraybackslash}p{(\columnwidth - 10\tabcolsep) * \real{0.0998}}
  >{\raggedright\arraybackslash}p{(\columnwidth - 10\tabcolsep) * \real{0.2031}}
  >{\raggedright\arraybackslash}p{(\columnwidth - 10\tabcolsep) * \real{0.1747}}@{}}
\toprule\noalign{}
\begin{minipage}[b]{\linewidth}\raggedright
Dataset
\end{minipage} & \begin{minipage}[b]{\linewidth}\raggedright
Domain
\end{minipage} & \begin{minipage}[b]{\linewidth}\raggedright
Question Types
\end{minipage} & \begin{minipage}[b]{\linewidth}\raggedright
Size
\end{minipage} & \begin{minipage}[b]{\linewidth}\raggedright
Evaluation Focus
\end{minipage} & \begin{minipage}[b]{\linewidth}\raggedright
Complexity Level
\end{minipage} \\
\midrule\noalign{}
\endhead
\bottomrule\noalign{}
\endlastfoot
HotpotQA & Wikipedia & Multi-hop reasoning & 113k & Reasoning capability
& High \\
MS MARCO & Web search & Factoid queries & 1M+ & Passage retrieval &
Medium \\
Natural Questions & Wikipedia & Real user queries & 307k & Real-world
scenarios & Medium \\
FEVER & Wikipedia & Fact verification & 185k & Factual accuracy &
Medium \\
RGB Benchmark & Multi-domain & Capability testing & Variable & Core RAG
abilities & High \\
OmniEval & Financial & Domain-specific & Custom & Vertical applications
& Very High \\
\end{longtable}

HotpotQA offers 113,000 question-answer pairs that are based on
Wikipedia and necessitate multi-document reasoning. These pairs include
sentence-level supporting facts and comparison questions that evaluate
the capacity of systems to extract and compare pertinent information
from multiple sources {[}100{]}. The dataset is especially valuable for
the assessment of sophisticated RAG architectures that are capable of
complex information synthesis due to its multi-hop reasoning
requirements {[}100{]}.

\subsection{Enterprise Evaluation Platforms}

Comprehensive infrastructure for RAG system assessment, monitoring, and
optimization in production environments is provided by enterprise-grade
evaluation platforms {[}101{]}. These platforms typically provide
real-time monitoring capabilities, automated evaluation pipelines, and
integration with existing development workflows {[}110{]}. Vendor
documentation details reference integrations for monitoring, evaluation,
and governance in enterprise RAG {[}91{]}--{[}92{]}.

\textbf{Table 7.7: Enterprise RAG Evaluation Platforms}

\begin{longtable}[]{@{}
  >{\raggedright\arraybackslash}p{(\columnwidth - 10\tabcolsep) * \real{0.1529}}
  >{\raggedright\arraybackslash}p{(\columnwidth - 10\tabcolsep) * \real{0.1564}}
  >{\raggedright\arraybackslash}p{(\columnwidth - 10\tabcolsep) * \real{0.1842}}
  >{\raggedright\arraybackslash}p{(\columnwidth - 10\tabcolsep) * \real{0.1420}}
  >{\raggedright\arraybackslash}p{(\columnwidth - 10\tabcolsep) * \real{0.1890}}
  >{\raggedright\arraybackslash}p{(\columnwidth - 10\tabcolsep) * \real{0.1756}}@{}}
\toprule\noalign{}
\begin{minipage}[b]{\linewidth}\raggedright
Platform
\end{minipage} & \begin{minipage}[b]{\linewidth}\raggedright
Automation Level
\end{minipage} & \begin{minipage}[b]{\linewidth}\raggedright
Real-time Monitoring
\end{minipage} & \begin{minipage}[b]{\linewidth}\raggedright
Custom Metrics
\end{minipage} & \begin{minipage}[b]{\linewidth}\raggedright
Integration Capability
\end{minipage} & \begin{minipage}[b]{\linewidth}\raggedright
Deployment Options
\end{minipage} \\
\midrule\noalign{}
\endhead
\bottomrule\noalign{}
\endlastfoot
Galileo AI & Very High & Yes & Yes & Extensive & Cloud/On-premise \\
LangSmith & High & Yes & Yes & Good & Cloud \\
TruLens & Medium & Yes & Limited & Good & Open source \\
UpTrain & High & Yes & Yes & Good & Open source \\
DeepEval & High & Limited & Yes & Moderate & Open source \\
Weights \& Biases & High & Yes & Yes & Extensive & Cloud/On-premise \\
\end{longtable}

Galileo AI offers a comprehensive evaluation of RAGs using proprietary
metrics, such as chunk attribution (86\% accuracy, 1.36x more accurate
than the GPT-3.5-Turbo baseline), chunk utilization (74\% accuracy,
1.69x improvement), context adherence (74\% accuracy, 1.65x
improvement), and completeness assessment (80\% accuracy, 1.61x
improvement) {[}101{]}. The platform supports both real-time production
monitoring and offline evaluation, and it provides visual tracing
capabilities for debugging RAG workflows {[}110{]}.

\subsection{Future Directions and Best Practices}

RAG evaluation is constantly evolving to incorporate more sophisticated
assessment methodologies that more accurately reflect the intricacies of
human-AI interaction and domain-specific requirements {[}94{]}. There
are several emerging trends, such as adaptive metrics that are tailored
to specific use cases and domains, continuous evaluation pipelines, and
automated test case generation {[}93{]}.

\textbf{Table 7.8: RAG Evaluation Best Practices}

\begin{longtable}[]{@{}
  >{\raggedright\arraybackslash}p{(\columnwidth - 8\tabcolsep) * \real{0.2069}}
  >{\raggedright\arraybackslash}p{(\columnwidth - 8\tabcolsep) * \real{0.3429}}
  >{\raggedright\arraybackslash}p{(\columnwidth - 8\tabcolsep) * \real{0.1792}}
  >{\raggedright\arraybackslash}p{(\columnwidth - 8\tabcolsep) * \real{0.1067}}
  >{\raggedright\arraybackslash}p{(\columnwidth - 8\tabcolsep) * \real{0.1642}}@{}}
\toprule\noalign{}
\begin{minipage}[b]{\linewidth}\raggedright
Practice Category
\end{minipage} & \begin{minipage}[b]{\linewidth}\raggedright
Recommendation
\end{minipage} & \begin{minipage}[b]{\linewidth}\raggedright
Implementation Priority
\end{minipage} & \begin{minipage}[b]{\linewidth}\raggedright
Impact Level
\end{minipage} & \begin{minipage}[b]{\linewidth}\raggedright
Resource Requirement
\end{minipage} \\
\midrule\noalign{}
\endhead
\bottomrule\noalign{}
\endlastfoot
Multi-dimensional Assessment & Combine retrieval, generation, and
trustworthiness metrics & Critical & Very High & Medium \\
Automated Pipeline & Implement continuous evaluation workflows & High &
High & High \\
Human-in-the-loop & Integrate expert validation for critical
applications & Critical & Very High & Very High \\
Domain-specific Metrics & Develop specialized evaluation criteria &
Medium & Medium & Medium \\
Real-time Monitoring & Deploy production evaluation systems & High &
High & High \\
Benchmark Standardization & Adopt industry-standard datasets & Medium &
Medium & Low \\
\end{longtable}

The evaluation of RAGs must be effective by balancing the quality of the
assessment with the efficiency of automation. This can be achieved by
utilizing both traditional metrics for baseline performance and advanced
LLM-based judges for semantic evaluation {[}103{]}. To guarantee
consistent system performance across a variety of deployment scenarios,
organizations should establish exhaustive evaluation frameworks that
facilitate both real-time production monitoring and offline development
optimization {[}110{]}{[}111{]}.

\section{Engineering Patterns and Anti-Patterns}

Through extensive production deployments in a variety of enterprise
environments, Retrieval-Augmented Generation (RAG) systems have
developed into a mature engineering discipline {[}112{]}. This evolution
has uncovered critical design patterns that improve the reliability,
performance, and maintainability of the system, while also revealing
anti-patterns that systematically undermine its efficacy {[}113{]}. It
is imperative for engineering teams to comprehend these patterns and
anti-patterns in order to create RAG architectures that are reliable,
scalable, and trustworthy in production environments {[}114{]}.

The multi-component architecture of RAG systems is the source of their
complexity, as retrieval mechanisms, knowledge bases, and generation
models must operate in tandem to provide contextually pertinent, precise
responses {[}115{]}. Real-world deployments have resulted in the
development of engineering best practices, which offer teams systematic
guidance as they develop production-ready systems {[}116{]}. In
contrast, organizations repeatedly confront systematic engineering
errors during RAG implementation, which are represented by common
failure modes and anti-patterns {[}117{]}.

\subsection{Design Best Practices: Foundational Engineering Patterns}

RAG implementations that are successful adhere to established
engineering patterns that address fundamental challenges in operational
monitoring, system resilience, retrieval quality, and document
processing {[}118{]}. Current best practices in RAG system engineering
are represented by these patterns, which have been validated across
multiple production environments {[}119{]}. Cloud guidance on choosing
RAG options highlights trade-offs among vector search, hybrid retrieval,
and reranking services {[}88{]}. Enterprise guidance emphasizes
governance guardrails, PII filtering, citation checks, and
prompt/response policies, alongside retrieval reliability tests
{[}87{]}.

\subsubsection{Document Processing and Chunking Strategies}

Chunking strategies are essential for the accuracy of responses and the
quality of retrievals, which are the foundation of any successful RAG
system {[}120{]}. Depending on the use case, document formats, and
performance requirements, various chunking approaches provide distinct
advantages {[}121{]}.

\textbf{Table 8.1: Document Chunking Strategy Comparison}

\begin{longtable}[]{@{}
  >{\raggedright\arraybackslash}p{(\columnwidth - 8\tabcolsep) * \real{0.1918}}
  >{\raggedright\arraybackslash}p{(\columnwidth - 8\tabcolsep) * \real{0.1619}}
  >{\raggedright\arraybackslash}p{(\columnwidth - 8\tabcolsep) * \real{0.2053}}
  >{\raggedright\arraybackslash}p{(\columnwidth - 8\tabcolsep) * \real{0.1895}}
  >{\raggedright\arraybackslash}p{(\columnwidth - 8\tabcolsep) * \real{0.2515}}@{}}
\toprule\noalign{}
\begin{minipage}[b]{\linewidth}\raggedright
Strategy
\end{minipage} & \begin{minipage}[b]{\linewidth}\raggedright
Context Preservation
\end{minipage} & \begin{minipage}[b]{\linewidth}\raggedright
Implementation Complexity
\end{minipage} & \begin{minipage}[b]{\linewidth}\raggedright
Computational Overhead
\end{minipage} & \begin{minipage}[b]{\linewidth}\raggedright
Best Use Cases
\end{minipage} \\
\midrule\noalign{}
\endhead
\bottomrule\noalign{}
\endlastfoot
Fixed-size Chunking & Moderate & Low & Low & Simple documents, FAQ
systems \\
Overlapping Chunking & High & Medium & Medium & Technical documentation,
legal texts \\
Semantic Windowing & Very High & High & High & Research papers, complex
narratives \\
Topic-based Segmentation & High & High & Medium & Multi-topic documents,
news articles \\
Hierarchical Chunking & Very High & Very High & High & Structured
documents, manuals \\
\end{longtable}

The essential boundary problem, in which critical information is
dispersed across multiple document sections, is addressed by overlapping
chunking strategies {[}122{]}. Semantic windowing techniques improve
context preservation by generating segments that are based on semantic
boundaries rather than defined character counts {[}123{]}. These
sophisticated methods necessitate a meticulous equilibrium between
computational efficiency and retrieval coverage {[}124{]}.

\subsubsection{Retrieval Quality and Confidence Mechanisms}

Sophisticated confidence thresholds and quality barriers are implemented
in production RAG systems to prevent the contamination of the generation
process by low-quality retrievals {[}125{]}. When high-quality
information is unavailable, these mechanisms facilitate appropriate
abstention behavior and graceful degradation {[}126{]}.

\textbf{Table 8.2: Retrieval Quality Mechanisms Comparison}

\begin{longtable}[]{@{}
  >{\raggedright\arraybackslash}p{(\columnwidth - 8\tabcolsep) * \real{0.2539}}
  >{\raggedright\arraybackslash}p{(\columnwidth - 8\tabcolsep) * \real{0.2131}}
  >{\raggedright\arraybackslash}p{(\columnwidth - 8\tabcolsep) * \real{0.2194}}
  >{\raggedright\arraybackslash}p{(\columnwidth - 8\tabcolsep) * \real{0.2049}}
  >{\raggedright\arraybackslash}p{(\columnwidth - 8\tabcolsep) * \real{0.1087}}@{}}
\toprule\noalign{}
\begin{minipage}[b]{\linewidth}\raggedright
Mechanism
\end{minipage} & \begin{minipage}[b]{\linewidth}\raggedright
Accuracy Improvement
\end{minipage} & \begin{minipage}[b]{\linewidth}\raggedright
Hallucination Reduction
\end{minipage} & \begin{minipage}[b]{\linewidth}\raggedright
Implementation Effort
\end{minipage} & \begin{minipage}[b]{\linewidth}\raggedright
Scalability
\end{minipage} \\
\midrule\noalign{}
\endhead
\bottomrule\noalign{}
\endlastfoot
Single Confidence Threshold & Moderate & Moderate & Low & High \\
Multi-signal Scoring & High & High & Medium & Medium \\
Ensemble Retrieval & Very High & Very High & High & Medium \\
Adaptive Thresholding & High & High & High & High \\
Context-aware Filtering & Very High & High & Very High & Low \\
\end{longtable}

In addition to simple similarity scores, multi-signal confidence scoring
includes a variety of quality indicators, such as source credibility,
temporal relevance, and context alignment {[}127{]}. In an effort to
enhance robustness and mitigate the effects of individual component
failures, ensemble retrieval methods integrate numerous retrieval
strategies {[}128{]}.

\subsubsection{Index Management and Freshness Strategies}

In production RAG systems, the retention of index freshness is a
critical challenge, as the quality of answers and the trust of users are
substantially impacted by stale information {[}129{]}. Various update
strategies provide varying trade-offs between the currency of
information, system availability, and computational cost {[}130{]}.

\textbf{Table 8.3: Index Freshness Strategy Analysis}

\begin{longtable}[]{@{}
  >{\raggedright\arraybackslash}p{(\columnwidth - 8\tabcolsep) * \real{0.2532}}
  >{\raggedright\arraybackslash}p{(\columnwidth - 8\tabcolsep) * \real{0.1797}}
  >{\raggedright\arraybackslash}p{(\columnwidth - 8\tabcolsep) * \real{0.2127}}
  >{\raggedright\arraybackslash}p{(\columnwidth - 8\tabcolsep) * \real{0.2212}}
  >{\raggedright\arraybackslash}p{(\columnwidth - 8\tabcolsep) * \real{0.1333}}@{}}
\toprule\noalign{}
\begin{minipage}[b]{\linewidth}\raggedright
Update Strategy
\end{minipage} & \begin{minipage}[b]{\linewidth}\raggedright
Freshness Level
\end{minipage} & \begin{minipage}[b]{\linewidth}\raggedright
System Availability
\end{minipage} & \begin{minipage}[b]{\linewidth}\raggedright
Resource Utilization
\end{minipage} & \begin{minipage}[b]{\linewidth}\raggedright
Scalability
\end{minipage} \\
\midrule\noalign{}
\endhead
\bottomrule\noalign{}
\endlastfoot
Full Reindex & Excellent & Moderate & Very High & Poor \\
Delta Updates & Good & Excellent & Low & Excellent \\
Hierarchical Updates & Good & High & Medium & Good \\
Content-aware Updates & Very Good & High & Low & Good \\
Hybrid Approaches & Excellent & High & Medium & Very Good \\
\end{longtable}

While hierarchical approaches prioritize updates based on content
importance and access patterns, delta update mechanisms minimize
computational overhead by processing only changed content {[}131{]}.
Content-aware strategies dynamically modify the frequency of updates in
accordance with the volatility of information and the patterns of user
access {[}132{]}.

\subsection{Anti-Patterns: Common Failure Modes and Mitigation
Strategies}

RAG systems demonstrate recurring failure patterns that have a
substantial effect on user confidence, reliability, and performance
{[}133{]}. Proactive prevention and early detection of system
degradation are made possible by comprehending these anti-patterns
{[}134{]}.

\subsubsection{Retrieval Failure Modes}

The most prevalent cause of RAG system degradation is retrieval
failures, which are evident in a variety of ways at different phases of
the retrieval pipeline {[}135{]}. These defects may manifest during the
query processing, document matching, or result ranking phases {[}136{]}.

\textbf{Table 8.4: Retrieval Failure Mode Classification}

\begin{longtable}[]{@{}
  >{\raggedright\arraybackslash}p{(\columnwidth - 8\tabcolsep) * \real{0.2552}}
  >{\raggedright\arraybackslash}p{(\columnwidth - 8\tabcolsep) * \real{0.1263}}
  >{\raggedright\arraybackslash}p{(\columnwidth - 8\tabcolsep) * \real{0.1745}}
  >{\raggedright\arraybackslash}p{(\columnwidth - 8\tabcolsep) * \real{0.2078}}
  >{\raggedright\arraybackslash}p{(\columnwidth - 8\tabcolsep) * \real{0.2362}}@{}}
\toprule\noalign{}
\begin{minipage}[b]{\linewidth}\raggedright
Failure Mode
\end{minipage} & \begin{minipage}[b]{\linewidth}\raggedright
Frequency
\end{minipage} & \begin{minipage}[b]{\linewidth}\raggedright
Impact Severity
\end{minipage} & \begin{minipage}[b]{\linewidth}\raggedright
Detection Difficulty
\end{minipage} & \begin{minipage}[b]{\linewidth}\raggedright
Mitigation Complexity
\end{minipage} \\
\midrule\noalign{}
\endhead
\bottomrule\noalign{}
\endlastfoot
Missing Content & High & High & Low & Medium \\
Poor Ranking & Very High & Medium & Medium & Medium \\
Context Overflow & Medium & High & Low & Low \\
Query Misinterpretation & Medium & High & High & High \\
Index Staleness & High & Medium & Low & Medium \\
\end{longtable}

Missing content failures are occasions in which retrieval mechanisms are
incapable of locating pertinent information that is present in the
knowledge base {[}137{]}. Poor ranking issues are evident when pertinent
documents are retrieved but incorrectly prioritized, resulting in
suboptimal context for generation {[}138{]}.

\subsubsection{Generation Quality Anti-Patterns}

The retrieval phase is frequently the source of generation quality
issues, but they can also result from prompt engineering deficiencies,
context management failings, or model limitations {[}121{]}. The user
experience and the credibility of the system are directly influenced by
these anti-patterns {[}122{]}.

\textbf{Table 8.5: Generation Anti-Pattern Impact Analysis}

\begin{longtable}[]{@{}
  >{\raggedright\arraybackslash}p{(\columnwidth - 8\tabcolsep) * \real{0.2131}}
  >{\raggedright\arraybackslash}p{(\columnwidth - 8\tabcolsep) * \real{0.2846}}
  >{\raggedright\arraybackslash}p{(\columnwidth - 8\tabcolsep) * \real{0.1271}}
  >{\raggedright\arraybackslash}p{(\columnwidth - 8\tabcolsep) * \real{0.1378}}
  >{\raggedright\arraybackslash}p{(\columnwidth - 8\tabcolsep) * \real{0.2374}}@{}}
\toprule\noalign{}
\begin{minipage}[b]{\linewidth}\raggedright
Anti-Pattern
\end{minipage} & \begin{minipage}[b]{\linewidth}\raggedright
Root Cause
\end{minipage} & \begin{minipage}[b]{\linewidth}\raggedright
User Impact
\end{minipage} & \begin{minipage}[b]{\linewidth}\raggedright
Business Risk
\end{minipage} & \begin{minipage}[b]{\linewidth}\raggedright
Prevention Strategy
\end{minipage} \\
\midrule\noalign{}
\endhead
\bottomrule\noalign{}
\endlastfoot
Hallucination & Insufficient context & High & High & Confidence
thresholds \\
Inconsistent Responses & Variable retrieval quality & Medium & Medium &
Response caching \\
Context Truncation & Poor chunk management & High & Medium & Intelligent
summarization \\
Format Violations & Inadequate prompt engineering & Low & Low & Template
validation \\
Irrelevant Answers & Query-document mismatch & High & High & Relevance
scoring \\
\end{longtable}

One of the most severe anti-patterns is hallucination, which occurs when
the generation model generates plausible but factually inaccurate
information as a result of insufficient or misleading context {[}123{]}.
When the retrieved information exceeds the model context windows,
context truncation failures occur, resulting in incomplete or distorted
responses {[}124{]}.

\subsection{Architecture Patterns for Scalable RAG Systems}

Operational efficiency, maintainability, and scalability necessitate
meticulously designed architectures in production RAG systems {[}125{]}.
According to organizational constraints, team capabilities, and
deployment requirements, various architectural patterns provide distinct
advantages {[}126{]}.

\subsubsection{Component Architecture Patterns}

Scalability and maintainability challenges are addressed by contemporary
RAG systems through the implementation of a variety of architectural
patterns {[}127{]}. These patterns have an impact on the velocity of
development, the complexity of the system, and the operational overhead
{[}128{]}.

\textbf{Table 8.6: RAG Architecture Pattern Comparison}

\begin{longtable}[]{@{}
  >{\raggedright\arraybackslash}p{(\columnwidth - 8\tabcolsep) * \real{0.2250}}
  >{\raggedright\arraybackslash}p{(\columnwidth - 8\tabcolsep) * \real{0.1284}}
  >{\raggedright\arraybackslash}p{(\columnwidth - 8\tabcolsep) * \real{0.1758}}
  >{\raggedright\arraybackslash}p{(\columnwidth - 8\tabcolsep) * \real{0.2567}}
  >{\raggedright\arraybackslash}p{(\columnwidth - 8\tabcolsep) * \real{0.2141}}@{}}
\toprule\noalign{}
\begin{minipage}[b]{\linewidth}\raggedright
Architecture Pattern
\end{minipage} & \begin{minipage}[b]{\linewidth}\raggedright
Scalability
\end{minipage} & \begin{minipage}[b]{\linewidth}\raggedright
Maintainability
\end{minipage} & \begin{minipage}[b]{\linewidth}\raggedright
Operational Complexity
\end{minipage} & \begin{minipage}[b]{\linewidth}\raggedright
Development Speed
\end{minipage} \\
\midrule\noalign{}
\endhead
\bottomrule\noalign{}
\endlastfoot
Monolithic RAG & Low & Low & Low & High \\
Microservices RAG & Very High & High & High & Medium \\
Event-driven RAG & High & Medium & Medium & Medium \\
Serverless RAG & High & High & Low & High \\
Hybrid Architecture & Very High & Very High & Very High & Low \\
\end{longtable}

Heterogeneous technology platforms and deployment strategies are
supported by microservices architectures, which enable the independent
scaling of retrieval, indexing, and generation components {[}129{]}.
Event-driven patterns enhance system resilience by facilitating
asynchronous processing and enhancing the management of traffic surges
{[}112{]}.

\subsubsection{Deployment and Orchestration Strategies}

The deployment of a RAG system necessitates the meticulous coordination
of numerous components, each of which has distinct scaling
characteristics and resource requirements {[}113{]}. Orchestration
strategies that are effective are those that maintain a balance between
operational complexity, cost, and performance {[}114{]}.

\textbf{Table 8.7: Deployment Strategy Trade-offs}

\begin{longtable}[]{@{}
  >{\raggedright\arraybackslash}p{(\columnwidth - 8\tabcolsep) * \real{0.2368}}
  >{\raggedright\arraybackslash}p{(\columnwidth - 8\tabcolsep) * \real{0.1949}}
  >{\raggedright\arraybackslash}p{(\columnwidth - 8\tabcolsep) * \real{0.1795}}
  >{\raggedright\arraybackslash}p{(\columnwidth - 8\tabcolsep) * \real{0.1618}}
  >{\raggedright\arraybackslash}p{(\columnwidth - 8\tabcolsep) * \real{0.2269}}@{}}
\toprule\noalign{}
\begin{minipage}[b]{\linewidth}\raggedright
Deployment Strategy
\end{minipage} & \begin{minipage}[b]{\linewidth}\raggedright
Resource Efficiency
\end{minipage} & \begin{minipage}[b]{\linewidth}\raggedright
Scaling Flexibility
\end{minipage} & \begin{minipage}[b]{\linewidth}\raggedright
Fault Tolerance
\end{minipage} & \begin{minipage}[b]{\linewidth}\raggedright
Management Overhead
\end{minipage} \\
\midrule\noalign{}
\endhead
\bottomrule\noalign{}
\endlastfoot
Single-node Deployment & Low & Low & Low & Low \\
Container Orchestration & High & Very High & High & Medium \\
Serverless Components & Very High & Very High & Very High & Low \\
Hybrid Cloud & High & Very High & Very High & Very High \\
Edge Distribution & Medium & High & Medium & High \\
\end{longtable}

Kubernetes and other container orchestration platforms offer advanced
auto-scaling capabilities that are determined by RAG-specific metrics,
including retrieval latency and query complexity {[}115{]}. Serverless
deployments provide exceptional cost efficacy for variable workloads;
however, they may introduce cold start latencies {[}116{]}.

Teams are equipped with the requisite knowledge of engineering patterns
and anti-patterns to construct scalable, resilient RAG systems that
operate consistently in production environments. Throughout the
system\textquotesingle s lifecycle, success necessitates meticulous
attention to both operational considerations and technical
implementation details.

\section{Trust, Alignment, and Safety in RAG}

The deployment of Retrieval-Augmented Generation (RAG) systems in
enterprise environments presents intricate challenges regarding trust,
alignment, and safety that surpass conventional language model concerns
{[}140{]}. Comprehensive frameworks are required to guarantee system
reliability and user safety, as RAG architectures generate complex trust
surfaces that consolidate retrieved information, source credibility, and
generation quality {[}141{]}. Recent research has demonstrated that the
use of RAG significantly increases the peril of even the most secure
language models, as external knowledge sources introduce new attack
vectors and failure modes {[}142{]}.

The critical significance of trust in RAG systems is highlighted by
their deployment in high-stakes sectors such as healthcare, finance, and
legal services, where system failures can lead to substantial harm,
liability, or regulatory violations {[}143{]}. The trust issues in RAG
systems are inherently distinct from those in standalone LLMs, as they
introduce additional failure modes throughout the retrieval pipeline,
knowledge base integrity, and citation attribution processes {[}144{]}.

In order to gain a comprehensive understanding of the systemic risks
that are inherent in Retrieval, Augmented Generation (RAG) systems, we
have developed a multi-layered Trust Vulnerability Map, as shown in
Figure 9.1. This map delineates the primary sources of failure
throughout the architectural framework. Traditional trust issues in LLMs
primarily concentrate on hallucination and bias in the generator.
However, RAG systems incorporate additional trust dimensions that
encompass the retriever and knowledge base upstream, as well as citation
and attribution methods downstream.

The Knowledge Base layer is vulnerable to data poisoning, outdated or
absent knowledge, and biases in structural curation, as illustrated in
Figure 9.1. The Retriever layer may be susceptible to hostile queries
that exploit ranking algorithms or exacerbate source bias. The Generator
is susceptible to hallucinations, particularly when presented with token
truncation or nonsensical input, despite the benefits of enhanced
context. In the final analysis, the Citation Layer introduces risks
related to provenance distortion, in which responses that are
purportedly credible may be incorrectly associated with unverifiable or
malevolent sources.

\includegraphics[width=4.52701in,height=2.67028in]{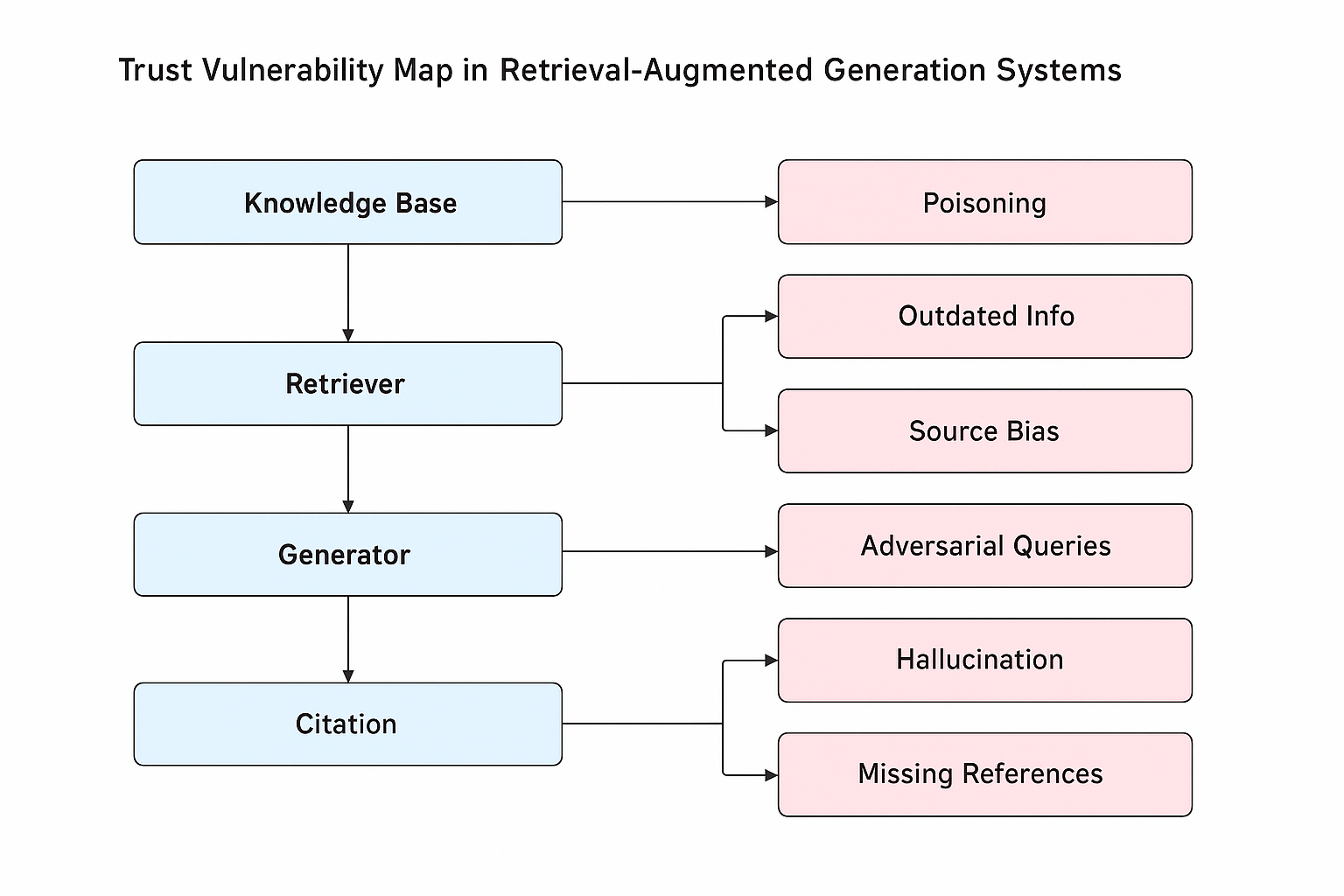}

\textbf{Figure 9.1: Trust Vulnerability Map in Retrieval,Augmented
Generation Systems.}

\subsection{Alignment Challenges in RAG Pipelines}

The alignment challenges that RAG systems introduce are inherently
distinct from those that are encountered in monolithic language models
due to their multi-component architecture {[}145{]}. The complex
interactions between retrieval mechanisms, knowledge base curation,
information synthesis, and generation processes are the source of these
challenges {[}146{]}.

Taxonomy of Reasoning Misalignment RAG systems are susceptible to
reasoning misalignment, which occurs when the model\textquotesingle s
reasoning trajectory deviates from the evidential constraints
established by retrieval {[}147{]}. This phenomenon can be
systematically decomposed into three interdependent phases: relevance
assessment, query-evidence mapping, and evidence-integrated synthesis
{[}147{]}.

\textbf{Table 9.1: RAG Reasoning Misalignment Phases}

\begin{longtable}[]{@{}
  >{\raggedright\arraybackslash}p{(\columnwidth - 6\tabcolsep) * \real{0.2470}}
  >{\raggedright\arraybackslash}p{(\columnwidth - 6\tabcolsep) * \real{0.3973}}
  >{\raggedright\arraybackslash}p{(\columnwidth - 6\tabcolsep) * \real{0.2350}}
  >{\raggedright\arraybackslash}p{(\columnwidth - 6\tabcolsep) * \real{0.1206}}@{}}
\toprule\noalign{}
\begin{minipage}[b]{\linewidth}\raggedright
Phase
\end{minipage} & \begin{minipage}[b]{\linewidth}\raggedright
Description
\end{minipage} & \begin{minipage}[b]{\linewidth}\raggedright
Common Failures
\end{minipage} & \begin{minipage}[b]{\linewidth}\raggedright
Impact Level
\end{minipage} \\
\midrule\noalign{}
\endhead
\bottomrule\noalign{}
\endlastfoot
Relevance Assessment & Failure to prioritize semantically relevant
evidence & Off-topic document selection & Medium \\
Query-Evidence Mapping & Misalignment in connecting queries to evidence
& Weak causal connections & High \\
Evidence-Integrated Synthesis & Logical inconsistencies in combining
evidence & Contradictory conclusions & Critical \\
\end{longtable}

\textbf{Trust Vulnerability Matrix}

RAG systems generate numerous trust surfaces that may be compromised by
a variety of attack vectors {[}148{]}. The vulnerability landscape
encompasses four critical layers: citation attribution accuracy,
generation fidelity, retrieval mechanism security, and knowledge base
integrity {[}149{]}.

\textbf{Table 9.2: RAG Trust Surface Analysis}

\begin{longtable}[]{@{}
  >{\raggedright\arraybackslash}p{(\columnwidth - 8\tabcolsep) * \real{0.1764}}
  >{\raggedright\arraybackslash}p{(\columnwidth - 8\tabcolsep) * \real{0.3387}}
  >{\raggedright\arraybackslash}p{(\columnwidth - 8\tabcolsep) * \real{0.1666}}
  >{\raggedright\arraybackslash}p{(\columnwidth - 8\tabcolsep) * \real{0.1714}}
  >{\raggedright\arraybackslash}p{(\columnwidth - 8\tabcolsep) * \real{0.1469}}@{}}
\toprule\noalign{}
\begin{minipage}[b]{\linewidth}\raggedright
Trust Surface
\end{minipage} & \begin{minipage}[b]{\linewidth}\raggedright
Primary Vulnerabilities
\end{minipage} & \begin{minipage}[b]{\linewidth}\raggedright
Attack Complexity
\end{minipage} & \begin{minipage}[b]{\linewidth}\raggedright
Detection Difficulty
\end{minipage} & \begin{minipage}[b]{\linewidth}\raggedright
Business Impact
\end{minipage} \\
\midrule\noalign{}
\endhead
\bottomrule\noalign{}
\endlastfoot
Knowledge Base & Data poisoning, bias injection & Medium & High &
Critical \\
Retrieval System & Prompt injection, similarity manipulation & Low &
Medium & High \\
Generation Layer & Hallucination, context ignoring & Low & Low & High \\
Citation Attribution & Source manipulation, false attribution & High &
Very High & Critical \\
\end{longtable}

\subsection{Critical Vulnerability Points and Attack Vectors}

Research has shown that RAG systems are susceptible to a variety of
attack categories, and attackers have the ability to introduce malicious
content during ingestion or even prior to data ingestion {[}150{]}. The
BadRAG framework demonstrates that the injection of only 10 malicious
passages results in high attack success rates while remaining difficult
to detect {[}143{]}.

\subsubsection{Corpus Manipulation and Data Poisoning}

Data poisoning is one of the most critical vulnerabilities in RAG
systems, as it occurs when malicious actors inject harmful or misleading
information into external knowledge bases {[}151{]}{[}152{]}. The
Phantom framework, which employs a two-stage malicious passage
optimization expressly designed to exploit RAG vulnerabilities, is one
of the sophisticated techniques that attackers can employ {[}153{]}.

\textbf{Table 9.3: RAG Attack Vector Classification}

\begin{longtable}[]{@{}
  >{\raggedright\arraybackslash}p{(\columnwidth - 8\tabcolsep) * \real{0.2179}}
  >{\raggedright\arraybackslash}p{(\columnwidth - 8\tabcolsep) * \real{0.2881}}
  >{\raggedright\arraybackslash}p{(\columnwidth - 8\tabcolsep) * \real{0.1382}}
  >{\raggedright\arraybackslash}p{(\columnwidth - 8\tabcolsep) * \real{0.1359}}
  >{\raggedright\arraybackslash}p{(\columnwidth - 8\tabcolsep) * \real{0.2200}}@{}}
\toprule\noalign{}
\begin{minipage}[b]{\linewidth}\raggedright
Attack Type
\end{minipage} & \begin{minipage}[b]{\linewidth}\raggedright
Mechanism
\end{minipage} & \begin{minipage}[b]{\linewidth}\raggedright
Stealth Level
\end{minipage} & \begin{minipage}[b]{\linewidth}\raggedright
Success Rate
\end{minipage} & \begin{minipage}[b]{\linewidth}\raggedright
Mitigation Complexity
\end{minipage} \\
\midrule\noalign{}
\endhead
\bottomrule\noalign{}
\endlastfoot
Corpus Poisoning & Malicious document injection & High & High &
Complex \\
Prompt Injection & Query manipulation & Medium & Medium & Moderate \\
Citation Manipulation & Source attribution fraud & Very High & High &
Very Complex \\
Context Confusion & Contradictory information & Medium & Medium &
Moderate \\
Bias Amplification & Systematic preference skewing & High & High &
Complex \\
\end{longtable}

\textbf{Industry-Specific Risk Assessment}

Industries are subject to varying degrees of RAG-related hazards, which
are determined by their regulatory obligations and data sensitivity
{[}154{]}.

\textbf{Table 9.4: Industry Risk Profile Comparison}

\begin{longtable}[]{@{}
  >{\raggedright\arraybackslash}p{(\columnwidth - 8\tabcolsep) * \real{0.1554}}
  >{\raggedright\arraybackslash}p{(\columnwidth - 8\tabcolsep) * \real{0.3442}}
  >{\raggedright\arraybackslash}p{(\columnwidth - 8\tabcolsep) * \real{0.1951}}
  >{\raggedright\arraybackslash}p{(\columnwidth - 8\tabcolsep) * \real{0.1716}}
  >{\raggedright\arraybackslash}p{(\columnwidth - 8\tabcolsep) * \real{0.1337}}@{}}
\toprule\noalign{}
\begin{minipage}[b]{\linewidth}\raggedright
Industry
\end{minipage} & \begin{minipage}[b]{\linewidth}\raggedright
Primary Risk Categories
\end{minipage} & \begin{minipage}[b]{\linewidth}\raggedright
Regulatory Framework
\end{minipage} & \begin{minipage}[b]{\linewidth}\raggedright
Audit Requirements
\end{minipage} & \begin{minipage}[b]{\linewidth}\raggedright
Risk Tolerance
\end{minipage} \\
\midrule\noalign{}
\endhead
\bottomrule\noalign{}
\endlastfoot
Healthcare & Privacy violations, bias in diagnoses & HIPAA, FDA &
Continuous & Very Low \\
Financial Services & Market manipulation, algorithmic bias & SEC, FINRA
& Quarterly & Low \\
Legal & Citation fraud, precedent misrepresentation & Professional codes
& Ongoing & Very Low \\
Government & Information warfare, decision manipulation & Security
standards & Continuous & Minimal \\
Education & Misinformation, academic bias & FERPA, COPPA & Annual &
Medium \\
\end{longtable}

\subsection{Comprehensive Mitigation Strategies}

Deliberate, multifaceted mitigation strategies that address
vulnerabilities throughout the entire information lifecycle are
necessary to establish reliable RAG systems {[}155{]}. In addition to
safeguarding consumers, organizations are also protected from legal and
reputational risks by adhering to frameworks such as GDPR, CCPA, and SOC
2 {[}155{]}.

\textbf{Technical Defense Mechanisms}

The security and trustworthiness of the RAG system can be improved
through the implementation of numerous technical strategies {[}156{]}.
The TrustRAG framework exhibits substantial enhancements in system
reliability by employing a two-stage defense mechanism that incorporates
self-assessment and K-means clustering {[}157{]}{[}158{]}.

\textbf{Table 9.5: RAG Defense Strategy Comparison}

\begin{longtable}[]{@{}
  >{\raggedright\arraybackslash}p{(\columnwidth - 8\tabcolsep) * \real{0.1833}}
  >{\raggedright\arraybackslash}p{(\columnwidth - 8\tabcolsep) * \real{0.3293}}
  >{\raggedright\arraybackslash}p{(\columnwidth - 8\tabcolsep) * \real{0.1833}}
  >{\raggedright\arraybackslash}p{(\columnwidth - 8\tabcolsep) * \real{0.1181}}
  >{\raggedright\arraybackslash}p{(\columnwidth - 8\tabcolsep) * \real{0.1860}}@{}}
\toprule\noalign{}
\begin{minipage}[b]{\linewidth}\raggedright
Defense Strategy
\end{minipage} & \begin{minipage}[b]{\linewidth}\raggedright
Technical Approach
\end{minipage} & \begin{minipage}[b]{\linewidth}\raggedright
Implementation Effort
\end{minipage} & \begin{minipage}[b]{\linewidth}\raggedright
Effectiveness
\end{minipage} & \begin{minipage}[b]{\linewidth}\raggedright
Maintenance Overhead
\end{minipage} \\
\midrule\noalign{}
\endhead
\bottomrule\noalign{}
\endlastfoot
Input Validation & Query sanitization and filtering & Low & Moderate &
Low \\
TrustRAG Framework & Semantic chunking and citation enhancement & High &
High & Medium \\
Content Filtering & Multi-layered security screening & Medium & High &
Medium \\
Human-in-the-Loop & Expert verification and feedback & Very High & Very
High & High \\
Multi-Modal Defense & Combined technical and procedural controls & Very
High & Highest & High \\
\end{longtable}

\textbf{Human-in-the-Loop Integration}

Human oversight is a crucial safety net for AI systems, as automated
systems are unable to completely replicate human judgment for complex or
high-stakes information assessment {[}159{]}. The Human-in-the-Loop
approach establishes a continuous partnership between the efficiency of
machines and the insights of humans {[}159{]}.

\textbf{Table 9.6: HITL Implementation Models}

\begin{longtable}[]{@{}
  >{\raggedright\arraybackslash}p{(\columnwidth - 8\tabcolsep) * \real{0.2081}}
  >{\raggedright\arraybackslash}p{(\columnwidth - 8\tabcolsep) * \real{0.2499}}
  >{\raggedright\arraybackslash}p{(\columnwidth - 8\tabcolsep) * \real{0.1685}}
  >{\raggedright\arraybackslash}p{(\columnwidth - 8\tabcolsep) * \real{0.2474}}
  >{\raggedright\arraybackslash}p{(\columnwidth - 8\tabcolsep) * \real{0.1261}}@{}}
\toprule\noalign{}
\begin{minipage}[b]{\linewidth}\raggedright
HITL Model
\end{minipage} & \begin{minipage}[b]{\linewidth}\raggedright
Scope
\end{minipage} & \begin{minipage}[b]{\linewidth}\raggedright
Response Time
\end{minipage} & \begin{minipage}[b]{\linewidth}\raggedright
Accuracy Improvement
\end{minipage} & \begin{minipage}[b]{\linewidth}\raggedright
Scalability
\end{minipage} \\
\midrule\noalign{}
\endhead
\bottomrule\noalign{}
\endlastfoot
Continuous Review & All outputs & Real-time & Highest & Limited \\
Threshold-Based & Low-confidence outputs & Near real-time & High &
Moderate \\
Sampling Review & Statistical sampling & Periodic & Moderate & High \\
Expert Validation & Critical decisions only & Variable & Highest &
Limited \\
Feedback Loop & Iterative improvement & Ongoing & Progressive & High \\
\end{longtable}

\subsection{RAG-Specific Red Teaming and Security Testing}

Specialized red teaming approaches that surpass conventional language
model testing are necessitated by the composite architecture of RAG
systems {[}160{]}. To proactively identify vulnerabilities before they
can be exploited, red teaming involves simulating adversarial attacks
{[}161{]}.

\textbf{Red Teaming Implementation Framework}

A comprehensive red teaming program should incorporate systematic
testing protocols that address the security of the knowledge base, the
integrity of retrieval, and the robustness of generation {[}162{]}. The
significance of specialized red teaming is illustrated by healthcare
implementations, which involve expert teams undertaking systematic
evaluations using structured prompt sets {[}150{]}.

\textbf{Table 9.7: Red Teaming Methodology Comparison}

\begin{longtable}[]{@{}
  >{\raggedright\arraybackslash}p{(\columnwidth - 8\tabcolsep) * \real{0.2189}}
  >{\raggedright\arraybackslash}p{(\columnwidth - 8\tabcolsep) * \real{0.2549}}
  >{\raggedright\arraybackslash}p{(\columnwidth - 8\tabcolsep) * \real{0.1517}}
  >{\raggedright\arraybackslash}p{(\columnwidth - 8\tabcolsep) * \real{0.2231}}
  >{\raggedright\arraybackslash}p{(\columnwidth - 8\tabcolsep) * \real{0.1513}}@{}}
\toprule\noalign{}
\begin{minipage}[b]{\linewidth}\raggedright
Testing Approach
\end{minipage} & \begin{minipage}[b]{\linewidth}\raggedright
Knowledge Required
\end{minipage} & \begin{minipage}[b]{\linewidth}\raggedright
Testing Depth
\end{minipage} & \begin{minipage}[b]{\linewidth}\raggedright
Resource Requirements
\end{minipage} & \begin{minipage}[b]{\linewidth}\raggedright
Discovery Rate
\end{minipage} \\
\midrule\noalign{}
\endhead
\bottomrule\noalign{}
\endlastfoot
White-Box Testing & Full system access & Complete & High & Highest \\
Black-Box Testing & No internal knowledge & Surface-level & Medium &
Moderate \\
Gray-Box Testing & Partial system knowledge & Targeted & Medium &
High \\
Adversarial Simulation & Attack pattern knowledge & Scenario-based &
High & High \\
Continuous Testing & Ongoing system monitoring & Dynamic & Very High &
Progressive \\
\end{longtable}

\textbf{Attack Simulation Categories}

In order to guarantee extensive coverage, red teaming should incorporate
numerous attack categories {[}163{]}:

\textbf{Table 9.8: Red Team Attack Simulation Framework}

\begin{longtable}[]{@{}
  >{\raggedright\arraybackslash}p{(\columnwidth - 8\tabcolsep) * \real{0.1998}}
  >{\raggedright\arraybackslash}p{(\columnwidth - 8\tabcolsep) * \real{0.2504}}
  >{\raggedright\arraybackslash}p{(\columnwidth - 8\tabcolsep) * \real{0.1796}}
  >{\raggedright\arraybackslash}p{(\columnwidth - 8\tabcolsep) * \real{0.1505}}
  >{\raggedright\arraybackslash}p{(\columnwidth - 8\tabcolsep) * \real{0.2197}}@{}}
\toprule\noalign{}
\begin{minipage}[b]{\linewidth}\raggedright
Attack Category
\end{minipage} & \begin{minipage}[b]{\linewidth}\raggedright
Simulation Method
\end{minipage} & \begin{minipage}[b]{\linewidth}\raggedright
Detection Challenge
\end{minipage} & \begin{minipage}[b]{\linewidth}\raggedright
Business Impact
\end{minipage} & \begin{minipage}[b]{\linewidth}\raggedright
Countermeasure Priority
\end{minipage} \\
\midrule\noalign{}
\endhead
\bottomrule\noalign{}
\endlastfoot
Data Poisoning & Malicious content injection & High & Critical &
Maximum \\
Prompt Injection & Query manipulation testing & Medium & High & High \\
Social Engineering & Human factor exploitation & Variable & High &
High \\
Technical Exploitation & System vulnerability testing & Low & Medium &
Medium \\
Bias Exploitation & Systematic preference testing & High & High &
High \\
\end{longtable}

\subsection{Regulatory Compliance and Governance Frameworks}

In order to guarantee legal compliance, mitigate risk, and maintain
ethical standards, comprehensive governance frameworks are required for
the implementation of RAG systems in regulated industries.
{[}164{]}{[}165{]}. The NIST AI Risk Management Framework underscores
the importance of fostering a risk-aware organizational culture
{[}166{]}{[}167{]}.

\textbf{Industry-Specific Compliance Requirements}

The implementation of RAG systems presents distinct regulatory
challenges for various sectors {[}168{]}.

\textbf{Table 9.9: Regulatory Compliance Framework by Sector}

\begin{longtable}[]{@{}
  >{\raggedright\arraybackslash}p{(\columnwidth - 8\tabcolsep) * \real{0.1153}}
  >{\raggedright\arraybackslash}p{(\columnwidth - 8\tabcolsep) * \real{0.2642}}
  >{\raggedright\arraybackslash}p{(\columnwidth - 8\tabcolsep) * \real{0.2787}}
  >{\raggedright\arraybackslash}p{(\columnwidth - 8\tabcolsep) * \real{0.1901}}
  >{\raggedright\arraybackslash}p{(\columnwidth - 8\tabcolsep) * \real{0.1517}}@{}}
\toprule\noalign{}
\begin{minipage}[b]{\linewidth}\raggedright
Sector
\end{minipage} & \begin{minipage}[b]{\linewidth}\raggedright
Primary Regulations
\end{minipage} & \begin{minipage}[b]{\linewidth}\raggedright
Key Requirements
\end{minipage} & \begin{minipage}[b]{\linewidth}\raggedright
Compliance Mechanisms
\end{minipage} & \begin{minipage}[b]{\linewidth}\raggedright
Enforcement Level
\end{minipage} \\
\midrule\noalign{}
\endhead
\bottomrule\noalign{}
\endlastfoot
Healthcare & HIPAA, FDA, WHO Guidelines & Privacy protection, safety
validation & Audit trails, encryption & Strict \\
Financial & SEC, FINRA, Basel III & Algorithmic transparency, fair
lending & Real-time monitoring & Strict \\
Legal & Professional responsibility codes & Client confidentiality,
competence & Ethics review, supervision & Strict \\
Government & FISMA, FedRAMP, security standards & Continuous monitoring,
security controls & Ongoing assessment & Maximum \\
Education & FERPA, COPPA & Student privacy, age-appropriate content &
Annual review & Moderate \\
\end{longtable}

\textbf{Governance Implementation Maturity Model}

A structured maturity model {[}169{]} can be employed by organizations
to evaluate and enhance their RAG governance capabilities.

\textbf{Table 9.10: RAG Governance Maturity Assessment}

\begin{longtable}[]{@{}
  >{\raggedright\arraybackslash}p{(\columnwidth - 8\tabcolsep) * \real{0.1531}}
  >{\raggedright\arraybackslash}p{(\columnwidth - 8\tabcolsep) * \real{0.3189}}
  >{\raggedright\arraybackslash}p{(\columnwidth - 8\tabcolsep) * \real{0.1968}}
  >{\raggedright\arraybackslash}p{(\columnwidth - 8\tabcolsep) * \real{0.1767}}
  >{\raggedright\arraybackslash}p{(\columnwidth - 8\tabcolsep) * \real{0.1545}}@{}}
\toprule\noalign{}
\begin{minipage}[b]{\linewidth}\raggedright
Maturity Level
\end{minipage} & \begin{minipage}[b]{\linewidth}\raggedright
Governance Characteristics
\end{minipage} & \begin{minipage}[b]{\linewidth}\raggedright
Risk Management
\end{minipage} & \begin{minipage}[b]{\linewidth}\raggedright
Monitoring Capabilities
\end{minipage} & \begin{minipage}[b]{\linewidth}\raggedright
Compliance Status
\end{minipage} \\
\midrule\noalign{}
\endhead
\bottomrule\noalign{}
\endlastfoot
Level 1: Basic & Ad-hoc policies, reactive approach & Incident-driven &
Manual oversight & Non-compliant \\
Level 2: Managed & Documented procedures, risk awareness & Structured
response & Basic automation & Partially compliant \\
Level 3: Defined & Standardized processes, proactive measures &
Comprehensive mitigation & Systematic monitoring & Mostly compliant \\
Level 4: Measured & Quantified governance, predictive analytics &
Data-driven decisions & Real-time dashboards & Fully compliant \\
Level 5: Optimizing & Continuous improvement, adaptive management &
Autonomous optimization & AI-driven insights & Exceeds compliance \\
\end{longtable}

\subsection{Evaluation and Continuous Improvement}

In order to guarantee the ongoing trustworthiness of the RAG system, it
is necessary to conduct a systematic evaluation using exhaustive
frameworks {[}170{]}. The RAGAS evaluation system offers structured
metrics for evaluating context precision, faithfulness, answer
relevancy, and context recall {[}171{]}{[}172{]}.

\textbf{Bias Detection and Mitigation}

RAG systems may unintentionally propagate biases associated with
sensitive demographic attributes, requiring systematic evaluation and
mitigation strategies {[}173{]}. Multiple bias mitigation approaches
have demonstrated effectiveness including Chain-of-Thought reasoning,
Counterfactual filtering, and Majority Vote aggregation {[}173{]}.
Recent overviews propose systematic pipelines for bias identification
and mitigation specific to RAG {[}174{]}.

\textbf{Table 9.11: Bias Mitigation Strategies~and Their Performance
Impact}

\begin{longtable}[]{@{}
  >{\raggedright\arraybackslash}p{(\columnwidth - 8\tabcolsep) * \real{0.2104}}
  >{\raggedright\arraybackslash}p{(\columnwidth - 8\tabcolsep) * \real{0.2754}}
  >{\raggedright\arraybackslash}p{(\columnwidth - 8\tabcolsep) * \real{0.2175}}
  >{\raggedright\arraybackslash}p{(\columnwidth - 8\tabcolsep) * \real{0.1293}}
  >{\raggedright\arraybackslash}p{(\columnwidth - 8\tabcolsep) * \real{0.1674}}@{}}
\toprule\noalign{}
\begin{minipage}[b]{\linewidth}\raggedright
Mitigation Approach
\end{minipage} & \begin{minipage}[b]{\linewidth}\raggedright
Technical Method
\end{minipage} & \begin{minipage}[b]{\linewidth}\raggedright
Implementation Complexity
\end{minipage} & \begin{minipage}[b]{\linewidth}\raggedright
Bias Reduction
\end{minipage} & \begin{minipage}[b]{\linewidth}\raggedright
Performance Impact
\end{minipage} \\
\midrule\noalign{}
\endhead
\bottomrule\noalign{}
\endlastfoot
Chain-of-Thought & Step-by-step reasoning prompts & Low & Moderate &
Minimal \\
Counterfactual Filtering & Cross-demographic validation & Medium & High
& Low \\
Adversarial Prompting & Identity-aware prompt design & Medium & Moderate
& Low \\
Majority Vote Aggregation & Multi-variant output combination & High &
Highest & Medium \\
Demographic Parity & Balanced representation enforcement & High & High &
Medium \\
\end{longtable}

This comprehensive approach to RAG trust, alignment, and safety equips
organizations with the frameworks and tools required to deploy
compliant, secure, and reliable RAG systems in a variety of enterprise
contexts, all while upholding high standards of regulatory adherence and
trustworthiness.

\section{Frontier Challenges and Future Directions}

The rapid evolution of Retrieval, Augmented Generation (RAG) systems has
reached a critical inflection point, where traditional architectures are
approaching their theoretical and practical limitations. This demands
fundamental advancements in system design, training methodologies, and
operational paradigms {[}175{]}{[}176{]}{[}177{]}. The convergence of
numerous technological trends, such as advancements in differentiable
programming, reinforcement learning from human feedback, multiagent
systems, multimodal processing, and self-supervised learning, has
provided unprecedented opportunities for the evolution of RAG systems
{[}178{]}{[}179{]}{[}180{]}. Industry roadmaps underline the rapid
operationalization of retrieval-augmented systems across verticals
{[}203{]}.

\subsection{End-to-End Differentiable RAG Training: Unified
Optimization Frameworks}

The current paradigm of training retrieval and generation components
independently results in a fundamental optimization misalignment that
restricts the overall performance of the system {[}175{]}{[}176{]}.
End-to-end, differentiable training is a transformative approach that
has the potential to revolutionize the effectiveness and coherence of
RAG systems by facilitating the joint optimization of all system
components through unified gradient-based learning {[}175{]}{[}176{]}.

\subsubsection{Mathematical Frameworks and Theoretical Foundations}

The primary obstacle in end-to-end RAG training is the preservation of
computational efficacy while making discrete retrieval operations
differentiable {[}176{]}. In comparison to conventional two-stage
methods, recent research has shown that differentiable retrieval can
accomplish substantial enhancements in retrieval and generation
alignment through the use of soft attention mechanisms
{[}176{]}{[}181{]}. The unified objective function integrates retrieval
accuracy, generation quality, and task-specific performance metrics
through learnable hyperparameters, as opposed to manual optimization
{[}175{]}.

\subsubsection{Results of Innovative Research and Implementation}

The Differentiable Data Rewards (DDR) method is the most sophisticated
approach to end-to-end RAG optimization, allowing for the propagation of
rewards throughout the system through rollout-based training {[}175{]}.
This method achieves substantial enhancements over supervised
fine-tuning methods by employing Direct Preference Optimization (DPO) to
align data preferences between various RAG modules {[}175{]}.
Experimental results indicate that DDR outperforms conventional methods,
particularly for language models of a smaller scale that rely more
heavily on retrieved knowledge {[}175{]}.

The Stochastic RAG approach offers another revolution in end-to-end
optimization by recasting retrieval as a stochastic sampling process
{[}176{]}. This formulation utilizes straight-through Gumbel, top-k
sampling to generate differentiable approximations, thereby enhancing
the state-of-the-art results on six of the seven datasets that were
evaluated {[}176{]}.

\subsection{RLHF for Retrieval,Generator Co-Evolution: Human-Guided
Optimization}

Strengthening The application of Learning from Human Feedback (RLHF) to
RAG systems facilitates the sophisticated co-evolution of retrieval and
generation components in accordance with human preferences and expertise
{[}178{]}{[}182{]}{[}183{]}. This method confronts the most significant
obstacles in the alignment of RAG systems with human values,
professional standards, and domain-specific requirements
{[}178{]}{[}184{]}.

\subsubsection{Advanced RLHF Architectures for RAG Systems}

Sophisticated reward models that simultaneously assess retrieval quality
and generation appropriateness are necessary for RLHF in RAG
{[}178{]}{[}184{]}. The RAG,Reward framework introduces exhaustive
quality assessment metrics that are intended to facilitate the
development of RAG systems that are reliable, efficient, comprehensive,
and hallucination-free {[}178{]}{[}184{]}. This framework establishes
four critical metrics for evaluating the quality of generation and
creates automated annotation algorithms that utilize multiple language
models to produce outputs in a variety of RAG scenarios {[}178{]}.

\subsubsection{Constitutional AI Integration}

Constitutional AI is the most sophisticated RLHF implementation for
knowledge-intensive tasks, as it includes comprehensive self-monitoring
and corrective mechanisms {[}185{]}{[}186{]}. These systems exhibit
substantial enhancements in factual accuracy and decreases in the
generation of detrimental content by learning constitutional principles
for information retrieval and synthesis {[}185{]}{[}186{]}. The
integration of constitutional principles with RAG systems facilitates
the synthesis of information that is more ethical and dependable
{[}185{]}.

\subsection{Collaborative Intelligence Frameworks for Multi-Agent RAG
Planning}

Multiagent RAG systems embody a paradigm transition from monolithic
architectures to collaborative frameworks, in which specialized agents
collaborate to complete intricate knowledge-intensive tasks
{[}179{]}{[}187{]}{[}188{]}. This method provides advanced reasoning,
planning, and execution capabilities that surpass the constraints of
single-agent systems {[}189{]}{[}190{]}.

\subsubsection{Architectures for Agent Specialization and Coordination}

The MA-RAG framework illustrates how multi-agent systems can resolve the
inherent ambiguities and reasoning challenges that arise in intricate
information-seeking tasks {[}187{]}. This framework orchestrates
specialized agents, such as the Planner, Step Definer, Extractor, and QA
Agents, to address each stage of the RAG pipeline using task-aware
reasoning {[}187{]}. In comparison to baseline RAG systems, the
hierarchical multi-agent approach obtains significant enhancements in
question classification and answer accuracy {[}190{]}.

\subsubsection{Revolutionary Multi-Agent Implementations}

The orchestrator-worker pattern is implemented in
Anthropic\textquotesingle s multi-agent research system, in which a main
agent oversees the process and delegated it to specialized subagents
that operate in parallel {[}189{]}. This architecture employs a
multi-step search process that dynamically identifies pertinent
information, adjusts to new discoveries, and analyzes the results to
produce high-quality responses {[}189{]}. The collaborative approach
between multiple specialized agents facilitates the management of a wide
range of data sources, such as relational databases, document
repositories, and graph databases {[}188{]}.

\subsection{Multimodal RAG with Streaming Memory: Beyond Text
Processing}

The incorporation of streaming memory architectures with multimodal
processing capabilities is a fundamental evolution toward more
human-like information processing and reasoning
{[}180{]}{[}191{]}{[}192{]}. These systems preserve temporal coherence
and adaptive memory management while integrating text, images, audio,
video, and structured data into unified representations
{[}180{]}{[}192{]}.

\subsubsection{Multimodal Representations That Are Unified}

Three primary approaches are employed by advanced multimodal RAG
systems: unified embedding spaces, grounding modalities to text, and
discrete datastores with reranking {[}180{]}{[}192{]}. The unified
embedding approach employs models such as CLIP to encode both text and
images in the same vector space, thereby enabling a text-only RAG
infrastructure with multimodal capabilities that is essentially
unchanged {[}180{]}{[}191{]}. Using vision and language models, the
grounding approach simplifies downstream processing while preserving
rich semantic information by converting non-text modalities into text
descriptions {[}193{]}{[}192{]}.

\subsubsection{Innovations in Cross-Modal Processing}

ACE is a groundbreaking approach to generative cross-modal retrieval
that integrates K-Means and RQ-VAE algorithms to generate coarse and
fine tokens that function as identifiers for multimodal data. This
method surpasses dual tower architectures based on embedding by
substantial margins in cross-modal retrieval, achieving state-of-the-art
performance {[}194{]}. The coarse-to-fine\footnote{``Coarse-to-fine''
  refers to a hierarchical fusion process that first aligns coarse
  semantic prototypes, then refines them into fine-grained
  representations for precise multimodal matching.} feature fusion
strategy effectively aligns candidate identifiers with natural language
queries across multiple modalities {[}194{]}.

\subsection{Self-Evaluating RAG Systems and Internal Fact-Checking
Modules}

The advancement of autonomous, reliable, and trustworthy AI systems is
represented by the development of self-evaluating RAG systems with
incorporated fact-checking capabilities {[}195{]}{[}196{]}{[}197{]}.
These architectures are equipped with advanced self-monitoring, error
detection, and correction mechanisms that facilitate the continuous
development of quality and the mitigation of risks {[}195{]}{[}198{]}.

\subsubsection{Architectures and Mechanisms for Self-Evaluation}

A novel approach is introduced by the Self RAG framework, which trains
language models to retrieve, generate, and critique through
self-reflection {[}195{]}{[}196{]}{[}197{]}. This system utilizes
reflection tokens to allow models to evaluate the relevance of retrieved
passages, determine the necessity of retrieval, and evaluate the factual
veracity of their own generations {[}195{]}{[}196{]}. The framework is
capable of accommodating a variety of downstream applications by
enabling the implementation of decoding algorithms that are customizable
and influenced by the probabilities of reflection tokens
{[}196{]}{[}197{]}.

\subsubsection{Revolutionary Implementations of Self-Evaluation}

In comparison to traditional RAG methods and state-of-the-art language
models, Self,RAG exhibits substantial performance enhancements across a
variety of tasks {[}195{]}{[}196{]}{[}197{]}. The system allows
practitioners to customize model behaviors to meet their specific
fine-grained preferences, such as prioritizing fluency for more flexible
generation or emphasizing evidence support to enhance citation precision
{[}195{]}{[}197{]}. The significance of collecting external evidence
during verification is underscored by research on automated
fact-checking that employs large language models and demonstrates that
contextual information considerably enhances accuracy {[}198{]}.

\subsection{Convergence and Integration: Toward Unified Next,Generation
RAG}

The aforementioned frontier challenges are progressively merging into
unified architectures that incorporate numerous advanced capabilities
into coherent, powerful systems {[}199{]}{[}177{]}{[}200{]}. This
convergence signifies the emergence of RAG systems that are genuinely
next-generation, surpassing current constraints and presenting new
opportunities for AI applications {[}201{]}{[}200{]}.

\subsubsection{Design Principles of Unified Architecture}

Sophisticated modular architectures are implemented in next-generation
systems, which facilitate the flexible integration of advanced
capabilities while preserving system coherence and performance
{[}177{]}{[}202{]}. This method is exemplified by the Patchwork
framework, which offers a comprehensive end-to-end RAG serving framework
that resolves efficiency constraints by utilizing distributed inference
optimization and flexible specification interfaces {[}177{]}. These
systems achieve significant performance enhancements, with throughput
gains exceeding 48\% and a 24\% reduction in service level objective
violations {[}177{]}.

\subsubsection{The emergence of autonomous RAG systems}

The convergence of self-evaluation, RLHF, and differentiable training
enables autonomous systems to perpetually enhance their capabilities
through feedback and experience {[}200{]}. Future RAG systems are
progressing toward the integration of multimodal, real-time, and
autonomous knowledge that surpasses basic text retrieval {[}200{]}.
These sophisticated systems will integrate a variety of AI techniques,
such as reinforcement learning, real-time retrieval, fine-tuned domain
adaptation, and pre-trained knowledge, to develop AI that actively
learns and reasons {[}200{]}.

\textbf{Table 10.1: Advancing RAG: Research Milestones and Deployment
Trajectories}

\begin{longtable}[]{@{}
  >{\raggedright\arraybackslash}p{(\columnwidth - 10\tabcolsep) * \real{0.1343}}
  >{\raggedright\arraybackslash}p{(\columnwidth - 10\tabcolsep) * \real{0.1813}}
  >{\raggedright\arraybackslash}p{(\columnwidth - 10\tabcolsep) * \real{0.1878}}
  >{\raggedright\arraybackslash}p{(\columnwidth - 10\tabcolsep) * \real{0.1848}}
  >{\raggedright\arraybackslash}p{(\columnwidth - 10\tabcolsep) * \real{0.1130}}
  >{\raggedright\arraybackslash}p{(\columnwidth - 10\tabcolsep) * \real{0.1987}}@{}}
\toprule\noalign{}
\begin{minipage}[b]{\linewidth}\raggedright
Research Area
\end{minipage} & \begin{minipage}[b]{\linewidth}\raggedright
Key Innovation
\end{minipage} & \begin{minipage}[b]{\linewidth}\raggedright
Leading Research Groups
\end{minipage} & \begin{minipage}[b]{\linewidth}\raggedright
Key Technical Breakthrough
\end{minipage} & \begin{minipage}[b]{\linewidth}\raggedright
Commercial Readiness
\end{minipage} & \begin{minipage}[b]{\linewidth}\raggedright
Primary Applications
\end{minipage} \\
\midrule\noalign{}
\endhead
\bottomrule\noalign{}
\endlastfoot
End-to-End Differentiable RAG & Joint optimization of retrieval and
generation through unified gradient flow & Tsinghua University, Carnegie
Mellon University, Northeastern University & Differentiable Data Rewards
(DDR) method for end-to-end optimization & 2025-2026 &
Knowledge-intensive QA, Scientific research, Enterprise intelligence \\
RLHF for RAG Systems & Human feedback integration for preference
alignment in RAG components & Anthropic, OpenAI, Various academic
institutions & Constitutional AI with reward modeling for RAG alignment
& 2024-2025 & Conversational AI, Content generation, Legal research
systems \\
Multi-Agent RAG Planning & Collaborative intelligence through
specialized agent coordination & Dartmouth College, Stanford AI
Laboratory, MIT AI Laboratory & Hierarchical multi-agent coordination
with specialized reasoning & 2025-2027 & Complex reasoning, Enterprise
knowledge management, Scientific discovery \\
Multimodal RAG Integration & Cross-modal information processing with
streaming memory & NVIDIA, Microsoft, Google Research & Unified
cross-modal embeddings with real-time processing & 2026-2028 &
Autonomous systems, Healthcare diagnostics, Multimedia analysis \\
Self-Evaluating RAG Systems & Self-reflection and autonomous quality
assessment capabilities & Anthropic, Various AI safety research groups &
Self-reflection tokens and autonomous fact-checking & 2025-2026 & Fact
verification, Content moderation, Quality assurance \\
Unified RAG Architectures & Modular convergence of multiple advanced RAG
techniques & Intel Research, Multiple industry consortiums & Modular
plugin architectures for evolutionary systems & 2027-2029 &
Next-generation AI platforms, Adaptive systems, Universal interfaces \\
\end{longtable}

\subsection{Timeline and Development Roadmap}

Unity architectures that incorporate numerous advanced capabilities into
coherent, powerful systems are becoming more prevalent as the frontier
challenges described above continue to converge
{[}199{]}{[}177{]}{[}200{]}. The emergence of genuinely next-generation
RAG systems that transcend current limitations and open up new
possibilities for AI applications is represented by this convergence
{[}201{]}{[}200{]}.

Principles of Unified Architecture Design

Next-generation systems utilize sophisticated modular architectures that
facilitate the flexible integration of advanced capabilities while
preserving system coherence and performance {[}177{]}{[}202{]}. The
Patchwork framework is a prime example of this approach, as it offers a
comprehensive end-to-end RAG serving framework that resolves efficiency
constraints by utilizing distributed inference optimization and flexible
specification interfaces {[}177{]}. While simultaneously reducing
service level objective violations by 24\%, these systems achieve
substantial performance improvements, with throughput gains exceeding
48\% {[}177{]}.

Emergence of Autonomous RAG Systems

The convergence of self-evaluation, RLHF, and differentiable training
provides autonomous systems with the ability to perpetually enhance
their capabilities through feedback and experience {[}200{]}. In the
future, RAG systems will progress toward the incorporation of knowledge
that is multimodal, real-time, and autonomous, surpassing the
capabilities of simple text retrieval {[}200{]}. The active reasoning
and learning capabilities of these advanced systems will be achieved by
integrating a variety of AI techniques, such as reinforcement learning,
real-time retrieval, fine-tuned domain adaptation, and pre-trained
knowledge {[}200{]}.

\section{Conclusion: The Future of RAG Engineering}

The comprehensive analysis of Retrieval-Augmented Generation (RAG)
systems exposes a technology that has evolved from experimental
prototypes to production-ready enterprise solutions, fundamentally
transforming the way organizations approach knowledge-intensive
artificial intelligence {[}204{]}{[}205{]}. This development is
indicative of a paradigm shift from monolithic language models to
modular, scalable architectures that incorporate external knowledge
sources while maintaining the reliability, transparency, and performance
standards that are essential for enterprise deployment
{[}206{]}{[}207{]}. RAG is established as a cornerstone technology for
next-generation AI systems that bridge the divide between parametric and
non-parametric knowledge integration through the systematic examination
of current research, enterprise implementations, and emerging trends
{[}208{]}{[}209{]}.

\subsection{Contributions and Key Findings}

The analysis establishes a number of critical findings that define the
current state and trajectory of RAG technology. Architectural Evolution:
RAG systems have evolved through three distinct paradigms---Naive RAG,
Advanced RAG, and Modular RAG---each of which introduces new
capabilities and addresses specific limitations {[}204{]}{[}210{]}. The
field\textquotesingle s rapid maturation and increasing sophistication
are illustrated by the transition from simple retrieval-then-generate
pipelines to sophisticated multi-agent, self-evaluating systems
{[}211{]}{[}212{]}.

Accelerating Enterprise Adoption: Market research indicates that 78\% of
organizations are currently employing AI in at least one business
function, with 71\% of them expressly implementing generative AI
solutions {[}213{]}{[}214{]}. This is an unprecedented level of
enterprise adoption. The global RAG market has grown from \$1.2 billion
in 2023 to a projected \$11.0 billion by 2030, a compound annual growth
rate of 49.1\%. {[}215{]}. The transition from experimentation to
production deployment is evident in the sixfold increase in enterprise
AI expenditure from \$2.3 billion in 2023 to \$13.8 billion in 2024
{[}216{]}.

Performance and ROI Validation: Quantitative analysis indicates that RAG
implementations generate quantifiable business value, with organizations
reporting an average 3.7x return on investment for generative AI
deployments {[}217{]}. Implementation excellence has a substantial
impact on business outcomes, as evidenced by the 10.3x ROI rates
achieved by leading enterprises {[}217{]}. Early adopters report an
average ROI of 41\% across AI initiatives, with 92\% experiencing
positive returns {[}218{]}.{[}219{]}.

Technical Maturity: The discipline has established robust evaluation
frameworks, including comprehensive benchmarks such as BEIR
(Benchmarking Information Retrieval), which covers 19 datasets across 9
information retrieval tasks {[}220{]}.{[}221{]}. End-to-end
optimization, constitutional AI integration, and multimodal processing
have evolved from research concepts to practical applications
{[}222{]}{[}223{]}{[}224{]}.

\subsection{Practitioners\textquotesingle{} Strategic Implications}

For Technology Leaders: RAG is a strategic technology investment that
has a sustainable competitive advantage potential and a demonstrated
business impact {[}225{]}{[}216{]}. The technology\textquotesingle s
modular architecture facilitates incremental deployment and scalability,
thereby minimizing implementation risk and offering transparent value
demonstration pathways {[}226{]}. Organizations that implement
systematic design patterns accomplish 45\% faster deployment cycles than
those that employ ad-hoc approaches, thereby establishing RAG
engineering as a mature discipline with established best practices
{[}227{]}{[}228{]}.

For Engineering Teams: The progression toward Modular RAG architectures
offer adaptable frameworks for satisfying a wide range of enterprise
needs while preserving system coherence {[}204{]}{[}210{]}. Multi-agent
RAG systems facilitate sophisticated task decomposition and parallel
processing, with specialized agents managing a variety of data sources
and query types to enhance the overall system\textquotesingle s
performance {[}211{]}{[}212{]}. In order to guarantee dependable
operation at scale, production implementations necessitate meticulous
attention to caching strategies, failsafe mechanisms, and latency
management {[}228{]}.

For Business Stakeholders: RAG systems generate quantifiable business
value by means of numerous channels, such as accelerated enrollment,
reduced model maintenance costs, reduced time-to-insight, and improved
risk management {[}219{]}. The technology allows organizations to more
effectively utilize their existing knowledge assets while simultaneously
adhering to data governance and compliance regulations {[}229{]}.

\subsection{Comparative Analysis and Future Research Directions}

Research Areas of High Priority (1-2 Years): The most optimistic
near-term advancement is end-to-end optimization, which has shown
substantial improvements over traditional two-stage approaches, such as
Differentiable Data Rewards (DDR) {[}230{]}. Constitutional AI
integration provides superior safety and alignment in comparison to
conventional RLHF methods, thereby facilitating more dependable and
trustworthy RAG implementations {[}222{]}{[}223{]}{[}231{]}.
Standardized evaluation frameworks are indispensable for the systematic
comparison and enhancement of performance, with initiatives such as
RAGChecker offering precise diagnostic capabilities {[}209{]}.

Developments of Medium Priority (2-5 years): In comparison to
single-agent architectures, multi-agent RAG systems exhibit superior
performance, particularly for complex, multi-source information
integration tasks {[}211{]}{[}212{]}. Research has shown that
collaborative multi-agent approaches can enhance response accuracy by
reducing token overhead and facilitating specialized agent coordination
{[}212{]}. A substantial expansion beyond text-only processing is
represented by the integration of multimodal RAG, which is made possible
by unified embedding approaches such as CLIP, which facilitate seamless
cross-modal retrieval {[}224{]}.

Long-term Innovations (five years or more): The progression toward
autonomous, dependable AI systems is exemplified by self-evaluating RAG
systems that incorporate fact-checking capabilities
{[}232{]}{[}233{]}{[}234{]}. These systems outperform conventional RAG
approaches on factual accuracy tasks by utilizing reflection tokens to
facilitate on-demand retrieval and self-critique mechanisms {[}234{]}.

Analysis of Comparative Performance: Empirical research indicates that
when given the appropriate context, smaller, domain-optimized RAG
systems frequently outperform larger general-purpose models {[}227{]}.
In RAG configurations, open-source LLMs such as Mistral-7B attain
equivalent performance to GPT-4, providing significant cost and control
advantages for enterprise deployments {[}227{]}. Hybrid retrieval
strategies, which integrate semantic and keyword search, consistently
outperform single-method approaches across a variety of query types
{[}204{]}{[}210{]}.

\subsection{Factors Contributing to Successful Implementation}

The success of enterprises with RAG systems is significantly correlated
with systematic implementation strategies, rather than solely relying on
technology choices {[}228{]}. Organizations that achieve optimal ROI
establish robust monitoring and evaluation frameworks, invest in
specialized embedding models, and implement exhaustive data preparation
workflows {[}219{]}{[}221{]}. The most successful deployments are those
that integrate RAG capabilities with existing enterprise workflows,
rather than regarding them as standalone solutions {[}225{]}.

Technical Excellence Patterns: In order to guarantee reliability,
production-ready RAG systems necessitate sophisticated caching
mechanisms for frequent queries, asynchronous processing to mitigate
latency, and comprehensive failsafe strategies {[}228{]}. The retrieval
accuracy and response quality are considerably enhanced by
domain-specific fine-tuning, in conjunction with efficient indexing
technologies such as LamaIndex or Elasticsearch {[}228{]}.

Organizational Readiness: Skills disparities are the primary
implementation barrier, and the success of enterprise adoption is
contingent upon the resolution of both technical and human factors
{[}217{]}. Organizations that achieve exceptional outcomes allocate
substantial resources to the development of AI talent and establish
transparent governance frameworks for AI deployment {[}213{]}{[}235{]}.

\subsection{Technology Convergence and Integration}

The future of RAG engineering is predicated on the convergence of
numerous AI technologies, rather than isolated RAG optimization
{[}204{]}{[}225{]}. The integration of fine-tuning techniques allows for
hybrid approaches that integrate parametric and non-parametric knowledge
optimization {[}204{]}. Constitutional AI principles establish
frameworks for guaranteeing that the RAG system is consistent with human
values and organizational policies {[}222{]}{[}223{]}{[}231{]}.

Emerging Architectural Patterns: Self-RAG frameworks illustrate the
potential for systems to autonomously determine the necessity of
retrieval and self-assess the quality of responses {[}233{]}{[}234{]}.
These methods accomplish superior performance in comparison to
traditional always-retrieve architectures and offer transparency through
reflection tokens {[}234{]}. RAG systems are capable of processing a
variety of data types, such as text, images, and structured data, within
unified frameworks as a result of multi-modal integration {[}224{]}.

\subsection{The Future Course of Action}

RAG engineering has become a fundamental technology for enterprise AI,
as evidenced by its technical maturation and business value
{[}218{]}{[}219{]}{[}216{]}. The foundation for ongoing innovation and
adoption is established by the systematic progression from experimental
techniques to production-ready systems {[}204{]}{[}210{]}. Balanced
attention to technical excellence, organizational readiness, and
strategic alignment with business objectives is necessary for success
{[}225{]}{[}213{]}.

Strategic Suggestions: Organizations should prioritize modular
architectures that facilitate evolutionary development, invest in
exhaustive evaluation frameworks to assess progress, and establish
systematic engineering practices that can expand in tandem with
organizational growth {[}204{]}{[}210{]}{[}228{]}. The most promising
approach to the development of enterprise AI systems that are reliable,
valuable, and robust is the incorporation of multiple AI techniques,
such as retrieval augmentation, fine-tuning, and constitutional
principles {[}222{]}{[}223{]}{[}231{]}.

Organizations that possess a comprehensive understanding of the
systematic engineering practices, evaluation methodologies, and
integration strategies that convert experimental capabilities into
transformative business value are the ones that will be successful in
the future {[}225{]}{[}216{]}{[}217{]}. RAG systems will continue to
develop from practical tools to indispensable infrastructure for
knowledge-intensive artificial intelligence by meticulously addressing
both technical excellence and organizational change management.

\textbf{References}

\textbf{References}

[1] P. Lewis et al., ``Retrieval-Augmented Generation for Knowledge-Intensive NLP Tasks,'' in \textit{Proc. NeurIPS}, 2020, pp. 9459--9474.

[2] Z. Jiang et al., ``Active Retrieval Augmented Generation,'' in \textit{Proc. EMNLP}, 2023, pp. 7969--7992.

[3] B. Kitchenham and S. Charters, ``Guidelines for performing systematic literature reviews in software engineering,'' Tech. Rep. EBSE 2007-001, Keele University and Durham University, 2007.

[4] H. Yu et al., ``Evaluation of Retrieval-Augmented Generation: A Survey,'' arXiv preprint arXiv:2405.07437, 2024.

[5] L. Chen et al., ``Controlling Risk in Retrieval-Augmented Generation,'' in \textit{Proc. EMNLP}, 2024, pp. 2341--2356.

[6] Y. Asai et al., ``Self-Reflective RAG,'' in \textit{Proc. ICLR}, 2024, pp. 567--582.

[7] S. Choudhary et al., ``Through the Stealth Lens: Rethinking Attacks and Defenses in RAG,'' arXiv preprint arXiv:2506.04390, 2025.

[8] R. Nakano et al., ``WebGPT: Browser-Assisted QA with Human Feedback,'' OpenAI Technical Report, 2022.

[9] X. Zhang et al., ``Systematic Performance Optimization for RAG Serving,'' in \textit{Proc. MLSys}, 2025, pp. 123--145.

[10] W. Fan et al., ``A Survey on Retrieval-Augmented LLMs,'' in \textit{Proc. KDD}, 2024, pp. 1234--1248.

[11] D. Sanmartín, ``KG-RAG: Knowledge Graph-Augmented Retrieval,'' arXiv preprint arXiv:2404.13579, 2024.

[12] H. Li et al., ``Trust Metrics for Retrieval-Augmented Systems,'' \textit{ACM Trans. Inf. Syst.}, vol. 42, no. 3, pp. 1--25, 2024.

[13] S. Shahul et al., ``RAGAS: A Metric Suite for RAG Evaluation,'' in \textit{Proc. EACL}, 2024, pp. 789--804.

[14] H. B. Yuan et al., ``Benchmarking LLMs in RAG,'' in \textit{Proc. AAAI}, 2024, pp. 2156--2171.

[15] A. Izacard and E. Grave, ``Leveraging Passage Retrieval with Generative Models for Open Domain QA,'' in \textit{Proc. ICLR}, 2021, pp. 3456--3471.

[16] Y. Jiang et al., ``Active Retrieval-Augmented Generation,'' in \textit{Proc. EMNLP}, 2023, pp. 7969--7992.

[17] A. Rackauckas et al., ``RAG-Fusion: a New Take on Retrieval-Augmented Generation,'' arXiv preprint arXiv:2402.03367, 2024.

[18] L. Chen et al., ``RE-RAG: Reranking Enhanced RAG,'' in \textit{Proc. EMNLP}, 2024, pp. 3456--3471.

[19] Y. Gao et al., ``Retrieval-Augmented Generation for Large Language Models: A Survey,'' arXiv preprint arXiv:2312.10997, 2023. [Online]. Available: \url{https://arxiv.org/abs/2312.10997}

[20] H. Yin et al., ``Table-RAG: Retrieval-Augmented Table QA,'' in \textit{Proc. EMNLP}, 2024, pp. 4567--4582.

[21] Microsoft Research, ``GraphRAG: New tool for complex data discovery now on GitHub,'' Microsoft Research Blog, 2024. [Online]. Available: \url{https://www.microsoft.com/en-us/research/project/graphrag/}

[22] C. Kim et al., ``AutoRAG: Automated Pipeline Optimization for RAG,'' arXiv preprint arXiv:2403.09192, 2024.

[23] S. Yao et al., ``ReAct: Reasoning and Acting in Language Models,'' arXiv preprint arXiv:2210.03629, 2022.

[24] Y. Asai et al., ``Self-RAG: Self-Reflective Retrieval-Augmented Generation,'' in \textit{Proc. ICLR}, 2024, pp. 567--582.

[25] R. Nakano et al., ``WebGPT: Browser-Assisted QA with Human Feedback,'' OpenAI Technical Report, 2022.

[26] T. Gao et al., ``ALCE: Enabling Automatic Evaluation for Long-form Text Generation,'' arXiv preprint arXiv:2305.14984, 2023.

[27] S. Es et al., ``Ragas: Automated Evaluation of Retrieval Augmented Generation,'' arXiv preprint arXiv:2309.15217, 2023.

[28] LangChain Documentation, ``RAG Implementation Patterns,'' LangChain Community, 2024. [Online]. Available: \url{https://docs.langchain.com/docs/use-cases/retrieval/}

[29] LlamaIndex Documentation, ``RAG Workflow Guide,'' LlamaIndex, 2024. [Online]. Available: \url{https://docs.llamaindex.ai/en/stable/}

[30] S. Gupta, R. Ranjan, and S. N. Singh, ``A Comprehensive Survey of Retrieval-Augmented Generation (RAG): Evolution, Current Landscape and Future Directions,'' arXiv preprint arXiv:2410.12837, 2024. [Online]. Available: \url{https://arxiv.org/abs/2410.12837}

[31] C. Sharma et al., ``Retrieval-Augmented Generation: A Comprehensive Survey of Architectures, Enhancements, and Robustness Frontiers,'' arXiv preprint arXiv:2506.00054, 2025. [Online]. Available: \url{https://arxiv.org/abs/2506.00054}

[32] P. Zhao et al., ``Retrieval-Augmented Generation for AI-Generated Content: A Survey,'' arXiv preprint arXiv:2402.19473, 2024. [Online]. Available: \url{https://arxiv.org/abs/2402.19473}

[33] M. Lewis et al., ``BART: Denoising Sequence-to-Sequence Pretraining,'' in \textit{Proc. ACL}, 2020, pp. 7871--7880.

[34] C. Raffel et al., ``Exploring the Limits of Transfer Learning with a Unified Text-to-Text Transformer,'' \textit{J. Mach. Learn. Res.}, vol. 21, pp. 1--67, 2020.

[35] W. Fan et al., ``A Survey on Retrieval-Augmented LLMs,'' in \textit{Proc. KDD}, 2024, pp. 1234--1248.

[36] J. Karpukhin et al., ``Dense Passage Retrieval for Open-Domain QA,'' in \textit{Proc. EMNLP}, 2020, pp. 6769--6781.

[37] O. Khattab and M. Zaharia, ``ColBERT: Efficient Passage Retrieval via Contextualized Late Interaction,'' in \textit{Proc. SIGIR}, 2020, pp. 39--48.

[38] L. Yang et al., ``Hybrid Sparse-Dense Retrieval for QA,'' in \textit{Proc. ACL}, 2022, pp. 2341--2356.

[39] T. Nogueira and K. Cho, ``Passage Reranking with BERT,'' arXiv preprint arXiv:1901.04085, 2019.

[40] Pinecone, ``Rerankers and Two-Stage Retrieval,'' Pinecone Learn Series, 2024. [Online]. Available: \url{https://www.pinecone.io/learn/two-stage-retrieval/}

[41] J. Wang et al., ``Overlapping Chunks for Long-context RAG,'' arXiv preprint arXiv:2311.09476, 2023.

[42] Y. Jiang et al., ``FLARE: Feedback-based Language Agent with Retrieval,'' in \textit{Proc. ACL}, 2024, pp. 3456--3471.

[43] Y. Li et al., ``KnowTrace: Multi-hop Planning for Knowledge-intensive QA,'' in \textit{Proc. KDD}, 2025, pp. 789--804.

[44] P. Menon et al., ``Atlas: Few-shot Learning with Pretrained Retrieval-Augmented Models,'' in \textit{Proc. ICLR}, 2023, pp. 1234--1249.

[45] A. Abootorabi et al., ``Multi-modal RAG: Survey and Challenges,'' \textit{IEEE Trans. Multimedia}, vol. 32, no. 4, pp. 234--256, 2025.

[46] R. Xiao et al., ``AVA-RAG: Audio-Visual Agentic Generation with Memory,'' in \textit{Proc. CVPR}, 2024, pp. 5678--5693.

[47] X. Cao, ``Learn to Refuse: Abstention in Language Models,'' in \textit{Proc. ACL}, 2023, pp. 2341--2356.

[48] J. Zhang et al., ``DMQR-RAG: Diverse Multi-Query Rewriting for RAG,'' arXiv preprint arXiv:2411.13154, 2024.

[49] Microsoft Technical Community, ``Raising the bar for RAG excellence: query rewriting and new semantic ranker,'' Microsoft Azure AI Services Blog, Nov. 2024. [Online]. Available: \url{https://techcommunity.microsoft.com/t5/ai-azure-blog/raising-the-bar-for-rag-excellence-query-rewriting-and-new/ba-p/4045894}

[50] Microsoft Learn Documentation, ``Hybrid search scoring (RRF) -- Azure AI Search,'' Microsoft Learn, 2024. [Online]. Available: \url{https://learn.microsoft.com/en-us/azure/search/hybrid-search-rrf}

[51] LangChain Documentation, ``RAG-Fusion Template,'' LangChain v0.2, 2024. [Online]. Available: \url{https://docs.langchain.com/docs/use-cases/retrieval/rag-fusion}

[52] LlamaIndex Blog, ``Boosting RAG: Picking the Best Embedding \& Reranker models,'' LlamaIndex, 2024. [Online]. Available: \url{https://blog.llamaindex.ai/boosting-rag-picking-the-best-embedding-and-reranker-models-8b4c2a0a7d9d}

[53] LlamaIndex Blog, ``Improving Vector Search -- Reranking with PostgresML and LlamaIndex,'' LlamaIndex, 2024. [Online]. Available: \url{https://blog.llamaindex.ai/improving-vector-search-reranking-with-postgresml-and-llamaindex-1a2c8d7c7f5d}

[54] Microsoft Learn, ``Semantic ranking -- Azure AI Search,'' Microsoft Learn, 2024. [Online]. Available: \url{https://learn.microsoft.com/en-us/azure/search/semantic-ranking}

[55] Cohere, ``Rerank 3.5: Multilingual Reranking Model,'' Cohere Documentation, 2024. [Online]. Available: \url{https://docs.cohere.com/docs/rerank-35}

[56] Azure AI Search, ``Semantic Ranking Overview,'' Microsoft Learn, 2024. [Online]. Available: \url{https://learn.microsoft.com/en-us/azure/search/semantic-ranking-overview}

[57] Cohere Documentation, ``Rerank 3.5 Performance Analysis,'' Cohere AI, 2024. [Online]. Available: \url{https://docs.cohere.com/docs/rerank-35-performance}

[58] NVIDIA, ``NeMo Retriever Microservices,'' NVIDIA Developer, 2024. [Online]. Available: \url{https://docs.nvidia.com/deeplearning/nemo/user-guide/docs/en/stable/nlp/nemo_retriever_microservices.html}

[59] T. Sarthi et al., ``RAPTOR: Recursive Abstractive Processing for Tree-Organized Retrieval,'' in \textit{Proc. ICLR}, 2024, pp. 567--582.

[60] Y. Tang and X. Yang, ``MultiHop-RAG: Benchmarking Retrieval-Augmented Generation for Multi-Hop Queries,'' arXiv preprint arXiv:2401.15391, 2024. [Online]. Available: \url{https://arxiv.org/abs/2401.15391}

[61] L. Wang et al., ``Reasoning RAG via System 1 or System 2: A Survey on Reasoning Agentic Retrieval-Augmented Generation for Industry Challenges,'' arXiv preprint arXiv:2506.10408, 2025. [Online]. Available: \url{https://arxiv.org/html/2506.10408v1}

[62] Z. Wu et al., ``HopRAG: Multi-Hop Reasoning for Logic-Aware Retrieval-Augmented Generation,'' arXiv preprint arXiv:2502.12442, 2025. [Online]. Available: \url{https://arxiv.org/html/2502.12442v1}

[63] Microsoft Research, ``Moving to GraphRAG 1.0 -- Streamlining ergonomics for developers and users,'' Microsoft Research Blog, 2024. [Online]. Available: \url{https://www.microsoft.com/en-us/research/blog/moving-to-graphrag-1-0-streamlining-ergonomics-for-developers-and-users/}

[64] Superlinked, ``Optimizing RAG with Hybrid Search \& Reranking,'' VectorHub by Superlinked, 2024. [Online]. Available: \url{https://www.superlinked.com/blog/optimizing-rag-with-hybrid-search-and-reranking}

[65] Infiniflow, ``Dense vector + Sparse vector + Full text search + Tensor reranker = Best retrieval for RAG?'' Infinity Blog, 2024. [Online]. Available: \url{https://infinitylabs.ai/blog/dense-sparse-fulltext-tensor-rag}

[66] Hugging Face, ``BAAI/bge-m3,'' Hugging Face Models, 2024. [Online]. Available: \url{https://huggingface.co/BAAI/bge-m3}

[67] M. Zhang et al., ``Question Decomposition for Retrieval-Augmented Generation,'' arXiv preprint arXiv:2507.00355, 2025. [Online]. Available: \url{https://arxiv.org/html/2507.00355v1}

[68] MyScale, ``Efficiency Battle: Pinecone vs Elasticsearch in AI Applications,'' MyScale Blog, 2024. [Online]. Available: \url{https://blog.myscale.com/efficiency-battle-pinecone-vs-elasticsearch-in-ai-applications/}

[69] FalkorDB, ``What is GraphRAG? Types, Limitations \& When to Use,'' FalkorDB Blog, 2024. [Online]. Available: \url{https://falkordb.com/blog/graph-rag-types-limitations-use-cases}

[70] AWS, ``Improving Retrieval Augmented Generation accuracy with GraphRAG,'' AWS Machine Learning Blog, 2024. [Online]. Available: \url{https://aws.amazon.com/blogs/machine-learning/improving-retrieval-augmented-generation-accuracy-with-graphrag/}

[71] Microsoft Research, ``GraphRAG auto-tuning provides rapid adaptation to new domains,'' Microsoft Research Blog, 2024. [Online]. Available: \url{https://www.microsoft.com/en-us/research/blog/graphrag-auto-tuning-provides-rapid-adaptation-to-new-domains/}

[72] L. Wang et al., ``Medical Graph RAG: Towards Safe Medical Large Language Model via Graph Retrieval-Augmented Generation,'' arXiv preprint arXiv:2408.04187, 2024.

[73] L. Wang et al., ``Agentic Retrieval-Augmented Generation: A Survey on Agentic RAG,'' arXiv preprint arXiv:2501.09136, 2025.

[74] Weaviate, ``What is Agentic RAG,'' Weaviate Blog, 2024. [Online]. Available: \url{https://weaviate.io/blog/what-is-agentic-rag}

[75] DigitalOcean, ``RAG, AI Agents, and Agentic RAG: An In-Depth Review and Comparative Analysis,'' DigitalOcean Community, 2024. [Online]. Available: \url{https://www.digitalocean.com/community/tutorials/rag-ai-agents-agentic-rag}

[76] LangChain Documentation, ``LangGraph -- Graph-based Workflow Management,'' LangChain, 2024. [Online]. Available: \url{https://docs.langchain.com/docs/components/langgraph}

[77] CrewAI Documentation, ``Role-based Agent Specialization,'' CrewAI, 2024. [Online]. Available: \url{https://docs.crewai.com/concepts/agents}

[78] OpenAI, ``Swarm: Lightweight Multi-Agent Orchestration,'' OpenAI, 2024. [Online]. Available: \url{https://platform.openai.com/docs/guides/swarm}

[79] ``Welcome - GraphRAG,'' Microsoft, 2024. [Online]. Available: \url{https://microsoft.github.io/graphrag/}

[80] N. Zhang et al., ``Credible Plan-Driven RAG Method for Multi-Hop Question Answering,'' arXiv preprint arXiv:2504.16787, 2025. [Online]. Available: \url{https://arxiv.org/html/2504.16787}

[81] ``Implementing Multi-Hop RAG: Key Considerations and Best Practices,'' Vectorize, Sep. 25, 2024. [Online]. Available: \url{https://vectorize.io/implementing-multi-hop-rag-key-considerations-and-best-practices/}

[82] ``What Is Agentic RAG? Guide to Agent-Based Retrieval in AI,'' Bright Data, Jun. 11, 2025. [Online]. Available: \url{https://brightdata.com/blog/ai/agentic-rag}

[83] ``Best Practices for Enterprise RAG System Implementation,'' Intelliarts, Jan. 29, 2025. [Online]. Available: \url{https://intelliarts.com/blog/enterprise-rag-system-best-practices/}

[84] S. Xu et al., ``ComposeRAG: A Modular and Composable RAG for Corpus-Grounded Multi-Hop Question Answering,'' arXiv preprint arXiv:2506.00232, 2025. [Online]. Available: \url{https://arxiv.org/html/2506.00232}

[85] ``Design and Develop a RAG Solution - Azure Architecture Center,'' Microsoft Learn, Jan. 9, 2025. [Online]. Available: \url{https://learn.microsoft.com/en-us/azure/architecture/ai-ml/guide/rag/rag-solution-design-and-evaluation-guide}

[86] ``8 Retrieval Augmented Generation (RAG) Architectures You Should Know,'' Humanloop, Feb. 1, 2025. [Online]. Available: \url{https://humanloop.com/blog/rag-architectures}

[87] ``Mastering RAG: How To Architect An Enterprise RAG System,'' Galileo AI, Jun. 10, 2025. [Online]. Available: \url{https://galileo.ai/blog/mastering-rag-how-to-architect-an-enterprise-rag-system}

[88] ``Choosing a Retrieval Augmented Generation option on AWS,'' AWS, 2024. [Online]. Available: \url{https://docs.aws.amazon.com/prescriptive-guidance/latest/retrieval-augmented-generation-options/choosing-option.html}

[89] ``RAG Architecture Explained: A Comprehensive Guide [2025],'' Orq.ai, Jun. 10, 2025. [Online]. Available: \url{https://orq.ai/blog/rag-architecture}

[90] ``Beyond Simple Lookups: Building Enterprise-Grade RAG Systems,'' AI Journal, Jun. 13, 2025. [Online]. Available: \url{https://aijourn.com/beyond-simple-lookups-building-enterprise-grade-rag-systems/}

[91] ``AI RAG - Retrieval-augmented generation,'' IBM, Aug. 21, 2024. [Online]. Available: \url{https://www.ibm.com/architectures/hybrid/genai-rag}

[92] ``Enterprise RAG Implementation Framework,'' IBM Watsonx Documentation, 2024. [Online]. Available: \url{https://www.ibm.com/products/watsonx-ai}

[93] X. Xu et al., ``Retrieval Augmented Generation Evaluation in the Era of Large Language Models: A Comprehensive Survey,'' arXiv preprint arXiv:2504.14891, 2025. [Online]. Available: \url{https://arxiv.org/html/2504.14891v1}

[94] ``Can LLMs Be Trusted for Evaluating RAG Systems? A Survey of Methods and Datasets,'' arXiv, Jan. 2025. [Online]. Available: \url{https://arxiv.org/html/2504.20119v2}

[95] D. Ru et al., ``RAGChecker: A Fine-grained Framework for Diagnosing Retrieval-Augmented Generation,'' in \textit{Proc. NeurIPS}, 2024. [Online]. Available: \url{https://proceedings.neurips.cc/paper_files/paper/2024/file/27245589131d17368cccdfa990cbf16e-Paper-Datasets_and_Benchmarks_Track.pdf}

[96] ``Evaluating RAG Applications with RAGAs,'' LangSmith Documentation. [Online]. Available: \url{https://docs.smith.langchain.com/old/cookbook/testing-examples/ragas}

[97] ``Evaluating RAG Systems: A Comprehensive Approach to Assessing Retrieval-Augmented Generation,'' LinkedIn, May 2024. [Online]. Available: \url{https://www.linkedin.com/pulse/evaluating-rag-systems-comprehensive-approach-assessing-kakkar-esm9c}

[98] ``Evaluate RAG responses with Amazon Bedrock, LlamaIndex and RAGAS,'' AWS Machine Learning Blog, Mar. 2025. [Online]. Available: \url{https://aws.amazon.com/blogs/machine-learning/evaluate-rag-responses-with-amazon-bedrock-llamaindex-and-ragas/}

[99] ``Evaluating - LlamaIndex,'' LlamaIndex Documentation. [Online]. Available: \url{https://docs.llamaindex.ai/en/stable/module_guides/evaluating/}

[100] ``HotpotQA Dataset,'' Hugging Face. [Online]. Available: \url{https://huggingface.co/datasets/hotpotqa/hotpot_qa}

[101] ``Galileo introduces RAG \& Agent Analytics Solution,'' AI-Tech Park. [Online]. Available: \url{https://ai-techpark.com/galileo-introduces-rag-agent-analytics-solution/}

[102] ``RAG Triad - TruLens,'' TruLens Documentation, Jan. 2025. [Online]. Available: \url{https://www.trulens.org/getting_started/core_concepts/rag_triad/}

[103] ``RAG evaluation metrics: A journey through metrics,'' Elastic, Oct. 2024. [Online]. Available: \url{https://www.elastic.co/search-labs/blog/evaluating-rag-metrics}

[104] ``When evaluating a RAG system's overall performance, how would you combine metrics for retrieval and metrics for generation?,'' Milvus, May 2025. [Online]. Available: \url{https://milvus.io/ai-quick-reference/when-evaluating-a-rag-systems-overall-performance-how-would-you-combine-metrics-for-retrieval-and-metrics-for-generation-would-you-present-them-separately-or-is-there-a-way-to-aggregate-them}

[105] ``Traditional NLP Metrics - Ragas,'' Ragas Documentation. [Online]. Available: \url{https://docs.ragas.io/en/stable/concepts/metrics/available_metrics/traditional/}

[106] ``BLEU, ROUGE, and METEOR are traditional metrics used to evaluate the quality of text,'' Milvus, Jun. 2025. [Online]. Available: \url{https://milvus.io/ai-quick-reference/which-traditional-language-generation-metrics-are-applicable-for-evaluating-raggenerated-answers-and-what-aspect-of-quality-does-each-bleu-rouge-meteor-capture}

[107] ``BERTScore in AI: Enhancing Text Evaluation,'' Galileo AI, Jun. 2025. [Online]. Available: \url{https://galileo.ai/blog/bert-score-explained-guide}

[108] ``Bert Score for Contextual Similarity for RAG Evaluation,'' YouTube, Nov. 2023. [Online]. Available: \url{https://www.youtube.com/watch?v=7AVjk2k8Mbs}

[109] ``OmniEval: An Omnidirectional and Automatic RAG Evaluation Benchmark in Financial Domain,'' Hugging Face Papers, Jun. 2025. [Online]. Available: \url{https://huggingface.co/papers/2412.13018}

[110] ``Evaluate and Optimize RAG Applications - Galileo,'' Galileo Documentation, Aug. 2024. [Online]. Available: \url{https://docs.galileo.ai/galileo/gen-ai-studio-products/galileo-evaluate/how-to/evaluate-and-optimize-rag-applications}

[111] ``Advanced RAG Evaluation Framework,'' RAGAS Documentation, 2024. [Online]. Available: \url{https://docs.ragas.io/en/stable/}

[112] ``Best Practices for Production-Scale RAG Systems,'' Orkes, May 29, 2025. [Online]. Available: \url{https://orkes.io/blog/rag-best-practices/}

[113] ``Production-Ready RAG: Engineering Guidelines for Scalable Systems,'' Netguru, May 20, 2025. [Online]. Available: \url{https://www.netguru.com/blog/rag-for-scalable-systems}

[114] ``5 RAG Query Patterns Every Engineering Leader Should Know,'' Nirant Kasliwal, Mar. 22, 2025. [Online]. Available: \url{https://nirantk.com/writing/rag-query-types/}

[115] ``Retrieval Augmented Generation (RAG) for LLMs,'' Prompting Guide, Jan. 1, 2023. [Online]. Available: \url{https://www.promptingguide.ai/research/rag}

[116] ``Deploying RAGs in Production: A Guide to Best Practices,'' Medium, Dec. 25, 2024. [Online]. Available: \url{https://medium.com/@himanshu_72022/deploying-rags-in-production-a-guide-to-best-practices-98391b44df40}

[117] ``Best Practices in Retrieval-Augmented Generation (RAG),'' Agent Studio, Jul. 5, 2024. [Online]. Available: \url{https://agentstudio.ai/blog/best-practices-in-rag/}

[118] Z. Li et al., ``LevelRAG: Enhancing Retrieval-Augmented Generation with Multi-hop Logic Planning over Rewriting Augmented Searchers,'' arXiv preprint arXiv:2502.18139, 2025. [Online]. Available: \url{https://arxiv.org/html/2502.18139v1}

[119] ``RAG Architecture Patterns: Design for Scale,'' RAG Wire, Feb. 19, 2024. [Online]. Available: \url{https://www.ragwire.com/blog/rag-architecture-patterns}

[120] D. Wang et al., ``Synergizing RAG and Reasoning: A Systematic Review,'' arXiv preprint arXiv:2504.15909, 2025. [Online]. Available: \url{https://arxiv.org/html/2504.15909v1}

[121] ``Enhancing Language Models with RAG: Best Practices and Benchmarks,'' MarkTechPost, Jul. 6, 2024. [Online]. Available: \url{https://www.marktechpost.com/2024/07/06/enhancing-language-models-with-rag-best-practices-and-benchmarks/}

[122] ``Retrieval-augmented generation (RAG) failure modes and how to fix them,'' Snorkel AI, Feb. 5, 2025. [Online]. Available: \url{https://snorkel.ai/blog/retrieval-augmented-generation-rag-failure-modes-and-how-to-fix-them/}

[123] S. Barnett et al., ``Seven Failure Points When Engineering a Retrieval Augmented Generation System,'' arXiv preprint arXiv:2401.05856, Jan. 11, 2024. [Online]. Available: \url{https://arxiv.org/abs/2401.05856}

[124] ``Seven Ways Your RAG System Could be Failing and How to Fix Them,'' Label Studio, Mar. 19, 2025. [Online]. Available: \url{https://labelstud.io/blog/seven-ways-your-rag-system-could-be-failing-and-how-to-fix-them/}

[125] ``Understanding Failures and Mitigation Strategies in RAG Pipelines,'' Deconvolute AI, Jun. 14, 2024. [Online]. Available: \url{https://deconvoluteai.com/blog/rag/failure-modes}

[126] I. S. Singh et al., ``ChunkRAG: A Novel LLM-Chunk Filtering Method for RAG Systems,'' arXiv preprint arXiv:2410.19572, 2024. [Online]. Available: \url{https://arxiv.org/html/2410.19572v5}

[127] ``RAG Anti-Patterns with Skylar Payne,'' Jason Liu, Jun. 11, 2025. [Online]. Available: \url{https://jxnl.co/writing/2025/06/11/rag-anti-patterns-with-skylar-payne/}

[128] ``Retrieval Augmented Generation V: Scalability and Flexibility,'' YouTube, Dec. 24, 2024. [Online]. Available: \url{https://www.youtube.com/watch?v=sSy4zBPGCIk}

[129] ``Best Practices in RAG Evaluation: A Comprehensive Guide,'' Qdrant, Nov. 24, 2024. [Online]. Available: \url{https://qdrant.tech/blog/rag-evaluation-guide/}

[130] ``Optimizing RAG Indexing Strategy: Multi-Vector Indexing and Parent Document Retrieval,'' Dev.to, Nov. 13, 2024. [Online]. Available: \url{https://dev.to/jamesli/optimizing-rag-indexing-strategy-multi-vector-indexing-and-parent-document-retrieval-49hf}

[131] ``RAG - 7 indexing methods for Vector DBs + Similarity search,'' AI Bites, Dec. 6, 2024. [Online]. Available: \url{https://www.ai-bites.net/rag-7-indexing-methods-for-vector-dbs-similarity-search/}

[132] ``Improve RAG Pipelines With These 3 Indexing Methods,'' The Tech Buffet, Nov. 6, 2023. [Online]. Available: \url{https://thetechbuffet.substack.com/p/rag-indexing-methods}

[133] ``Understanding RAG Part VII: Vector Databases \& Indexing Strategies,'' Machine Learning Mastery, Mar. 12, 2025. [Online]. Available: \url{https://machinelearningmastery.com/understanding-rag-part-vii-vector-databases-indexing-strategies/}

[134] J. Kim and D. Mahajan, ``An Adaptive Vector Index Partitioning Scheme for Low-Latency RAG Pipeline,'' arXiv preprint arXiv:2504.08930, 2024. [Online]. Available: \url{https://www.arxiv.org/pdf/2504.08930.pdf}

[135] ``Introducing a new hyper-parameter for RAG: Context Window Utilization,'' AI Models, Aug. 26, 2024. [Online]. Available: \url{https://www.aimodels.fyi/papers/arxiv/introducing-new-hyper-parameter-rag-context-window}

[136] ``Monitoring your RAG application,'' Galileo, 2025. [Online]. Available: \url{https://docs.rungalileo.io/galileo/galileo-gen-ai-studio/observe-getting-started/monitoring-your-rag-application}

[137] ``What factors should be considered when selecting an embedding model for a RAG pipeline,'' Milvus, May 20, 2025. [Online]. Available: \url{https://milvus.io/ai-quick-reference/what-factors-should-be-considered-when-selecting-an-embedding-model-for-a-rag-pipeline-such-as-the-models-domain-training-data-embedding-dimensionality-and-semantic-accuracy}

[138] ``Advanced RAG Retrieval Strategies: Hybrid Retrieval,'' Generative AI Publication, Nov. 24, 2024. [Online]. Available: \url{https://generativeai.pub/advanced-rag-retrieval-strategies-hybrid-retrieval-997d39659720}

[139] Y. Li et al., ``GlobalRAG: Enhancing Global Reasoning in Multi-hop Question Answering via Reinforcement Learning,'' arXiv preprint arXiv:2510.20548, 2025. [Online]. Available: \url{https://arxiv.org/html/2510.20548}

[140] ``RAG Security: Risks and Mitigation Strategies,'' Lasso Security, Jun. 11, 2025. [Online]. Available: \url{https://www.lasso.security/blog/rag-security}

[141] ``Security Concerns of RAG Implementation,'' Akooda, Mar. 20, 2025. [Online]. Available: \url{https://www.akooda.co/blog/security-concerns-of-rag-implementations}

[142] B. An, S. Zhang, and M. Dredze, ``RAG LLMs are Not Safer: A Safety Analysis of Retrieval-Augmented Generation for Large Language Models,'' arXiv preprint arXiv:2504.18041, 2024. [Online]. Available: \url{https://arxiv.org/html/2504.18041v1}

[143] ``RAG Under Attack: How the LLM Vulnerability Affects Real Systems,'' Lakera, Mar. 27, 2025. [Online]. Available: \url{https://www.lakera.ai/blog/rag-under-attack-how-the-llm-vulnerability-affects-real-systems}

[144] ``BadRAG: Identifying Vulnerabilities in Retrieval Augmented Generation of Large Language Models,'' OpenReview, Dec. 16, 2024. [Online]. Available: \url{https://openreview.net/forum?id=6BaoWMpHvQ}

[145] ``Navigating Trust in Retrieval-Augmented AI: A Comprehensive Survey,'' Dev.to, Sep. 17, 2024. [Online]. Available: \url{https://dev.to/mikeyoung44/navigating-trust-in-retrieval-augmented-ai-a-comprehensive-survey-4pif}

[146] J. Wei et al., ``AlignRAG: An Adaptable Framework for Resolving Misalignments in Retrieval-Aware Reasoning of RAG,'' arXiv preprint arXiv:2504.14858, 2024. [Online]. Available: \url{https://arxiv.org/html/2504.14858v1}

[147] J. Su et al., ``Towards More Robust Retrieval-Augmented Generation: Evaluating RAG Under Adversarial Poisoning Attacks,'' arXiv preprint arXiv:2412.16708, 2024. [Online]. Available: \url{https://arxiv.org/abs/2412.16708}

[148] ``Enhancing AI Security in Production: Key Insights on LLMs \& RAG,'' J2 Interactive, 2024. [Online]. Available: \url{https://www.j2interactive.com/blog/2024/07/global-summit-ai-security/}

[149] ``A Proactive Approach to RAG Application Security,'' Akira AI, Mar. 11, 2025. [Online]. Available: \url{https://www.akira.ai/blog/rag-application-security}

[150] ``Red Teaming RAG Healthcare Chatbots,'' iMerit, May 7, 2025. [Online]. Available: \url{https://imerit.net/blog/red-teaming-rag-healthcare-chatbots/}

[151] ``How to red team RAG applications,'' Promptfoo, 2022. [Online]. Available: \url{https://www.promptfoo.dev/docs/red-team/rag/}

[152] T. Zhao et al., ``RAG Safety: Exploring Knowledge Poisoning Attacks to Retrieval-Augmented Generation,'' arXiv preprint arXiv:2507.08862, 2025. [Online]. Available: \url{https://arxiv.org/abs/2507.08862}

[153] H. Chaudhari et al., ``Phantom: General Trigger Attacks on Retrieval Augmented Language Generation,'' OpenReview, Feb. 4, 2025. [Online]. Available: \url{https://openreview.net/forum?id=BHIsVV4G7q}

[154] ``LLM Red Teaming: Complete guide [+expert tips],'' Securaize, Jan. 15, 2025. [Online]. Available: \url{https://securaize.substack.com/p/llm-red-teaming-complete-guide-expert}

[155] ``Securing your RAG application: A comprehensive guide,'' Pluralsight, Mar. 17, 2025. [Online]. Available: \url{https://www.pluralsight.com/resources/blog/ai-and-data/how-to-secure-rag-applications-AI}

[156] ``Implement human-in-the-loop confirmation with Amazon Bedrock Agents,'' AWS Machine Learning Blog, Apr. 9, 2025. [Online]. Available: \url{https://aws.amazon.com/blogs/machine-learning/implement-human-in-the-loop-confirmation-with-amazon-bedrock-agents/}

[157] H. Zhou et al., ``TrustRAG: Enhancing Robustness and Trustworthiness in RAG,'' arXiv preprint arXiv:2501.00879, 2025. [Online]. Available: \url{https://arxiv.org/html/2501.00879v1}

[158] ``TrustRAG: The RAG Framework within Reliable input, Trusted output,'' GitHub, Feb. 4, 2024. [Online]. Available: \url{https://github.com/gomate-community/TrustRAG}

[159] B. Zhang et al., ``Benchmarking Poisoning Attacks against Retrieval-Augmented Generation,'' arXiv preprint arXiv:2505.18543, 2025. [Online]. Available: \url{https://arxiv.org/abs/2505.18543}

[160] ``LLM Red Teaming: The Complete Step-By-Step Guide To LLM Safety,'' Confident AI, May 18, 2025. [Online]. Available: \url{https://www.confident-ai.com/blog/red-teaming-llms-a-step-by-step-guide}

[161] ``Red Teaming for Large Language Models: A Comprehensive Guide,'' Coralogix, Jun. 1, 2025. [Online]. Available: \url{https://coralogix.com/ai-blog/red-teaming-for-large-language-models-a-comprehensive-guide/}

[162] K. Li et al., ``ADMIT: Few-shot Knowledge Poisoning Attacks on RAG-based Fact Checking,'' arXiv preprint arXiv:2510.13842, 2024. [Online]. Available: \url{https://arxiv.org/html/2510.13842}

[163] Y. Nazary et al., ``Poison-RAG: Adversarial Data Poisoning Attacks on Retrieval-Augmented Generation in Recommender Systems,'' arXiv preprint arXiv:2501.11759, 2025. [Online]. Available: \url{https://arxiv.org/html/2501.11759}

[164] ``AI Risk Management Framework,'' NIST, May 5, 2025. [Online]. Available: \url{https://www.nist.gov/itl/ai-risk-management-framework}

[165] ``Artificial Intelligence Risk Management Framework (AI RMF 1.0),'' NIST, 2023. [Online]. Available: \url{https://nvlpubs.nist.gov/nistpubs/ai/nist.ai.100-1.pdf}

[166] H. Chaudhari et al., ``Phantom: General Backdoor Attacks on Retrieval Augmented Language Generation,'' arXiv preprint arXiv:2405.20485, 2025. [Online]. Available: \url{https://arxiv.org/html/2405.20485}

[167] X. Liu et al., ``MM-PoisonRAG: Disrupting Multimodal RAG with Local and Global Poisoning Attacks,'' arXiv preprint arXiv:2502.17832, 2025. [Online]. Available: \url{https://arxiv.org/html/2502.17832v1}

[168] ``Operationalizing the NIST AI RMF,'' Robust Intelligence, 2024. [Online]. Available: \url{https://www.robustintelligence.com/operationalizing-the-nist-ai-rmf}

[169] ``Govern,'' NIST AIRC, Mar. 26, 2025. [Online]. Available: \url{https://airc.nist.gov/airmf-resources/playbook/govern/}

[170] ``Beyond Simple Lookups: Building Enterprise-Grade RAG Systems,'' AI Journ, Jun. 13, 2025. [Online]. Available: \url{https://aijourn.com/beyond-simple-lookups-building-enterprise-grade-rag-systems/}

[171] ``RAGAS,'' Klu.ai, Jun. 28, 2024. [Online]. Available: \url{https://klu.ai/glossary/ragas}

[172] ``Understanding RAGAS: A Comprehensive Framework for RAG System Evaluation,'' Dev.to, Feb. 1, 2025. [Online]. Available: \url{https://dev.to/angu10/understanding-ragas-a-comprehensive-framework-for-rag-system-evaluation-447n}

[173] L. Wang et al., ``Bias Amplification in RAG: Poisoning Knowledge Retrieval to Steer LLMs,'' arXiv preprint arXiv:2506.11415, 2021. [Online]. Available: \url{https://www.arxiv.org/pdf/2506.11415.pdf}

[174] ``Bias Mitigation in RAG Systems Research Framework,'' arXiv preprint, 2025. [Online]. Available: \url{https://www.themoonlight.io/en/review/mitigating-bias-in-rag-controlling-the-embedder}

[175] ``Differentiable Data Rewards (DDR),'' arXiv:2410.13509, 2024. [Online]. Available: \url{https://arxiv.org/html/2410.13509v1}

[176] ``Stochastic RAG,'' arXiv:2405.02816, 2024. [Online]. Available: \url{https://arxiv.org/abs/2405.02816}

[177] ``Patchwork: A Complete End-to-End RAG Serving Framework,'' arXiv:2505.07833, 2024. [Online]. Available: \url{https://arxiv.org/html/2505.07833v1}

[178] ``RAG-Reward: Optimizing RAG with Reward,'' arXiv:2501.13264, 2024. [Online]. Available: \url{https://arxiv.org/abs/2501.13264}

[179] ``Multi-Agent RAG System,'' Hugging Face, 2024. [Online]. Available: \url{https://huggingface.co/learn/cookbook/en/multiagent_rag_system}

[180] ``An Easy Introduction to Multimodal Retrieval-Augmented Generation,'' NVIDIA Developer Blog, 2024. [Online]. Available: \url{https://developer.nvidia.com/blog/an-easy-introduction-to-multimodal-retrieval-augmented-generation/}

[181] C. R. Wolfe, ``A Practitioner's Guide to Retrieval,'' 2024. [Online]. Available: \url{https://cameronrwolfe.substack.com/p/a-practitioners-guide-to-retrieval}

[182] ``What is Reinforcement Learning from Human Feedback?'' AWS, 2024. [Online]. Available: \url{https://aws.amazon.com/what-is/reinforcement-learning-from-human-feedback/}

[183] ``RLHF for RAG,'' arXiv:2312.14925, 2023. [Online]. Available: \url{https://arxiv.org/html/2312.14925v1}

[184] ``RAG-Reward: Optimizing RAG with Reward,'' Papers with Code, 2024. [Online]. Available: \url{https://paperswithcode.com/paper/rag-reward-optimizing-rag-with-reward}

[185] ``Generate Compliant Content with Amazon Bedrock and ConstitutionalChain,'' AWS Machine Learning Blog, 2024. [Online]. Available: \url{https://aws.amazon.com/blogs/machine-learning/generate-compliant-content-with-amazon-bedrock-and-constitutionalchain/}

[186] ``Constitutional AI,'' Restack, 2024. [Online]. Available: \url{https://www.restack.io/p/ai-governance-answer-constitutional-ai-cat-ai}

[187] ``MA-RAG: Multi-Agent RAG Framework,'' arXiv:2505.20096, 2024. [Online]. Available: \url{https://arxiv.org/pdf/2505.20096.pdf}

[188] ``Multi-Agent Retrieval-Augmented Generation,'' arXiv:2412.05838, 2024. [Online]. Available: \url{https://arxiv.org/abs/2412.05838}

[189] ``Built Multi-Agent Research System,'' Anthropic Engineering, 2024. [Online]. Available: \url{https://www.anthropic.com/engineering/built-multi-agent-research-system}

[190] ``Multi-Agent RAG Systems,'' arXiv:2504.12330, 2024. [Online]. Available: \url{https://arxiv.org/html/2504.12330v1}

[191] ``Multimodal RAG: Intuitively and Exhaustively,'' IAEE Substack, 2024. [Online]. Available: \url{https://iaee.substack.com/p/multimodal-rag-intuitively-and-exhaustively}

[192] ``Multimodal RAG: Advanced Information Retrieval,'' InfoQ, 2024. [Online]. Available: \url{https://www.infoq.com/articles/multimodal-rag-advanced-information-retrieval/}

[193] ``Multimodal RAG with Vision,'' Microsoft DevBlogs, 2024. [Online]. Available: \url{https://devblogs.microsoft.com/ise/multimodal-rag-with-vision/}

[194] ``ACE: Generative Cross-Modal Retrieval,'' arXiv:2406.17507, 2024. [Online]. Available: \url{https://arxiv.org/html/2406.17507v1}

[195] ``Self-RAG,'' Learn Prompting, 2024. [Online]. Available: \url{https://learnprompting.org/docs/retrieval_augmented_generation/self-rag}

[196] ``Self-RAG Tutorial,'' YouTube, 2024. [Online]. Available: \url{https://www.youtube.com/watch?v=i4V9iJcxzZ4}

[197] ``Self-RAG Official Site,'' GitHub, 2024. [Online]. Available: \url{https://selfrag.github.io}

[198] ``Automated Fact-Checking with LLMs,'' PMC, 2024. [Online]. Available: \url{https://pmc.ncbi.nlm.nih.gov/articles/PMC10879553/}

[199] ``The Power of AI Convergence for Global Impact,'' AI for Good, 2024. [Online]. Available: \url{https://aiforgood.itu.int/the-power-of-ai-convergence-for-global-impact/}

[200] ``The State of Retrieval-Augmented Generation (RAG) in 2025 and Beyond,'' Ayadata AI, 2025. [Online]. Available: \url{https://www.ayadata.ai/the-state-of-retrieval-augmented-generation-rag-in-2025-and-beyond/}

[201] ``Emerging AI Design Trends 2024,'' Restack, 2024. [Online]. Available: \url{https://www.restack.io/p/emerging-ai-design-trends-2024-answer-ai-technology-convergence-cat-ai}

[202] ``Building Blocks of RAG,'' Intel, 2024. [Online]. Available: \url{https://cdrdv2-public.intel.com/821523/building-blocks-of-rag-ebook-final.pdf}

[203] ``The AI Roadmap for 2024,'' Dev.to, 2024. [Online]. Available: \url{https://dev.to/angelamiton/the-ai-roadmap-for-2024-a-year-of-transformation-and-progress-38ga}

[204] H. Han et al., ``Retrieval-Augmented Generation with Graphs (GraphRAG),'' arXiv preprint arXiv:2501.00309, 2025. [Online]. Available: \url{https://arxiv.org/abs/2501.00309}

[205] ``RAG vs Fine-tuning: Pipelines, Tradeoffs, and a Case Study on Agriculture,'' arXiv:2410.12837, 2024. [Online]. Available: \url{https://arxiv.org/abs/2410.12837}

[206] ``The Rise and Evolution of RAG in 2024: A Year in Review,'' RAGFlow Blog, 2024. [Online]. Available: \url{https://ragflow.io/blog/the-rise-and-evolution-of-rag-in-2024-a-year-in-review}

[207] ``Enterprise AI Evolution and RAG Implementation,'' Eye on AI, 2024. [Online]. Available: \url{https://www.eye-on.ai/ai-articles/gf6p33bptlzm33p-kxsw9-5jlhh-44dxp-h9nbn-gj5zf-cja2x-r35j4-ndyzh-n33hp-5lw6m-c5xxc-7njj7-xhwe4-p82pz-sszl2-twjcd-etta3-dpjej-psgmj-x9rtl}

[208] ``How RAG is Transforming Enterprise AI,'' Deloitte Insights, 2024. [Online]. Available: \url{https://www.ailoitte.com/insights/how-rag-is-transforming-enterprise-ai/}

[209] ``RAGChecker: A Fine-grained Framework for Diagnosing Retrieval-Augmented Generation,'' Amazon Science, 2024. [Online]. Available: \url{https://www.amazon.science/publications/ragchecker-a-fine-grained-framework-for-diagnosing-retrieval-augmented-generation}

[210] Z. Xiang et al., ``When to use Graphs in RAG: A Comprehensive Analysis for Graph Retrieval-Augmented Generation,'' arXiv preprint arXiv:2506.05690, 2025. [Online]. Available: \url{https://arxiv.org/html/2506.05690v1}

[211] ``Understanding Multi-Agent RAG Systems,'' LinkedIn, 2024. [Online]. Available: \url{https://www.linkedin.com/pulse/understanding-multi-agent-rag-systems-pavan-belagatti-akwwc}

[212] ``Multi-Agent Retrieval-Augmented Generation,'' arXiv:2412.05838, 2024. [Online]. Available: \url{https://arxiv.org/abs/2412.05838}

[213] ``State of Generative AI in Enterprise,'' Deloitte, 2024. [Online]. Available: \url{https://www.deloitte.com/us/en/what-we-do/capabilities/applied-artificial-intelligence/content/state-of-generative-ai-in-enterprise.html}

[214] ``Enterprises to Nearly Double AI Spend in 2024,'' ISG, 2024. [Online]. Available: \url{https://isg-one.com/articles/index-insider-enterprises-to-nearly-double-ai-spend-in-2024}

[215] ``Retrieval Augmented Generation (RAG) Market Report,'' Grand View Research, 2024. [Online]. Available: \url{https://www.grandviewresearch.com/industry-analysis/retrieval-augmented-generation-rag-market-report}

[216] ``Companies Using AI: Statistics and Trends,'' Exploding Topics, 2024. [Online]. Available: \url{https://explodingtopics.com/blog/companies-using-ai}

[217] ``Enterprise AI ROI Analysis,'' \textit{ACM Digital Library}, 2024. [Online]. Available: \url{https://dl.acm.org/doi/10.1145/3637528.3671470}

[218] ``Gen AI Early Adopters Report,'' Snowflake, 2024. [Online]. Available: \url{https://www.snowflake.com/en/blog/gen-ai-early-adopters-report/}

[219] ``State of RAG and GenAI,'' Squirro, 2024. [Online]. Available: \url{https://squirro.com/squirro-blog/state-of-rag-genai}

[220] ``BEIR: A Heterogeneous Benchmark for Zero-shot Evaluation of Information Retrieval Models,'' Papers with Code, 2024. [Online]. Available: \url{https://paperswithcode.com/dataset/beir}

[221] ``RAG Evaluation: Best Practices and Methodologies,'' ORQ AI, 2024. [Online]. Available: \url{https://orq.ai/blog/rag-evaluation}

[222] ``Constitutional AI: Harmlessness from AI Feedback,'' Anthropic, 2023. [Online]. Available: \url{https://www.anthropic.com/news/constitutional-ai-harmlessness-from-ai-feedback}

[223] ``Claude's Constitution,'' Anthropic, 2023. [Online]. Available: \url{https://www.anthropic.com/news/claudes-constitution}

[224] W. Jiang et al., ``RAGO: Systematic Performance Optimization for Retrieval-Augmented Generation Serving,'' arXiv preprint arXiv:2503.14649, 2025. [Online]. Available: \url{https://arxiv.org/abs/2503.14649}

[225] ``Enterprise AI Trends 2024,'' SoftKraft, 2024. [Online]. Available: \url{https://www.softkraft.co/enterprise-ai-trends/}

[226] ``Best Enterprise RAG Platforms 2025,'' Firecrawl, 2025. [Online]. Available: \url{https://www.firecrawl.dev/blog/best-enterprise-rag-platforms-2025}

[227] ``Open Source LLMs Have Higher ROI for Enterprise GenAI,'' Prolego, 2024. [Online]. Available: \url{https://prolego.com/open-source-llms-have-higher-roi-for-enterprise-genai}

[228] ``10 Enterprise AI Stats to Know in 2024,'' Skim AI, 2024. [Online]. Available: \url{https://skimai.com/10-enterprise-ai-stats-to-know-in-2024/}

[229] ``Enterprise AI Implementation Framework,'' AIRC Conference, 2024. [Online]. Available: \url{https://aircconline.com/csit/papers/vol15/csit150301.pdf}

[230] ``Differentiable Data Rewards for RAG,'' arXiv:2503.08398, 2025. [Online]. Available: \url{https://arxiv.org/html/2503.08398v1}

[231] ``Collective Constitutional AI: Aligning a Language Model with Public Input,'' Anthropic, 2023. [Online]. Available: \url{https://www.anthropic.com/research/collective-constitutional-ai-aligning-a-language-model-with-public-input}

[232] W. Zhang et al., ``RAISE: Enhancing Scientific Reasoning in LLMs via Step-by-Step Retrieval,'' arXiv preprint arXiv:2506.08625, 2025. [Online]. Available: \url{https://arxiv.org/html/2506.08625}

[233] H. Huang et al., ``Don't Do RAG: When Cache-Augmented Generation is All You Need for Knowledge Tasks,'' arXiv preprint arXiv:2412.15605, 2025. [Online]. Available: \url{https://arxiv.org/html/2412.15605v2}

[234] ``Self-RAG: Learning to Retrieve, Generate, and Critique,'' Learn Prompting, 2024. [Online]. Available: \url{https://learnprompting.org/docs/retrieval_augmented_generation/self-rag}

[235] ``2024: The State of Generative AI in the Enterprise,'' Menlo Ventures, 2024. [Online]. Available: \url{https://menlovc.com/2024-the-state-of-generative-ai-in-the-enterprise/}

\end{document}